\documentclass[reprint]{revtex4-1}
\usepackage{amsmath,amssymb,natbib,bm,graphicx,url,epsf,color}
\usepackage{epstopdf}
\usepackage{tikz}
\usepackage{colortbl}
\usepackage{amssymb,amstext,amscd,amsthm,amsopn,amsmath,mathtools,amsbsy,mathrsfs}

\newcommand{\Cal}[1]{{\cal #1}}
\newcommand{\A}[1]{A^{\vphantom{\dagger}}_{#1}\hspace*{-0.5mm}}
\newcommand{\conf}{\mathscr C}
\newcommand{\drs}{{\rm dr}}
\newcommand{\lin}{_{\rm in}}
\newcommand{\lout}{_{\rm out}}
\newcommand{\ii}{\mathfrak{i}}
\newcommand{\OT}{\Cal O_\ind{T}}
\newcommand{\ZFS}{\ket{0}_\drs}
\newcommand{\citeOR}[1]{{\bf \Huge\point{\cite{#1}}}}
\renewcommand{\citeOR}[1]{\cite{#1}}
\newcommand{\qL}{{\fr q}}
\newcommand{\pL}{{\fr p}}
\newcommand{\barV}{{\overline V}}
\newcommand{\barT}{{\overline T}}
\renewcommand{\AA}{{ A}}
\renewcommand{\SS}[1]{{ \AA_{#1}}}
\renewcommand{\SS}[1]{{ \AA_{s^{#1}}  }}
\newcommand{\shS}[2]{\hspace*{-#2 mm}#1\hspace*{-#2 mm}}
\newcommand{\nst}{{m_\alpha}}
\newcommand{\Navo}{{\Cal N_A\vph}}
\newcommand{\vph}{^{\vphantom{\dagger}}}
\newcommand{\Hilb}{\mathscr{H}}
\newcommand{\XXX}[1]{}
\newcommand{\be}{\begin{equation}}
\newcommand{\ee}{\end{equation}}
\newcommand{\bea}{\begin{eqnarray}}
\newcommand{\eea}{\end{eqnarray}}
\newcommand{\fr}[1]{\mathfrak{#1}}
\newcommand{\p}{\partial}
\newcommand{\ind}[1]{{\mbox{\scriptsize #1}}}
\newcommand{\TB}{T_\B}
\newcommand{\B}{{\textsc{b}}}

\newcommand{\f}{{\textsc{f}}}
\newcommand{\ket}[1]{\left|#1 \right \rangle}
\newcommand{\too}{\longrightarrow}
\newcommand{\oto}{\leftrightarrow}
\newcommand{\kb}{k_\B}
\newcommand{\nl}{\nonumber\\}
\newcommand{\ens}[2]{\left\{#1\right\}_{#2}}
\newcommand{\Ket}[1]{|#1 \rangle}
\newcommand{\RR}{\mathbb{R}}
\newcommand{\VM}[1]{\big\langle#1\big\rangle}
\newcommand{\Vm}[1]{\langle#1\rangle}
\newcommand{\vm}[1]{\left\langle#1 \right \rangle}
\newcommand{\ZZ}{\mathbb{Z}}
\newcommand{\dr}[1]{\mbox{#1}}
\newcommand{\Ccurl}{\mathscr{C}}
\renewcommand{\Im}{\fr{Im}}

\newcommand{\sgn}{\mbox{sign}}
\newcommand{\kernelem}{\Phi\vph}
\newcommand{\kernel}{K}

\graphicspath {{Figs/}{../}}

\begin{document}

\title{
Full exact solution of the out-of-equilibrium\\ boundary sine Gordon model 
}
\author{Edouard Boulat}
\affiliation{Laboratoire Mat\'eriaux et Ph\'enom\`enes Quantiques, Universit\'e de Paris, CNRS-UMR 7162, Paris 75013, France.}

\begin{abstract}

The massless boundary sine-Gordon (SG) model is the only interacting impurity model 
with a  known exact solution
out-of-equilibrium, yet existing so far only for integer values of the sine Gordon coupling $\lambda$
[Phys. Rev. Lett. {\bf74}, 3005 (1995)].
We present here a full exact solution for arbitrary rational values of $\lambda$, at arbitrary voltage $V$ and temperature $T$.
We use the ``string" solutions of the bulk SG model, here regarded as genuine 
quasiparticles avoiding charge diffusion in momentum space. We carefully present the finite voltage and temperature
thermodynamics of this gas of interacting 
exotic quasiparticles, whose very nature depends on subtle
arithmetic properties of the rational SG parameter $\lambda$,
and explicitly check that the string representation is thermodynamically complete.
By considering a Loschmidt echo,
we  derive    the  exact transmission probability of  
strings on the impurity. 
We obtain the exact universal scaling function for the electrical current 
$I(V,T)$.
Our results are in excellent 
agreement with recent experimental out-of-equilibrium data and question
the reality of these exotic quasiparticles.
\end{abstract}

\maketitle

\section{Introduction}

Large assemblies of interacting entities can be  the siege of emergence, i.e. manifestation of more complex ``entities" with new features that were absent at the elementary level.
Restricting this  
vastly fertile concept of emergence to the study of  its  manifestations 
when elementary entities are the microscopic degrees of freedom of inert matter (electrons, atoms...)  is
certainly at the heart of condensed
matter physics, as expressed 
by the famous statement  by P.W. Anderson: ``More is different"\citeOR{MoreIsDifferent}.
In condensed matter 
a typical scale for ``more" is the number of involved particles $N=\Cal O( \Navo)$ with $\Navo\simeq 6\times10^{23}$
 the Avogadro number, or, best conveying the idea of the complexity in a quantum interacting system, the number of states given by the dimension of the many-body Hilbert space $\Hilb$:
 $\ln(\mbox{dim}\Hilb)=\Cal O(\Navo)$.
Understanding such correlated systems
defines 
notoriously-hard-to-solve 
``quantum many-body problem" (QMBP).
The other side of the coin of the QMBP is the 
vast phenomenology of the possible collective arrangement of
the microscopic degrees of freedom, leading to the large collection of observed, as well as yet-to-be-discovered, states of matter.

The new, emergent properties of such states of matter can be ultimately traced back to the
very nature of the low energy states of the system, which, very often, can be described
as collections of  ``quasiparticles" (QPs) emerging, as the result of interactions, as genuinely \emph{new} elementary excitations,
possessing new features like new quantum numbers\citeOR{Woelfle08}.

Historically the earliest occurence of QPs in the QMBP stems from the study of $^3\mbox{He}$, a strongly interacting liquid of fermions.
The theory that has been developed to describe this situation, the Landau-Fermi liquid theory\citeOR{FermiLiquids},
is also the paradigm for the description of ordinary metals:
even in the case where electron-electron interactions are strong,
low energy states in a metal are described in terms of emergent QPs, the quasi-electrons,
that are (weakly) interacting objects still carrying charge 
 $e$ but with  renormalized mass   $m^*$. 
Due to certain interactions however, this metal can turn into a BCS superconductor\citeOR{BCS}:  there the Fermi-liquid picture breaks down, but the concept of QP is still relevant: the groundstate can be described as a condensate of charge $e^*=2e$ Cooper pairs, and the QPs describing excitations above it
have no definite charge\citeOR{BookFetterWalecka}.

Interactions can shape in an even more dramatic way the 
nature of QPs:  
a striking example
is provided by Tomonaga-L\"uttinger liquids (TLL),  that replace the paradigm of Landau-Fermi 
 liquid for massless electrons
in one dimension with short-range interactions\citeOR{Luttinger63,BookGiamarchi}:
there, the notion of quasi-electron becomes ineffectual, QPs exhibit spin-charge separation \citeOR{Auslaender02,Jompol09},
and carry fractional charge -- in clean realization of chiral TLL in edge states of the $\nu=\frac 13$ 
Fractional Quantum Hall state, charge $e^*=\frac e3$ QPs could be observed by noise measurement in the tunnelling current
 \citeOR{Saminadayar1997,dePicciotto1997}. This example emphasizes that tunnelling experiments constitute a central
 tool to probe the nature of the QPs.

On the theoretical side 
it is important to keep in mind that 
the relationship between the original microscopic
entities and the emergent QPs is of many-body nature: due to interactions it consists 
in a change of basis in 
$\Hilb$ (roughly speaking, a $e^{\Cal N_A\vph}\times e^{\Cal N_A\vph}$ unitary matrix),
making it extremely complex and most of the time out of reach of any exact description. On top of this,
a second major difficulty in the theoretical approach stems from the fact 
that spectroscopic experiments  probing the QPs are typically carried on in non-equilibrium situations, 
whereas the available analytical tools for tackling
strong correlations are designed for equilibrium situations.
While on the one hand we still clearly lack a general efficient theoretical  framework for treating in a controlled way
both strong interactions and out-of-equilibrium physics, on the other hand such conditions are routinely
produced, controlled and measured with high precision in milli-Kelvin experiments, making any progress in the theoretical effort a valuable one.

With respect to the first difficulty, integrable systems constitute a remarkable exception:
they are interacting many-body systems
where a rich underlying structure (they enjoy an infinite number of conservation laws)
allows for an exact
description of the emergent QPs, in spite of strong interactions
-- see e.g. 
the spin-charge separation
phenomenon in the 1D Hubbard model\citeOR{BookEsslerHubbard}. 
As ideal 1D systems 
with no dissipation mechanism, although
integrable models can sometimes  apply to the description of 
realistic experimental situations, 
they cannot address the non-equilibrium regime where dissipation is at play --
they can be viewed as sophisticated interacting generalizations of the ideal perfect gas in 1D.
The situation at hand is however different for quantum \emph{impurity} integrable
systems, describing homogeneous, free systems,
except at one point in space where interaction is concentrated.
Since dissipation in a realistic system typically occurs far from the impurity,
they can be directly relevant to the quantitive description 
of coherent transport.
Therefore, QIIS are unique systems where one can hope
to capture exactly the physics of strong correlations in a non-equilibrium context.

Considerable  
efforts have been devoted to the description of quantum impurities in out-of-equilibrium situations
by numerical approaches implementing elaborate  approximate solutions of the QMBP --
 by using  e.g. functional renormalization group (RG) \citeOR{FRG-Review,Meden08},  real-time RG \citeOR{andergassen10},  time dependent  density matrix RG or time dependent numerical RG \citeOR{Schmitteckert14} or fermionic representation for transport through TLL \citeOR{Aristov14}.
 Yet, numerical approaches ultimately rely on some sort of approximation to tackle the QMBP. 
Moreover, those methods
are usually designed to compute specific physical quantities, so that the access
to the \emph{nature} of the QPs -- which we believe is part of the elucidation of the physics -- 
either requires their \emph{a priori} knowledge
or remains out of reach.

The boundary sine Gordon (BSG) model  is an archetypical 
QIIS
which, in several respects, plays a central role. 
First, it has a wide range of applications in condensed matter physics, and even beyond.
This variety of realizations 
has its root 
in the \emph{minimal} character 
of this model, which can be considered  as the simplest
non-linear impurity model: it describes a free boson $\phi(x)$ interacting via a 
 cosine potential  $\cos(\beta\phi(x\!\!=\!\!0))$. Introducing  the SG parameter:
\be
\lambda\equiv \frac{8\pi}{\beta^2}-1
\label{defLambda}
\ee 
the BSG describes, for example:
$(i)$ transport through an impurity in a 1D conducting wire  with Tomonaga-L\"uttinger parameter 
$K=\frac1{\lambda+1}$\citeOR{KaneFisher92} ;
$(ii)$  electron tunnelling between edge states of the Fractional Quantum Hall Effect (FQHE) at filling 
$\nu=\frac1{\lambda+1}$ 
with $\nu^{-1}$ an integer \citeOR{FLS-PRL,FLS-PRB} ;
$(iii)$ low-energy transport through a quantum coherent conductor coupled to an electromagnetic environnement 
with low energy 
impedance 
$Z=\lambda \frac h{e^2}$\citeOR{SafiSaleur} ; 
$(iv)$ quantum Brownian motion in a cosine potential,
 the out-of-equilibrium drive being a global tilt of the potential\citeOR{Schmid83,Fisher85,Guinea85,BookWeiss}.
 It also 
 appears in the  FQHE at more exotic fillings\citeOR{Kane95,Chang03}, 
in arrays of Josephson junctions\citeOR{Lukyanov2007},
 and even 
in high-energy physics in the context of string theory\citeOR{Callan90,Sen02}.

\begin{figure}[h]
\renewcommand{\figurename}{\textbf{Figure}}
\renewcommand{\thefigure}{\textbf{\arabic{figure}}}
\includegraphics[width=8.5cm]{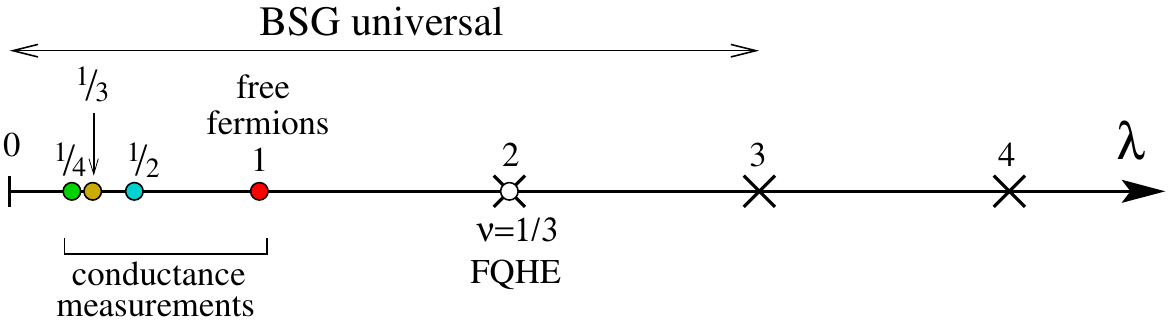}
\caption{
Out-of-equilibrium investigations of the BSG model.
  The BSG model describes the (low energy) universal features (i.e. independent of the microscopic details) of realistic situations when $\lambda\leq 3$.
  Crosses indicate the \emph{diagonal} BSG models $\lambda\in\mathbb{N}$ which were up to now the only out-of-equilibrium solvable points\cite{FLS-PRL} . 
 The present work 
addresses the off-diagonal case $\lambda\notin\mathbb{N}$.
 Coloured circles indicate values for which transport experimental data\cite{Anthore18} is available.
We mention that there are also available transport measurements at $\lambda=2$ in the $\nu=1/3$ FQHE 
\cite{Saminadayar1997}.
}
\label{figoverviewBSG}
\end{figure}

Second, is a remarkable fact that as of today, excluding  interacting systems that are 
unitarily related  to  free systems\citeOR{SchillerHershfieldFreeSystems,
KomnikGogolin03,SelaAffleck09},
the \emph{only} genuinely interacting system 
for which there is an exact solution  yielding explicitly the 
out-of-equilibrium universal scaling functions say for the current, 
  is the \emph{diagonal} BSG model 
\citeOR{FLS-PRL,FLS-PRB,Bazhanov99,NoteSDIRLM,Boulat08}
defined by very specific values of the SG parameter ($\lambda$ an integer) 
where the system enjoys additional symmetries.
More than 20 years after this breakthrough, the general solution of the out-of-equilibrium BSG away from diagonal points (defining the \emph{off-diagonal} regime), 
is still missing (see Fig.\ref{figoverviewBSG}).

\begin{figure}[h]
\renewcommand{\figurename}{\textbf{Figure}}
\renewcommand{\thefigure}{\textbf{\arabic{figure}}}
\includegraphics[width=8.1cm]{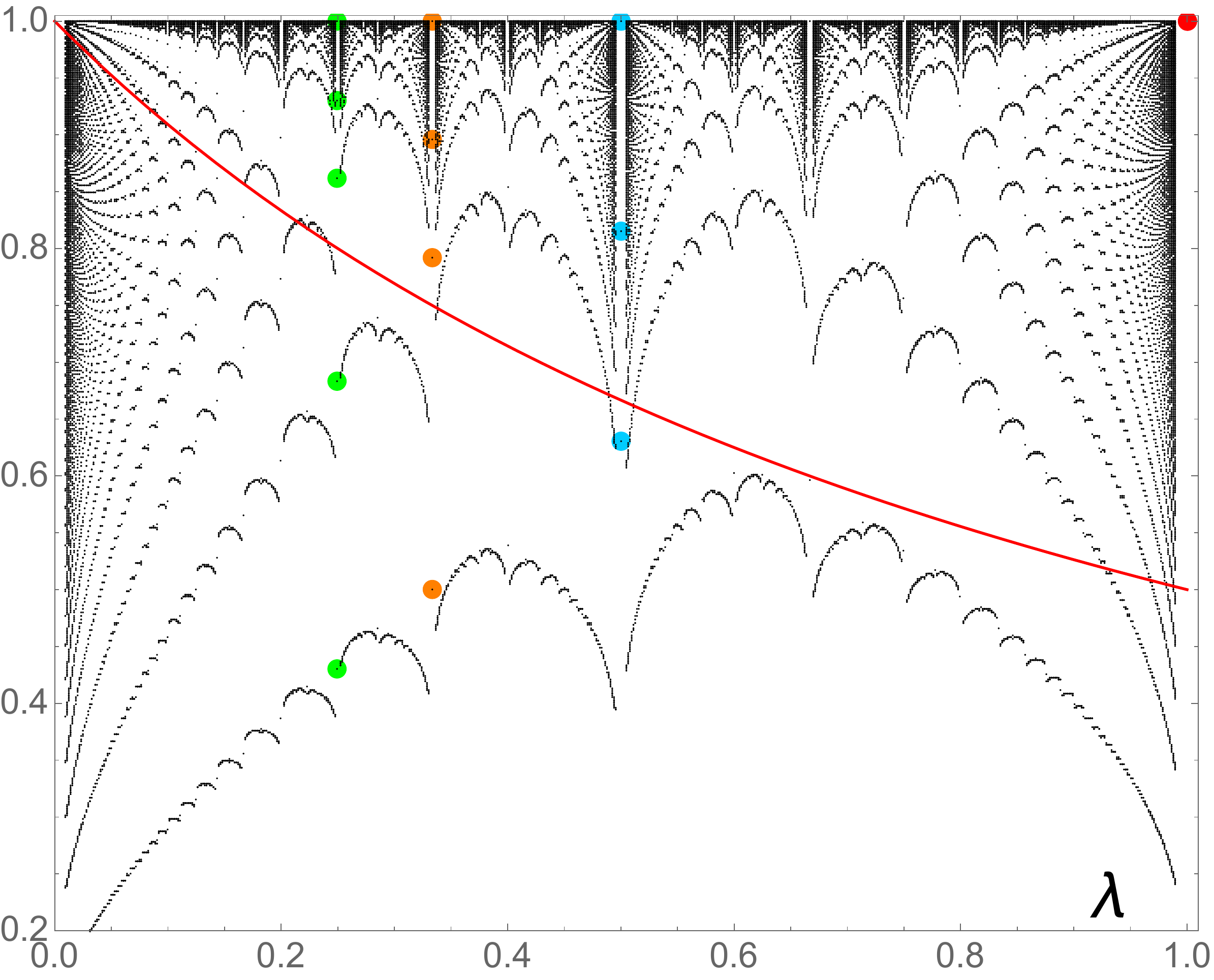}
\caption{
{\bf[Black]} 
Normalized cumulative plot ($x_a=\frac{N_a}{\sum_{b} N_b}$) of the number $N_a(V,T,\lambda)$ of quasiparticles of type ``$a$" involved in the exact description (shown here at vanishing bias voltage) as a function
of the rational SG parameter $\lambda=\frac \pL \qL \leq1$ with $\qL<100$. 
Note the very complex, self-similar structure that develops.
Colored bullets emphasize the data at values  $\lambda=\frac12,\frac13,\frac14$ where our
theory successfully accounts for the experimental  data of Ref.\cite{Anthore18}, see section \ref{sectionDiscussion}. 
In spite of this complicated structure, interactions and the charge $\qL$ of the quasiparticles
combine to yield  perfectly regular transport properties when $\lambda$ is varied, e.g.  
the large temperature conductance $\frac {h}{e^2}\,G_{\rm max}=\frac1{\lambda+1}$
of the BSG model
 {\bf [Red},  same dimensionless scale{\bf ]}. 
}
\label{figCumulCond}
\end{figure}

In this paper, motivated by recent high-precision measurements of the transport properties
of chiral TLL \citeOR{Anthore18},
 we present  an exact 
solution  when $\lambda$ is an arbitrary rational number giving access
to the exact universal scaling function for the conductance at arbitrary
voltage and temperature. Our solution proceeds in exploiting an equivalence
(see Fig.\ref{figGASES})
between the original gas of QPs with off-diagonal scattering (that leads to charge diffusion in momentum space),
and a gas 
involving exotic QPs diagonalizing charge transport, namely the string solutions 
of the XXZ model\citeOR{Takahashi72,TakahashiBook}
whose \emph{number, nature} and \emph{charge} depend in a  subtle way on the BSG parameter
$\beta$ (see Fig. \ref{figCumulCond}): for example the charge of the QPs is an \emph{everywhere discontinuous} function of $\beta$.

This article, in addition to exposing the exact solution to the out-of-equilibrium BSG model,
aims at  giving the opportunity to the non-specialist reader to grasp the spirit of the solution in spite of complicated technicalities.
The paper is organized as follows:
in Section \ref{sectionResults}, we sketch the main features
of the solution and give the main results, deferring all the technical details 
to the subsequent sections.
In Section \ref{sectionBulk}, after motivating their origin, we present the strings 
and  spend some time
gathering known results to present the (bare and dressed) basis of QPs and their thermodynamical properties.
In Section \ref{SectionImpurity} we establish
the main features of the impurity scattering in the string basis and we derive exactly the QP transmission probability
using the boundary Yang Baxter equation and a Loschmidt echo. Finally in Section \ref{sectionDiscussion} we discuss our solution and present
its remarkable agreement 
with recent experimental data
for the tunnelling in a resistive environnement.

\section{Description of main results}
\label{sectionResults}

\subsection{The model, obstacles and route towards its solution}

 The boundary sine Gordon (BSG) model is defined by its Hamiltonian:
\bea
&H_\ind{BSG}=H_0+H_\B
\label{HBSG}\\
&H_0=\hbar v_\f\int_{-\infty}^\infty dx(\p_x\phi)^2
\,;\;\; H_\B=\gamma \cos(\beta\phi(0))
\nonumber
\eea
It describes a free, chiral (right-moving) one-dimensional boson $\phi(x-t)$ defined on the whole line $x\in]-\infty,+\infty[$,  
the non-linear SG interaction $H_\B$ acting as an impurity at $x=0$. The model is forced out-of-equilibrium by a constant voltage $\frac{eV}2\int dx\, Q(x)$ and we choose to normalize the charge density as 
\be
Q(x)=\frac{\beta}{2\pi}\,\p_x\phi,
\label{chargedef}
\ee
ensuring that the fundamental soliton of the SG model carries charge unity\citeOR{NoteChargeNormalization}. 
The parameter $v_\f$ sets the velocity scale of the problem.
The parameter $\beta$, that fixes the period of the impurity potential, also determines the scaling dimension
$\frac{\beta^2}{8\pi}=\frac1{\lambda+1}$
of the impurity perturbing operator, the later being relevant when $\beta<\sqrt{8\pi}$ or $\lambda>0$. 
\newcommand{\NoteChargeNormalization}{
The charge $q_s$ of soliton, expressed in units of the electron charge $e$,
will of course depend on the precise situation modeled by the BSG model.
The general case $q_s\neq 1$ is obtained  by  rescaling $V\to q_s V$.
}
The strength $\gamma$
of the impurity coupling generates a typical energy scale, the ``impurity temperature"
$\TB\propto \gamma^{\frac{\lambda+1}{\lambda}}$
 that encapsulates
all the microscopic, non-universal details of the problem (coupling strengths of possible additional irrelevant operators, e.g. band curvature, high-energy cut-off...) in a realistic situation. Note that when $\beta<\sqrt{2\pi}$, 
the operator $\cos(2\beta\phi(0))$ becomes relevant ; moreover, as being essentially the square of $H_\B$, it is allowed by symmetries: therefore it is present, generates a new scale $\TB^{(2)}$ and universality is lost, defining the universal regime $\lambda\in[0,3]$.

It is clear that the BSG model (\ref{HBSG}) defines a scattering problem: given an arbitrary many-body state $\ket{\psi}_\ind{in}$, incoming  on the far left of the impurity,
what is the final, outgoing state $\ket{\psi}_\ind{out}$ on the far right?
The central object encoding the physics is the scattering matrix $\Cal R$, defined by $\ket{\psi}_\ind{out}=\Cal R\ket{\psi}_\ind{in}$.
The scattering matrix $\Cal R$ is a many-body object, and 
the non-linear character of $H_\B$ 
 forbids any simple description of the scattering in the basis of elementary excitations of the free boson $\phi(x)$
 (see left column of Table \ref{TableBA}).
  
\definecolor{lightgray}{gray}{0.8}
\begin{table}
\newcommand{\spLoc}{\hspace*{0.cm}}
\vspace*{.85cm}
$\color{red}
\hspace*{1.6cm}
\overset{\rm{\bf B.A.}}{\boldsymbol{\Longrightarrow}}\hspace*{1.54cm}
\overset{\rm{\bf A.B.A.}}{\boldsymbol{\Longrightarrow}}
\color{black}$
\vspace*{-.85cm}\\
\begin{tabular}{|c||c|c|c|}
\hline
Basis 
&\spLoc\begin{tabular}{c}free boson \\  $\phi(x)$\end{tabular}\spLoc
& \spLoc\begin{tabular}{c}(anti)solitons \\  $\SS{\pm}(\theta)$ \end{tabular}
\spLoc
&\spLoc \begin{tabular}{c}solitons/strings \\   $\AA_a(\theta)$ \end{tabular}\spLoc
\\\hline\hline
\begin{tabular}{c}Bulk \vspace*{-.1cm}\\ scattering  \end{tabular}
& trivial &
 \cellcolor{lightgray}
 \begin{tabular}{c}  factorized\vspace*{-.1cm}
\\
\cellcolor{lightgray}off-diagonal \end{tabular} \cellcolor{lightgray}
&
\begin{tabular}{c} factorized\vspace*{-.1cm} \\{\bf diagonal}  \end{tabular}
\\\hline
\begin{tabular}{c} Impurity \vspace*{-.1cm}\\scattering  \end{tabular}
& 
 \cellcolor{lightgray}
many-body & factorized &  {\bf factorized}  
\\\hline
\end{tabular}
\caption{Three different many-body basis for representing the BSG model. Gray boxes highlight an obstruction to the  exact solution. The last column allows for an exact solution of the finite temperature BSG model out-of-equilibrium for arbitrary rational SG parameter $\lambda$, whereas the central one suffices for the diagonal case $\lambda$ integer.}
\label{TableBA}
\end{table}

The central idea to solve
the problem
  is to identify a basis 
for many-body states built on QPs that diagonalizes 
the impurity scattering matrix $\Cal R$.
This will be  done 
by exploiting the integrability of the BSG model\citeOR{Ghoshal94} to identify QP modes\citeOR{Zamolodchikov79},
with specifically the following properties: 
$(i)$ 
the thermal  gas of bosons (i.e. the finite temperature and voltage density matrix) incoming towards the impurity, 
can be represented in the QP basis ;
$(ii)$ the QPs interact but the resulting scattering between QPs is \emph{factorized}, with no particle production
; $(iii)$ the scattering of QPs on the impurity is \emph{factorized}, with no particle production.

Introducing the symbol $\AA_a(\theta)$ to represent a QP mode with quantum number $a$, where the rapidity $\theta$ parametrizes the momentum $p\propto e^\theta$,
 point $(i)$ means that a many-body basis for the many-body Hilbert space $\Hilb$ of bosons is made of Fock states:
 $\ket{\psi}=\prod_{i}\AA_{a_i}(\theta_i)\ket{0}$ 
where $\ket{0}$ is the many-body vacuum and that
 the states $\ket{\psi}$
are suited for implementing the Thermodynamical Bethe Ansatz (TBA)\citeOR{Zamolodchikov90} allowing for a finite temperature $T$ and finite voltage $V$ description of the interacting QP gas.
Point $(ii)$ means that the interaction amongst arbitrary colliding many-body states can be decomposed into elementary
two-QP scattering processes 
$\AA_{a_1}(\theta_1)\AA_{a_2}(\theta_2)\too \AA_{b_2}(\theta_2)\AA_{b_1}(\theta_1) $ 
 the amplitude of this process being given
 by the bulk scattering matrix 
 $S_{a_1,a_2}^{b_1,b_2}(\theta_1-\theta_2)$.
Third $(iii)$, factorization of the impurity scattering
means that the latter can be described by a \emph{one-body} scattering matrix $R_{ab}$, 
in the sense that
$\Cal R\ket{\psi}=\sum_{b_1,b_2,...}\prod_{i}R_{a_i,b_i}(\theta_i)\AA_{b_i}(\theta_i)\ket{0}$, i.e. the QPs scatter one by one on the impurity with no QP production.

In the case of the BSG, the QPs fulfilling points $(\!ii\!\!-\!\!iii)$ have been identified by Goshal and Zamolodchikov \citeOR{Ghoshal94} 
as the solitons and antisolitons $\SS{\varepsilon}(\theta)$ of the SG model\citeOR{Zamolodchikov79} where  
$\varepsilon=\pm$ is a charge quantum number (center column of table \ref{TableBA}). This QP basis  has lead to the solution\citeOR{FLS-PRB,FLS-PRL} of the \emph{diagonal} BSG model ($\lambda\in\mathbb{N}$) out-of-equilibrium.
For those exceptional values, the scattering amongst the QPs is  diagonal or ``reflectionless" (the process $\SS{+}(\theta_1)\SS{-}(\theta_2)\too \SS{+}(\theta_2)\SS{-}(\theta_1)$ is forbidden) ensuring that  point $(i)$ is satisfied.

\begin{figure}[h]
\begin{center}
\includegraphics[width=5.5cm]{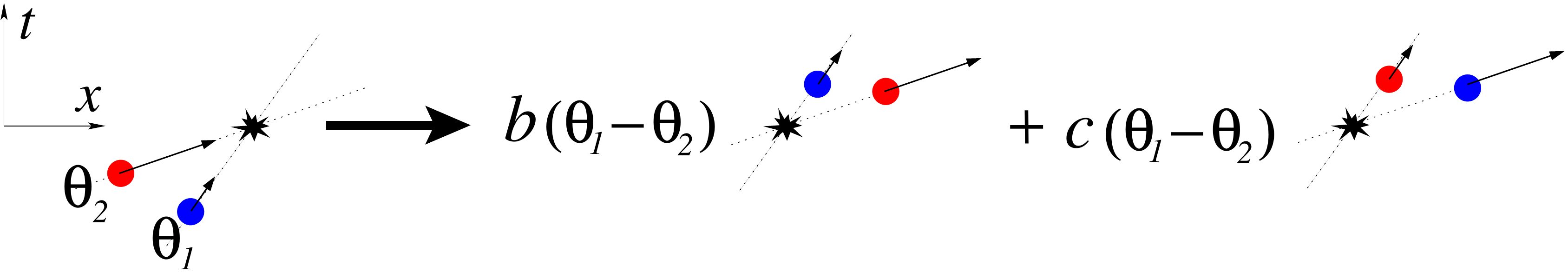}
\hspace*{-3mm}\raisebox{3mm}{OR:\;}
\raisebox{1mm}{\includegraphics[width=2.5cm]{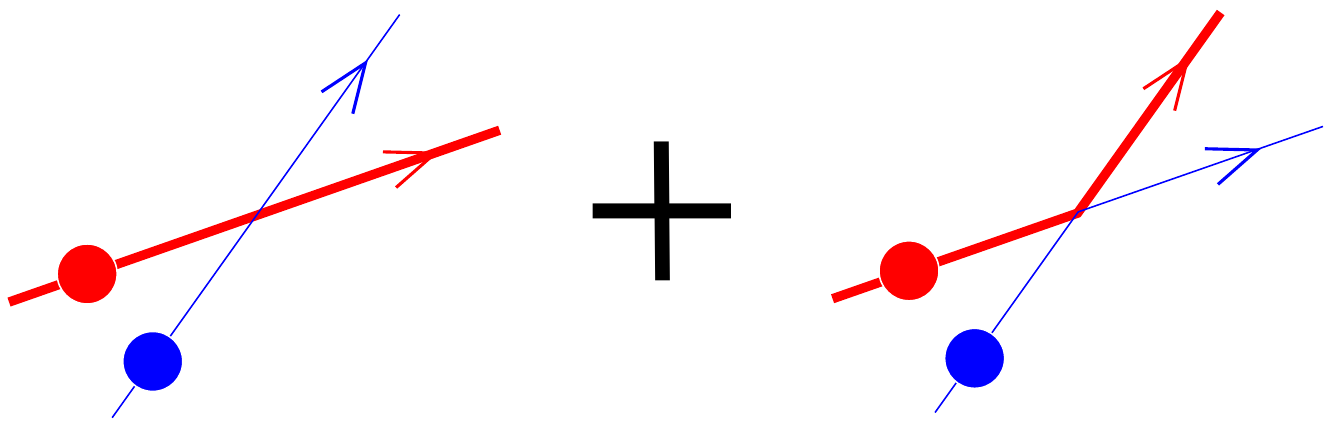}}
\vspace*{3mm}
\includegraphics[width=8.5cm]{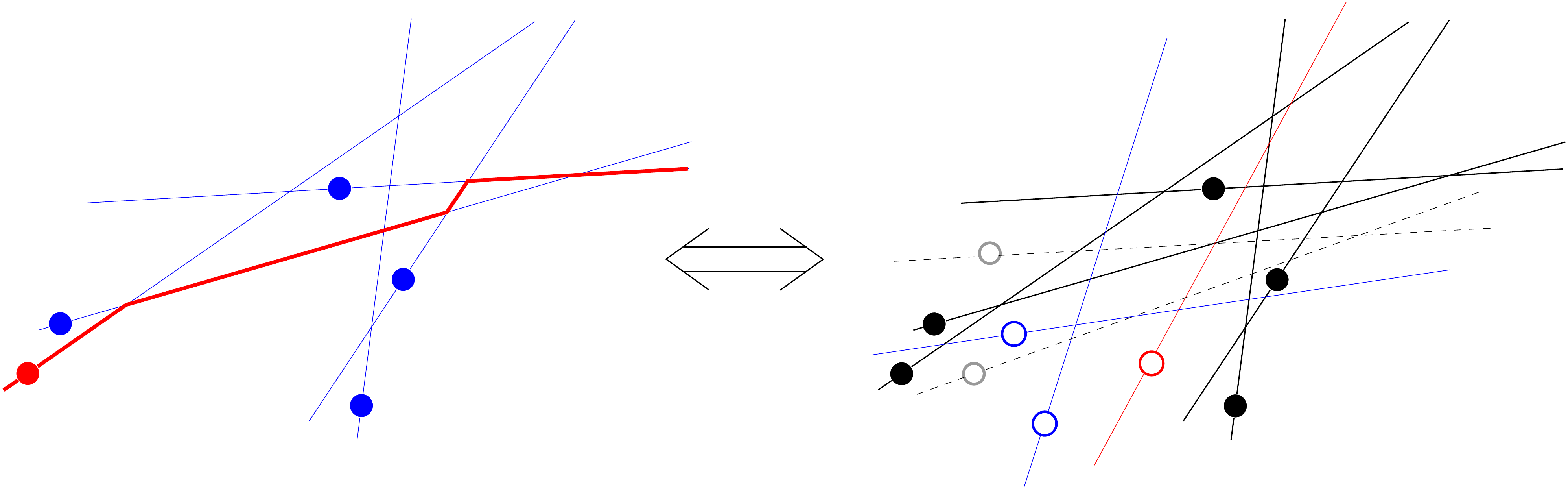}
\end{center}
\caption{
{\bf[Top]}
Illustration  in the ($x,t$) plane of the off-diagonal scattering
of solitons/antisolitons ({{\protect\includegraphics[width=2.5mm]{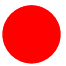}}}
\hspace*{-1.9mm} /\hspace*{-1mm}
{{\protect\includegraphics[width=2.5mm]{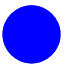}}})
$\SS{\pm}$
of the SG model with rational
parameter $\lambda=\frac \pL \qL $.
The trajectories have slopes $\propto e^\theta$ and in the short-hand notation [Right] the colour indicates the value of the charge quantum number $\varepsilon=\pm$.
$b$ and $c$ are scattering amplitudes.
{\bf[Bottom]} 
Equivalence between a gas [Left] of  
solitons/antisolitons with off-diagonal scattering
(note that only one of the many "diffusive" trajectories of the soliton
{{\protect\includegraphics[width=2.5mm]{SymbolSolPlus.pdf}}}
has been represented)
and a gas [Right] of new quasiparticles 
$\AA_a$
with \emph{diagonal} scattering, i.e. all quantum numbers $a$ (or colours) follow straight lines.
In this equivalence the 
original SG soliton/antisoliton becomes a \emph{neutral} soliton $\AA_s$ 
({{\protect\includegraphics[width=2.5mm]{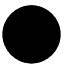}}})
carrying the energy, whereas new QPs emerge (open circles):
two strings $\AA_{c}^\pm$ carrying charge $\pm \qL$ 
({{\protect\includegraphics[width=2.5mm]{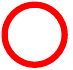}}}
\hspace*{-1.9mm} /\hspace*{-1mm}
{{\protect\includegraphics[width=2.5mm]{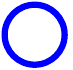}}}),
and additional neutral strings (for the case $\lambda=\frac13$ illustrated here, 
a single string $\AA_1$
({{\protect\includegraphics[width=2.5mm]{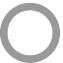}}}), see Appendix \ref{AppendixExampleLambda1over3} for details).
Whereas the total number $N_a$ of excited strings is fixed by the  thermodynamics of the gas, 
the equivalence preserves 
$N_{{\protect\includegraphics[width=1.8mm]{SymbolSolPlus.pdf}}}
+N_{{\protect\includegraphics[width=1.8mm]{SymbolSolMoins.pdf}}}
=N_{{\protect\includegraphics[width=1.8mm]{SymbolSolABA.pdf}}}
$
and 
$N_{{\protect\includegraphics[width=1.8mm]{SymbolSolPlus.pdf}}}
- N_{{\protect\includegraphics[width=1.8mm]{SymbolSolMoins.pdf}}}
=\qL \big(
N_{{\protect\includegraphics[width=1.8mm]{SymbolStringPlus.pdf}}}
-
N_{{\protect\includegraphics[width=1.8mm]{SymbolStringMoins.pdf}}}
\big)$.
}
\label{figGASES}
\end{figure}

However, in the generic case $\lambda\notin\mathbb{N}$
the scattering is \emph{off-diagonal}  meaning that  solitons and antisolitons can
exchange their quantum numbers during a scattering event, leading to  diffusion
of the soliton charge in momentum space (see Fig.\ref{figGASES} [Top] and [Bottom Left]). As a consequence the (anti)solitons do not fulfill point ($i$), and one faces a major difficulty: that of finding yet another many-body basis allowing for the finite $T,V$ description of the QP gas \emph{and} for the exact description of its scattering on the impurity.
The Algebraic Bethe Ansatz (ABA) technique \citeOR{Takahashi72,TakahashiBook,Fendley92,KorepinBook,FaddeevHouches} circumvents the issue of diffusion and furnishes a QP basis in which scattering becomes diagonal (third column of table \ref{TableBA}). 
ABA therefore yields an equivalence between  the original soliton/antisoliton gas with off-diagonal scattering, and a gas of new QPs with diagonal scattering, fulfilling point ($i$-$ii$)  (see Fig.\ref{figGASES}[Bottom Right]). We then show that this basis fulfils point ($iii$) and derive exactly the impurity scattering in this new basis to complete the toolbox for the exact out-of-equilibrium solution.

\subsection{Main results}

In order to solve the BSG model out-of-equilibrium,  two different kinds of data is needed:
first,  the QP content and the thermodynamics of those QPs representing the (voltage biased) 
free boson gas
incoming towards the impurity: this is reviewed in section \ref{sectionBulk} ; second,
the impurity scattering matrix $R_{ab}$ of those QPs ; our original derivation is presented  in section \ref{SectionImpurity}. 

The particular value $\lambda=1$ of the SG parameter (\ref{defLambda}), where the model can  mapped onto free fermions  \citeOR{KaneFisher92},
marks the boundary between the  repulsive ($\lambda<1$) and the attractive  ($\lambda>1$) regime
of the SG model. The attractive regime  is characterized by the appearance of neutral soliton-antisoliton
bound states (breathers), in the spectrum. In order to keep the presentation simple as possible, we focus here
on the repulsive 
case that retains the full richness of the solution, 
while deferring  the attractive case to Appendix \ref{AppendixAttractive}. 
We also illustrate what needs to be done in pratice to obtain the current $I(V,T)$ in two specific examples $\lambda=\frac13$ and $\lambda=\frac6{19}$ in Appendix \ref{AppendixExampleLambda}.

For arbitrary rational $\lambda$  one introduces the  continued fraction:
\be
\lambda =\frac \pL \qL = \frac1{\nu_1+\frac1{\nu_2+...+\frac1{\nu_\alpha}}},
\label{deffraccont}
\ee
where  $\nu_i$ are strictly positive integers ($\nu_\alpha>1$). The decomposition (\ref{deffraccont}) is unique \citeOR{ContinuedFractions}. Introducing the  integers 
\be
 m_0=0\;\;;\;\;m_i=\sum_{i'=1}^i\nu_{i'}\quad (i\leq \alpha),
\label{defmi}
\ee
there are  in total $\nst +1$ different kinds of QPs, whose creation operator
is denoted $\AA_a(\theta)$ where  the quantum number (or ``species'')  $a\in\{s,1,2,...,\nst\}$.

The first particle $\AA_s$ is a \emph{neutral} soliton, carrying kinetic energy, whereas the $\nst$ other particles are massless particles (carrying no energy) called strings.
All particles are neutral, except the last two ones $\AA_{\nst }\equiv \AA_c^{+}$ and $\AA_{\nst -1}\equiv \AA_c^{-}$,  carrying  $\pm \qL$  units of the SG-soliton charge.
The main features of the QP spectrum are summarized in Table \ref{TableQPs}.

\begin{table}[h]
\begin{tabular}{|c|c|c|c|c|}
\hline
\vphantom{\Big(}
particle&  symbol &  energy &  {\rm charge} & entropy
\\\hline\hline
\vphantom{\Big(}
 {\rm soliton}& $\AA_s\equiv \AA_0$ & \checkmark & 0 & \checkmark
 \\\hline
\renewcommand{\arraystretch}{0.4}
\begin{tabular}{c} neutral  \\ strings\end{tabular} &
 $ \renewcommand{\arraystretch}{0.0}
 \begin{array}{c}\AA_j\\ {\mbox {\scriptsize $(1\!\leq\! j\!< \!\nst \!\!-\!\!1)$}}
 \vspace*{.1cm}
 \end{array}$
 &  . & 
 0  & \checkmark
\\ \hline
\renewcommand{\arraystretch}{0.2}
\begin{tabular}{c} charged  \\ strings\end{tabular} &
$ \renewcommand{\arraystretch}{0.0}
 \begin{array}{l}\AA_c^-\equiv \AA_{\nst \!-1}\\ \AA_c^+\equiv \AA_{\nst }
 \vspace*{.1cm}
 \end{array}$
  & 
  $ \renewcommand{\arraystretch}{.8}
 \begin{array}{l}.\\ .
 \vspace*{.1cm}
 \end{array}$
   & 
  $ \renewcommand{\arraystretch}{0.8}
 \begin{array}{l}-\qL\\ +\qL 
 \vspace*{.1cm}
 \end{array}$
 & 
 $ \renewcommand{\arraystretch}{0.8}
 \begin{array}{l}  \checkmark \\ \checkmark 
 \vspace*{.1cm}
 \end{array}$
\\\hline
\end{tabular}
\caption{The dressed quasiparticle spectrum}
\label{TableQPs}
\end{table}

 The QPs 
can also be grouped in families $\big(\Cal F_i\big)_{i=1...\alpha}$ characterized by distinct scattering properties.
Each family  $\Cal F_i$ is made of the QPs $\big(\AA_j\big)_{m_{i-1}\leq j< m_i}$ (except for
the last $\Cal F_\alpha\!=\!\big(\AA_j\big)_{m_{\alpha-1}\leq j\leq \nst }$) and each QP $\AA_a\in\Cal F_i$  is  assigned a sign
 $\eta_a=(-1)^{i+1}$.

The QPs have diagonal scattering 
and the thermodynamics  of the QP gas  at finite $T,V$ --  i.e. precisely the density matrix describing 
our many-body system incoming towards the impurity -- 
  can be encoded 
by pseudoenergies $\epsilon_a(\theta)$, dimensionless functions that determine the total densities of QPs per unit length $\frac{\kb T}{h v_\f}P_a(\theta)$
with $P_a=\eta_a\p_\theta\epsilon_a$. 
The density of occupied particles per unit length is $\frac{\kb T}{h v_\f}\rho_a(\theta)$ with $\rho_a=P_a f_a$ that also expresses as $\rho_a=-\eta_a\p_\theta L_a$ with $L_a(\theta)=\ln(1+e^{\mu_a-\epsilon_a(\theta)})$,  and $f_a=\frac1{1+e^{\epsilon_a(\theta)-\mu_a}}$  a Fermi function ; in view of Table \ref{TableQPs} all (reduced) chemical potentials $\mu_a$ vanish except
$\mu_c^\pm=\pm \qL\frac{eV}{2\kb T}$. 

\begin{figure}[ht]
\renewcommand{\figurename}{\textbf{Figure}}
\renewcommand{\thefigure}{\textbf{\arabic{figure}}}
\begin{center}
(i)\hphantom{aaaaaaaaaaaaaaaaaaaaaaaaaaaaaaaaaaaaaaaaaaaaaaa}
\vspace*{-.5cm}
\raisebox{.1cm}{\includegraphics[width=8cm]{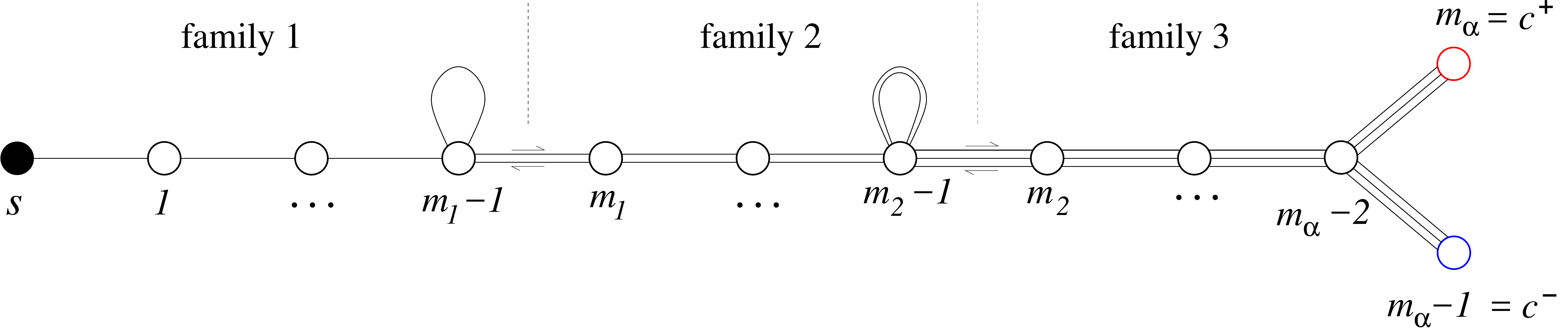}}
\\
\begin{tabular}{lll}
(ii.a) \quad&
\raisebox{-.4cm}{\includegraphics[width=1.5cm]{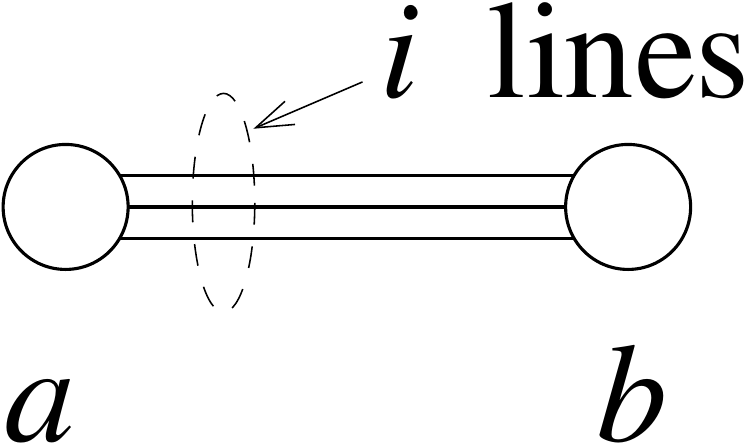}}
&
\hphantom{}$:\;\; \hat\kernel_{a,b}=\hat\kernelem_{p_i}\equiv \frac 1{2\cosh(p_i\frac{\pi \omega}2)}$
\vspace*{.3cm}
\\
(ii.b) &
\raisebox{-.4cm}{\includegraphics[width=1.5cm]{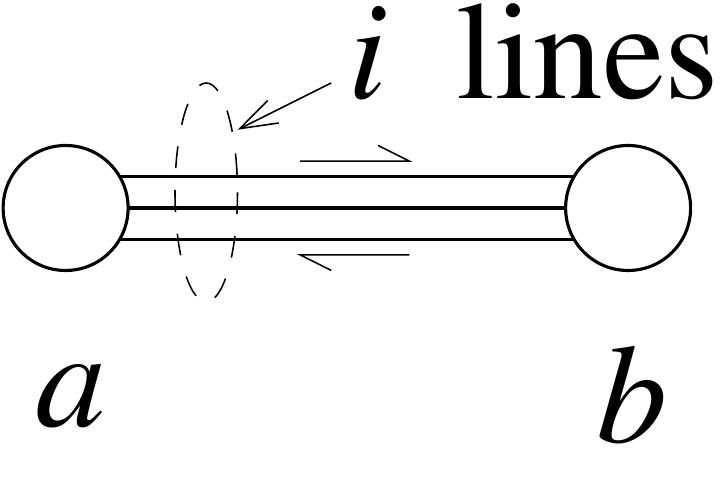}}
&
\vphantom{$\Big($}$:\;\; \hat\kernel_{b,a}=- \hat\kernel_{a,b}=\hat \kernelem_{p_i}$
\\
(ii.c) &
\hphantom{}\raisebox{-1.cm}{\includegraphics[width=1.8cm]{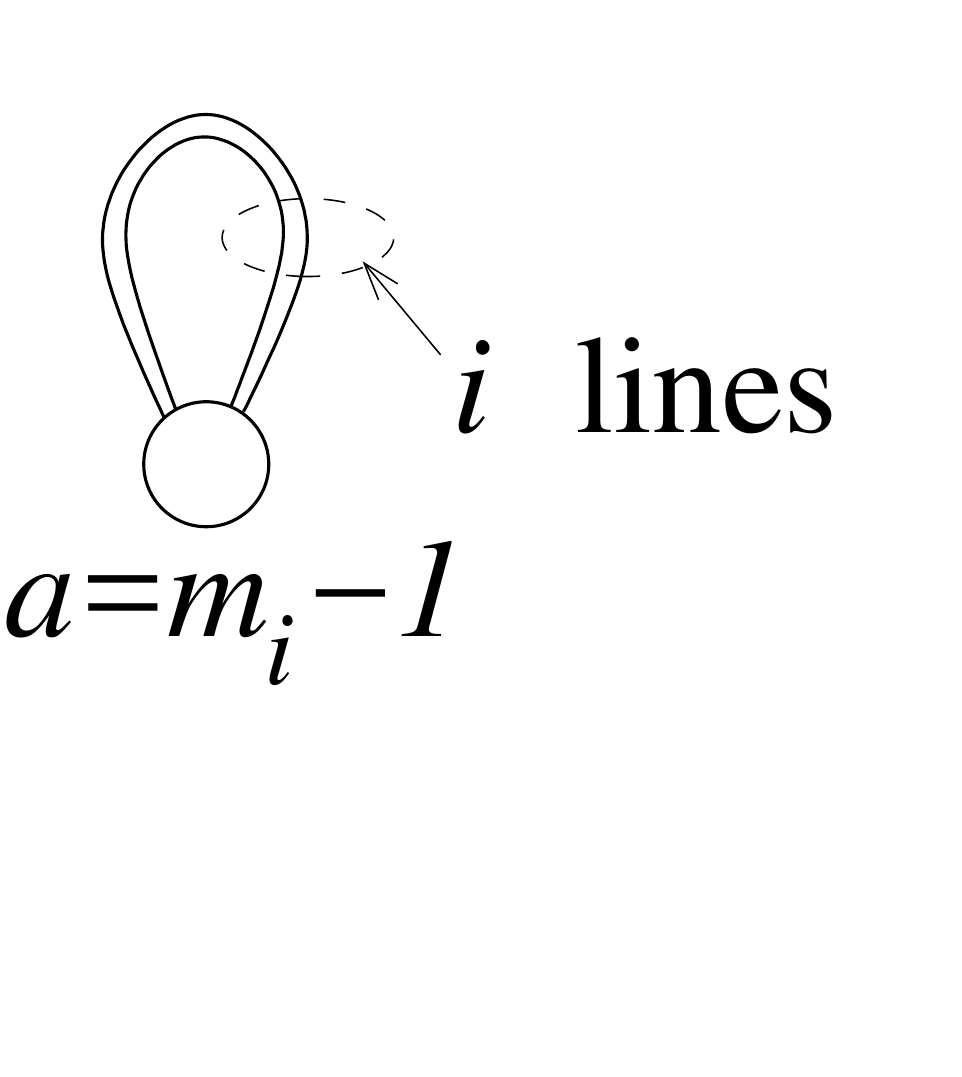}}
&
$:\;\;\hat\kernel_{a,a}
=\frac{\hat\kernelem_{p_i}\hat\kernelem_{p_{i+1}}}{\hat\kernelem_{p_i-p_{i+1}}}$
\end{tabular}
\vspace*{-1cm}
\end{center}
\caption{(i) Dynkin diagram for the kernel $\kernel_{ab}(\theta)$ entering the TBA equations, here depicted for $\alpha=3$, $(\nu_1,\nu_2,\nu_3)=(4,3,4)$ (i.e. 
$\lambda 
=\frac{13}{56}$). 
The nodes represent the different kinds of particles: the massive soliton (black node) and a collection of massless strings (open nodes). The lines connecting the nodes encode non-vanishing entries of the kernel $\kernel_{ab}$ made explicit in (ii). 
While most of the kernel is symmetric (ii.a), it also has non-symmetric entries (ii.b),
as well as diagonal entries (ii.c). 
}
\label{figDynkinGen}
\end{figure}

The pseudoenergies  obey a set of non-linear coupled integral equations, the TBA equations:
\be
\epsilon_a = \delta_{a,s}e^\theta - \frac1{2\pi}\kernel_{ba}\star L_b
\label{epsilonTBA}
\ee
where a sum over $b$ is implied, the convolution is defined as $(f\star g)(\theta)=\int_{-\infty}^\infty d\theta' f(\theta')g(\theta-\theta')$,  and the 
kernel entries $\kernel_{ab}(\theta)$ (see Fig.\ref{figDynkinGen}) are more conveniently given by their 
Fourier transforms
$\hat \kernel_{ab}(\omega)=\frac1{2\pi}\int_{-\infty}^\infty d\theta\,e^{-i\omega\theta}\kernel_{ab}(\theta)$ 
in terms of elementary functions: 
\be
\kernelem_{p_i}(\theta)=\frac{1/p_i}{\cosh(\theta/p_i)}
\quad\oto\quad
\hat\kernelem_{p_i}(\omega) = \frac1{2\cosh(\frac{\pi p_i\omega}2)}
\label{defphipi}
\ee
where the numbers $p_i$ are defined by:
\be
p_{0}=\lambda^{-1},\;\; p_1=1, \;\;  p\vph_{i>1}=p_{i-2}-\nu_{i-1} p_{i-1}.
\label{defpi}
\ee
 The 
symmetry of the kernel apparent in Fig.\ref{figDynkinGen} implies that the two charged QPs share the same pseudoenergy $ \epsilon_{c+}=\epsilon_{c^-}\equiv\epsilon_c$.

Our derivation  of the impurity scattering matrix $R_{ab}(\theta)$ for the QPs $\AA_a(\theta)$ (see Section \ref{SectionImpurity}), completes the description, and allows to establish  a universal formula for the current flowing through the impurity at arbitrary voltage and temperature, 
as a function of $\barT=\frac{T}{\TB}$ and $\barV=\frac{eV}{\kb\TB}$  (the boundary temperature $T_\B$ 
was introduced below Eq. (\ref{chargedef})) :
\bea
I(V,T)&=&\frac{(\qL e)\kb T}{h}\, \int_{-\infty} ^\infty \!\!\! d\theta \;\Cal T_\lambda(\theta)
 \big(
 \rho_c^+(\theta)-\rho_c^-(\theta)
 \big),
\label{CurrentTBA}\\
\Cal T_\lambda&=&   \frac{1}{1+\big(\barT e^\theta\big)^{-2\pL}}
\;\;;\;\;
\rho_c^\pm=\frac{(-1)^{\alpha+1}\frac{\p\epsilon_c(\theta)}{\p_\theta}}
{1+e^{\epsilon_c(\theta)\mp\qL\frac{\overline V}{2 \overline T}}},
\nonumber
\eea
where $\Cal T_\lambda(\theta)=|R_{c^\pm,c^\pm}(\theta)|^2$ 
is the transmission probability  of an incoming particle $\AA_{c}^\pm(\theta)$ through the impurity.

Although Eq. (\ref{CurrentTBA}), that solve the BSG model out-of-equilibrium by yielding exactly  the full universal scaling in the ($V,T$)-plane, look similar to those of the diagonal case $\lambda\in\mathbb{N}$ of Ref. \citeOR{FLS-PRB}, it  constitutes a novel prediction when $T\neq 0$. A first difference is that the involved QPs are now strings (and no longer solitons) carrying a non-trivial  charge $\pm \qL $. Second,
owing to our original derivation
of the boundary scattering of strings, 
the transmission probability $\Cal T_\lambda$ depends solely on $\pL$, the numerator  of $\lambda$.
Neither $\qL$ nor $\pL$ are smooth functions of $\lambda$, nor is the whole continued fraction representation (\ref{deffraccont}) of the SG parameter that fixes the QP content and its scattering. 
Therefore the structure of Equations (\ref{epsilonTBA},\ref{CurrentTBA})  for the pseudoenergies and the current is  highly irregular when $\lambda$ is varied in sharp
contrast with the behavior of physical quantities 
computed here that appear to be \emph{smooth} functions of $\lambda$ (see Fig.\ref{figCumulCond}). As the simplest illustration,  the 
expected high-energy  limit of the conductance $G(V,T\gg T_\B)=\frac{e^2}{h}\frac1{\lambda+1}$ appears to be the result
of a non-trivial calculation involving the complicated details of the QP spectrum
when starting from our Eq. (\ref{G0TBA}), see Appendix \ref{AppendixDeltaFc}.

At small voltage, the scaling function for the linear conductance $G_0(T)\equiv \frac{\p I(V,T)}{\p V}\Big|_{V=0}$
can be expressed as:
\be
G_0(T)=\frac{(\qL\: e)^2}{h}\,(-1)^{\alpha}\!\!\int_{-\infty}^\infty d\theta\;\Cal T_\lambda(\theta)\,\p_\theta f_c(\theta)
\label{G0TBA}
\ee
where $f_c=f_{c^+}=f_{c^-}$ is the $V=0$ Fermi factor for the strings $\A{c}{\!^\pm}$.
In particular, Eq.(\ref{G0TBA}) predicts a universal temperature dependance
in the linear regime $V\to 0$ that \emph{differs} from the universal voltage dependance at $T=0$. 

In the following, we work
within units where $\kb=e=\hbar=v_\f=1$, and sums 
over repeated indices are  omitted.

\section{The quasiparticle basis}
\label{sectionBulk}

The quasiparticles $\AA_a$ involved in the exact out-of-equilibrium solution
have their origin in the off-diagonal character of bulk scattering. 
To grasp how this can happen, 
let us consider the following simple
situation: 
the scattering of a single soliton $\SS{+}(\theta)$ through a gas of $N$ antisolitons at rapidities $(\theta_1,...,\theta_N)$. 
In the diagonal case the soliton $\SS{+}(\theta)$ goes through the gas and just pick up a phase factor $Z_{\rm }(\{\theta_i\},\theta)=\prod_i S_{+-}^{+-}(\theta-\theta_i)$, with $|Z|=1$.  
Off-diagonal scattering results in a final 
state which is a superposition of states where the positive charge can have moved at any rapidity $\theta_i$
simply by the off-diagonal process $\SS{+}(\theta)\SS{-}(\theta_i)\to \SS{+}(\theta_i)\SS{-}(\theta)$.  The probability $|Z|^2=\prod_i |S_{+-}^{+-}|^2(\theta-\theta_i)$ that the particle exiting the gas at rapidity $\theta$ is still the soliton is  exponentially small  in the number of antisolitons (since
$|S_{+-}^{+-}|<1$).
 Hence with exponential precision the final wave function consists in states where 
an antisoliton $\SS{-}(\theta)$ exits leaving a positive charge $\SS{+}(\theta_i)$ in the gas. This diffusion, in rapidity space,  of the   charge 
carried by solitons means that the  soliton positive charge \emph{does not propagate} in rapidity space
through the antisoliton gas.

Obtaining the QP basis that allows to solve
the off-diagonal BSG model requires the use of the Algebraic Bethe Ansatz  method, 
that can be viewed as a (considerable) generalization
of the preceding protocole, whereby ones  send solitons at well chosen (complex)  rapidities $\{\theta'_r\} $
so that 
 in the overall final state   the charge  distribution is unchanged and $Z( \{ \theta_i\},\{ \theta'_r\})$  is again a pure phase shift.
 The additional rapidities organize into ``strings" 
  (a bunch of complex rapidities with given length $k$ centered around a real rapidity, defined in sections \ref{SectionTransferMatrix} and \ref{SectionBareStrings})
 that can be considered
 as genuine QPs, diagonalizing the problem
 of the aforementioned diffusion in many-body rapidity space.
 To do so, the rapidities $\{\theta'_r\}$ are adjusted in a very precise way such that \emph{all} the
 diffusive trajectories in many-body rapidity space interfere completely destructively,
 leading to \emph{diagonal} bulk scattering in this new basis.   
 
 The string QPs are needed here to keep 
 track of the reorganization of charge in rapidity space after each scattering event.
 As a matter of fact, those QPs carry only entropy (no energy nor momentum), and charge.

We first present the soliton/antisoliton basis that solves the problem in the diagonal BSG model.
We then move on with the introduction of strings within ABA.
Since
some details will be crucial when introducing the impurity in Section \ref{SectionImpurity},
we review the main ingredients of this well known technique, 
by presenting the central idea of ABA  (section \ref{SectionTransferMatrix}),
then by introducing  the ``bare" strings (section \ref{SectionBareStrings}),
and by finally presenting the ``dressed" strings in Section \ref{SectionDressedStrings},
and we also discuss some important features of the finite ($V,T$) thermodynamics of the QP gas
in Section \ref{SectionThermoABA}.

\subsection{Solitons and antisolitons}
\label{SectionBulkSG}

The basis factorizing the scattering amongst QPs and on the impurity
can be identified by considering a massive generalization of the BSG model.
It is convenient for this purpose to ``fold" the model, defining a total (non-chiral) boson
living on the semi-infinite line $x<0$, $\Phi(x)=\phi(x)+\phi(-x)$, as a sum  
of right- and left-moving bosons $\phi(x)$ and $\phi(-x)$ respectively 
\citeOR{BookDifrancesco,BookTsvelikEtAl}, which together with the  boundary condition $\phi(0^+)=\phi(0^-)$ leads to: 
\bea
H_0&=&
\frac12\int_{-\infty}^0 dx\,\Cal H_0(x)\; ; \quad \Cal H_0(x)=(\p_x\Phi)^2+(\p_t\Phi)^2,\nl
H_\B&=&\gamma \cos(\frac\beta2\Phi(0)).
\label{HBSGfolded}
\eea
The well known trick \citeOR{FendleySaleurWarner}  is then to
 introduce a SG term $\Cal H_\ind{SG}(x)=\cos(\beta\Phi(x))$ in the bulk, 
 replacing $H_0$ in (\ref{HBSGfolded})
by $H_0^M=H_0 + \Gamma\int_{-\infty}^0 dx\, \Cal H_\ind{SG}(x)$.
The resulting total Hamiltonian $H_0^M+H_\B$, the massive BSG model, is integrable \citeOR{Ghoshal94}, with a QP content and bulk scattering matrix $S$ coinciding with that 
of the bulk  sine-Gordon model defined of the full line with  Hamiltonian $\int_{-\infty}^\infty (\Cal H_0+\Cal H_\ind{SG})$. In the repulsive
regime $\lambda<1$, the QPs consist in a pair of massive solitons and antisolitons $\SS{\pm}(\theta)$, 
where the rapidity $\theta$ parametrizes
the momentum $p=M\sinh\theta$ and energy $E=M\cosh\theta$ of the QPs, and $M$ is the mass gap of solitons.
Then, the limit of vanishing  $\Gamma$ (implying a vanishing mass gap $M\to 0$),  is taken
and the bulk part recovers its original nature of a free boson described by $H_0$.
In the meantime, the rapidity is redefined  $\theta\too \theta+\ln\frac{2E_0}{M}$ where $E_0$ 
is an arbitrary reference energy scale (which we will later choose as $E_0=T$) so that the dispersion relation
for QPs becomes that of massless particles, $E=p=E_0\,e^\theta$. 

At the end of this procedure, one obtains a representation of the many-body Hilbert space of the original  free boson (central column of Table \ref{TableBA}) with basis the Fock states built with  solitons $\SS{+}$
and antisolitons $\SS{-}$:
\be
 \ket{\ens{\theta_j,\varepsilon_j}{}}_N = \prod_{i=1}^N \SS{\varepsilon_i}(\theta_i)\ket{0}\quad \varepsilon_i\in\{+,-\}
\label{genericEigenstateSol}
\ee  
Our normalisation of the charge (\ref{chargedef}) ensures that the QPs $\SS{\pm}(\theta)$ carry charge $\pm 1$. 
The interaction amongst those QPs is encoded in the bulk scattering matrix $S$, defined 
by the relation 
$\A{\varepsilon_1}(\theta_1)\A{\varepsilon_2}(\theta_2) = S_{\varepsilon_1 \varepsilon_2}^{\varepsilon'_1 \varepsilon'_2}(\theta_1-\theta_2)\A{\varepsilon'_2}(\theta_2)\A{\varepsilon'_1}(\theta_1)$.
Charge symmetry leaves only few with non-zero entries \citeOR{Zamolodchikov79}:
\bea
S_{\pm\pm}^{\pm\pm}(\theta)&=&a(\theta) = e^{i\varphi_z^{\vphantom{\dagger}}(\theta)}
\label{defa}\\
S_{\pm\mp}^{\pm\mp}(\theta)&=&b(\theta) = \frac{-\sinh(\lambda\theta)}{\sinh(\lambda(\theta-i\pi))}e^{i\varphi_z^{\vphantom{\dagger}}(\theta)}
\label{defb}\\
S_{\pm\mp}^{\mp\pm}(\theta)&=&c(\theta) = \frac{-i\sin(\lambda\pi)}{\sinh(\lambda(\theta-i\pi))}e^{i\varphi_z^{\vphantom{\dagger}}(\theta)}
\label{defc}\\
\varphi_z^{\vphantom{\dagger}}(\theta) &=& \int_{-\infty}^\infty 
\frac{d\omega}{2\omega} \sin(\omega\theta)
\frac{\sinh\frac{\pi\omega(1-\lambda)}{2\lambda}}{\sinh \frac{\pi\omega}{2\lambda}\;
\cosh\frac{\pi\omega}{2}}
\label{defPhiZ}
\eea

For the integer SG model $\lambda\in\mathbb{N}$, the off-diagonal scattering amplitude vanishes, $c(\theta)=0$,
resulting in the existence of an additional symmetry: not only the 
charge, and the individual momenta of the QPs are conserved during a scattering event, but also the momentum-resolved charge. 
It results that the $N$-particle states   (\ref{genericEigenstateSol})
are stable under (bulk) scattering, and therefore allow for a thermodynamical treatment 
of the gas of solitons/antisolitons in the grand-canonical ensemble, with micro-states
of the form (\ref{genericEigenstateSol}). 

In the generic case with off-diagonal scattering $c(\theta)\neq 0$,
this  additional symmetry is absent:
 momentum and charge degrees of freedom are mixed
by scattering leading to the aforedmentionned diffusion and the states (\ref{genericEigenstateSol}) can no longer be used as the microstates of a thermodynamical treatement.

\subsection{Solitons and strings}
\label{SectionBulkABA}

We now present the ABA approach, that allows to circumvent the off-diagonal character of the scattering and leads to the identification of the correct states, i.e. that
are stable under bulk scattering (right-most column of table \ref{TableBA}).
Initially developed to solve the closely related XXZ model by Takahashi and coworkers 
\citeOR{Takahashi72,TakahashiBook,KorepinBook,FaddeevHouches}, ABA has been implemented with success in the bulk SG model
for various specific values of $\beta$ (see e.g. Refs \citeOR{Fowler82,Chung83,Fendley92,FendleySaleur94,Tateo95}), but the results are somehow scattered
in the literature. Moreover, the solution is  presented in a variety of forms, sometimes making use 
of a remarkable identity leading to simplifications \citeOR{Zamolodchikov91}. We aim here at gathering all the information relevant for the derivation of the impurity scattering in Section \ref{SectionImpurity}, and at presenting a unified TBA system for arbitrary $\lambda\in\mathbb{Q}$. We also carefully derive the  asymptotic 
behavior of the TBA equations when $\theta\to\pm\infty$.
  
 \subsubsection{Transfer matrix}
 \label{SectionTransferMatrix}

Let us consider $\Hilb^{\ens{\theta_i}{}}$, the Hilbert space with $N$ particles (indifferently solitons or antisolitons) at rapidities $\ens{\theta_i}{i=1...N}$. A basis for $\Hilb^{\ens{\theta_i}{}}$ are the states (\ref{genericEigenstateSol}) but due to non-diagonal scattering, the  later are not stable under scattering and eigenvectors are to be sought in the more generic form: 
\be
\ket{\Psi} = \sum_{\varepsilon_1,...,\varepsilon_N} \psi_{\varepsilon_1,...\varepsilon_N}(\theta_1,...\theta_N) \ket{\ens{\theta_j,\varepsilon_j}{}}.
\label{genericEigenstateT}
\ee  
The allowed values for the rapidities $\ens{\theta_i}{}$ are determined by a self-consistency condition: considering periodic boundary conditions and passing particle $k$ with rapidity $\theta_k$ through all others should leave the system invariant (up to a phase shift multiple of $2\pi$):
\be
e^{i p\vph_jL} \;\tau^{(s)}(\theta_j) = e^{2i\pi \Cal N_s(\theta_j)}
\label{quantCond}
\ee
where $p\vph_j=E_0 e^{\theta_j}$ is the momentum of particle $j$,  $L$ is the size of the system,  and the function $\Cal N_s(\theta)$ takes integer values when $\theta=\theta_j$, an allowed rapidity.
Equation (\ref{quantCond}) is the analog, for our gas of interacting particles where $\tau^{(s)}\neq 1$,  of the usual quantization condition $e^{ipL} = 1$ for a free gas in a box of size $L$. 

In our case, off-diagonal scattering results in that the \emph{soliton transfer matrix}  $\tau^{(s)}(\theta)$ entering the quantization equation (\ref{quantCond}), 
is a $2^{N}\times 2^{N}$ matrix.
First,
the  transfer matrix  can be elegantly related\citeOR{Fendley92} to the \emph{monodromy} matrix
$T_\varepsilon^{\; \varepsilon'}(\theta)=\left(\!\!{\small \begin{array}{cc}A&B\\C&D\end{array}}\!\!\right)$,
where $A,B,C,D$ are $2^N\times 2^N$ $\theta$-dependent matrices, via (we make use
of $\lim_{\theta\to0}S_{\varepsilon_1\varepsilon_2}^{\varepsilon'_1\varepsilon'_2}(\theta)=\delta_{\varepsilon_1}^{\varepsilon'_2}\delta_{\varepsilon_2}^{\varepsilon'_1}$):
\be
\tau^{(s)}(\theta)={\rm Tr}\big(T(\theta)\big)
=(A+D)(\theta)
\label{deftaus}
\ee
The monodromy matrix expresses the fate of a single particle with initial quantum number $\varepsilon\in\{+,-\}$ and rapidity $\theta$ passing 
through a bunch of $N$ particles with initial  quantum numbers $\ens{c_i}{i=1...N}$ and rapidities 
$\ens{\theta_i}{}$. It is a dynamical statement (time evolution of a particle scattering through a gas) 
 that directly translates at the level of the solitons/antisolitons modes' algebra as:
$\SS\vph_\varepsilon(\theta)\SS\vph_{c_1}(\theta_1)...\SS\vph_{c_{N}}(\theta_{N})
= 
\big[T_{\varepsilon}^{\;\varepsilon'}\big]^{\{d_i\}}_{\{c_i\}}\SS\vph_{d_1}\!(\theta_1)...\SS\vph_{d_{N}}\!(\theta_N)\SS\vph_{\varepsilon'}(\theta)$. Explicitly, one has: 
$[T_{{\varepsilon}}^{\;\varepsilon'}(\theta)]^{\ens{d_i}{}}_{\ens{c_i}{}}(\ens{\theta_i}{})=\sum_{k_1,...,k_{N-1}}S_{{\varepsilon}c_1}^{k_1d_1}(\theta\!-\!\theta_1) 
\times S_{k_1c_2}^{k_2d_2}(\theta\!-\!\theta_2)\hdots S_{k_{N-1}c_N}^{{\varepsilon'}d_N}(\theta-\theta_N)$.
Coming back to our initial problem, 
the task  of building the eigenfunctions $\psi$ in (\ref{genericEigenstateT}) requires diagonalizing 
the  transfer matrix $\tau^{(s)}(\theta)$ at least for rapidities $\theta=\theta_j$. 
Remarkably, this can be done at any $\theta$
following 
 the ABA method, yielding $\theta$-independent eigenvectors  but 
$\theta$-dependent eigenvalues.

\newcommand{\NoteCompleteness}{There are also exceptional solutions that do not obey the "string hypothesis"\citeOR{Hao13}. However, string solutions in the SG model appear to be complete in the thermodynamical sense, i.e. considering string states  only leads to the correct partition function in the thermodynamical limit.}  

\subsubsection{Bare strings}
\label{SectionBareStrings}

The essence of ABA is that eigenstates of $\tau^{(s)}(\theta)$ can be obtained by repeated applications of the $B$ matrix, evaluated at well chosen (complex) rapidities $\{\theta_r'\}$, on the lowest state in $\Hilb^{\ens{\theta_i}{}}$ consisting of $N$ antisolitons:
\bea
\ket{\Psi\{\theta'_r\}} &=& \prod_{r=1}^{N_f} B(\theta'_r)\ket{\Omega},
\label{EigenVect}
\\
\ket{\Omega}&=&\prod_{i=1}^N \A{-}(\theta_i)\ket{0}.
\label{defOmega}
\eea
The operator $B$, called the magnon operator, converts an antisoliton into a soliton and thus carries a charge $2$. Since we have $N_f$ flipped antisolitons, the charge of the state (\ref{EigenVect}) is $Q=-N_s+2N_f$.
The requirement that $\ket{\Psi\{\theta'_r\}}$ be an eigenstate of $\tau^{(s)}$ results in that the rapidities
$\{\theta'_r\}$ are organized in so-called strings \citeOR{NoteCompleteness,Hao13}. ``Bare" strings of length $k$ with rapidity $\theta$ (as opposed to dressed ones, see Section \ref{SectionThermoABA}) will be denoted by $B_k(\theta)$ and consist  in $k$ insertions of the magnon operator $B_k$ at $k$ different complex rapidities with common real part $\theta$ ; they are also  
 characterized by 
 a parity $\varepsilon=\pm 1$
\citeOR{Takahashi72,TakahashiBook}:
\be
\begin{array}{ll}
B_{k}(\theta) =\prod_{\ell=1}^{k} B\big(\theta+i\frac{\pi(2\ell-k)}{2}\big)\hphantom{\;+\lambda^{-1}}
& \varepsilon=+1\\
B_{k}(\theta) =\prod_{\ell=1}^{k} B\big(\theta+i\frac{\pi(2\ell-k+\lambda^{-1})}{2}\big)
& \varepsilon=-1
\end{array}
\label{defStrings}
\ee

The allowed lengths $k_j$ are constrained by the normalizability of the wave function that leaves $n=m_\alpha$ strings with lengths (in the first line $j>0$ and $m_{i-1}\leq j <m_i$):
\bea
 k_j&=&y_{i-2}+(j-m_{i-1})y_{i-1}  
 \label{defkj}
 \\
 k_{\nst } &=& y_{\alpha-1}
 \eea
where the integers $y_i$ read:
\be
 y_i=y_{i-2}+\nu_i y_{i-1}\quad;\quad y_{-1}=0\;,\;y_0=1.
 \label{defyi}
\ee
The parities are given by $\varepsilon_{m_1}=-1$ and otherwise
$\varepsilon_j=(-1)^{[\lambda(k_j-1)]}$.

Strings  do interact with each other and with solitons; the scattering matrix is \emph{diagonal} 
(to simplify notations we introduce $S_{ab}\equiv S_{ab}^{ab}$) 
and is given by ($j,j'\in\{1,...,\nst \}$ are string indices): 
\bea
S_{s,j}(\theta)&=&g_{-k_j}^{(\varepsilon_j)}(\theta)
\label{SmatSolString}
\\
S_{1,j}(\theta)&=&g_{k_j+1}^{(\varepsilon_j)}(\theta)g_{k_j-1}^{(\varepsilon_j)}(\theta)
\nl
S_{j,j'}(\theta)&=&\prod_{\ell=1}^{k_j}S_{1,j'}\big(\theta+i\frac\pi2(k_j+1-2\ell)\big)
\label{SmatStringString}
\eea
with the functions $g$:
\be
g^{\!(\!+\!)}_k\!(\theta)=\frac{\sinh\lambda(\theta-\frac{ik\pi}2)}{\sinh\lambda(\theta+\frac{ik\pi}2)} \;;\;  
g^{\!(\!-\!)}_k\!(\theta) = g^{\!(\!+\!)}_k\!(\theta+\textstyle{\frac {i\pi}{2\lambda}}).
\label{defgfunction}
\ee
The (now diagonal) scattering data (\ref{SmatSolString},\ref{SmatStringString}), together with (\ref{defa}), allows
to derive in a standard way the Bethe equations defining the allowed rapidities for solitons and strings 
(see Appendix \ref{AppendixBareABA}), which become after a continuum limit where
the number of particles per unit length is sent to infinity:
\be
\tilde\eta_a\,P_a = \delta_{a,s}e^\theta + \frac1{2\pi}\Phi_{a,b}\star \tilde\rho_b^{}
\label{contABAbare}
\ee
 where $\tilde\eta_a=-\eta_a \, (-1)^{\delta_{a,s}+\delta_{a,\nst}}$ are some signs required to have a positive total density. Here the density of occupied QP (solitons or bare string) of type $a$ per unit length is written $\frac T{2\pi}\rho_a$, and $\frac T{2\pi}P_a$ is the total (occupied and empty) density of allowed
 QP rapidities.

\subsubsection{Dressing}
\label{SectionDressedStrings}

The complicated structure of the scattering (\ref{SmatSolString},\ref{SmatStringString})
(all QPs scatter non-trivially with all QPs)
 can
be greatly simplified by performing a dressing operation consisting in a particle-hole transformation on all strings but the last one, defining the dressed modes $\AA_a(\theta)$ (third column of Table \ref{TableBA}). 
The dressing also has the advantage
to reveal reveal a simple implementation of the U(1) charge symmetry of the QP gas,
facilitating the coupling to an external bias voltage to later address non-equilibrium transport.
Technical details about the dressing can be found 
in Appendix \ref{AppendixABA}. 

We check that after dressing  the  continuous Bethe equations relating $\frac T{2\pi}P_a(\theta)$, the total density (per unit length) of allowed rapidities for the modes $\AA_a(\theta)$ ($a=s,1,...,\nst $), to $\frac T{2\pi} \rho_a(\theta)$, the QP density (per unit length) of \emph{occupied} allowed rapidities, become:
\be
P_a(\theta) =\delta_{a,s}e^\theta +\frac{1}{2\pi}\, \big(\kernel_{ab}\star \rho_b\big)(\theta),
\label{BAdensities}
\ee
where 
the kernel $K_{ab}(\theta)$ is made explicit in Fig. \ref{figDynkinGen}.
Note that the source term $\propto e^\theta$ is present for solitons only, indicating that strings
carry no momentum  nor kinetic energy.

The dressing operation affects
not only the scattering properties, but also the charge quantum number of the QPs. 
One checks (see Appendix \ref{AppendixChargeDressing}) that the dressed modes are neutral except
for the last two dressed strings $\AA_{\nst }\equiv \AA_c^+$ and $\AA_{\nst -1}\equiv \AA_c^-$,
carrying respectively $\pm \qL$ units of the original soliton $\SS{+}$.
 The features  of the QP spectrum at arbitrary $\lambda\in\mathbb{Q}$
 are summarized
in Table \ref{TableQPs}.

\subsubsection{Thermodynamics of the QP gas}
\label{SectionThermoABA}

The thermodynamics of this gas of interacting QPs is then derived in a standard way leading
 to the TBA
(Thermodynamical Bethe Ansatz)
equations \citeOR{Zamolodchikov90} that we already gave in Eq. (\ref{epsilonTBA}).
One can check explicitly (see Appendix \ref{AppendixFreeEnergy}) that the complicated description of the original free boson
gas at finite $(V,T)$ in terms of the new QP modes, is exact, in the sense that the finite 
$V,T$  partition functions do coincide in the thermodynamical limit,
providing an \emph{a posteriori} check 
of the  thermodynamical completeness of the soliton/string basis.

A few remarks are in order regarding those exotic  QP modes $\AA_{a\neq s}$.
Physically, those modes have their origin in the mixing of charge and momentum degrees 
of freedom that the original solitons/antisolitons experience due to off-diagonal scattering.
It results that the modes $\AA_{a\neq s}$ are purely entropic: they  do not carry energy (the later being carried only by solitons)
but solely entropy that we can interpret as connected to the redistribution of the charge degrees
of freedom after each off-diagonal scattering event. 

In the limit of vanishing temperature, where only original solitons modes $\SS{+}$
(for positive voltage $V$)  are occupied  and the scattering effectively becomes diagonal\citeOR{FLS-PRL,FLS-PRB} (only the matrix element $S_{++}^{++}$
is  involved, see Eq. (\ref{defa})), the entropic QPs become frozen and, we can check, as illustrated in Fig. \ref{figEntropy}, that the solitons
carry all the entropy (see Appendix \ref{AppendixEntropy}). Formally, the freezing of the entropic QPs in this limit results in 
all the densities $\rho_{a\neq s}(\theta)$ being linearly determined by the soliton density $\rho_s(\theta)$
(see Eq. (\ref{DensitiesTEq0})). 

In the opposite limit $V/T\ll 1$ the strings's entropy $S_a(V,T)$ is at 
its highest, and  reaches (see Appendix \ref{AppendixEntropy}) a universal limit that remarkably does not depend on $\lambda$: 
\be
\frac{h v_\f}{\kb T}
\sum_{a=1}^{\nst}\frac{S_a(0,T)}{\kb L}
= 2\mbox{Li}_2\Big(\frac14\Big)+\ln4\ln\frac43
\simeq 0.934...
\label{univEntropy}
\ee
where $\mbox{Li}_2(x)$ is the polylogarithm.

\begin{figure}[h]
\renewcommand{\figurename}{\textbf{Figure}}
\renewcommand{\thefigure}{\textbf{\arabic{figure}}}
\begin{center}
\includegraphics[width=6.35cm]{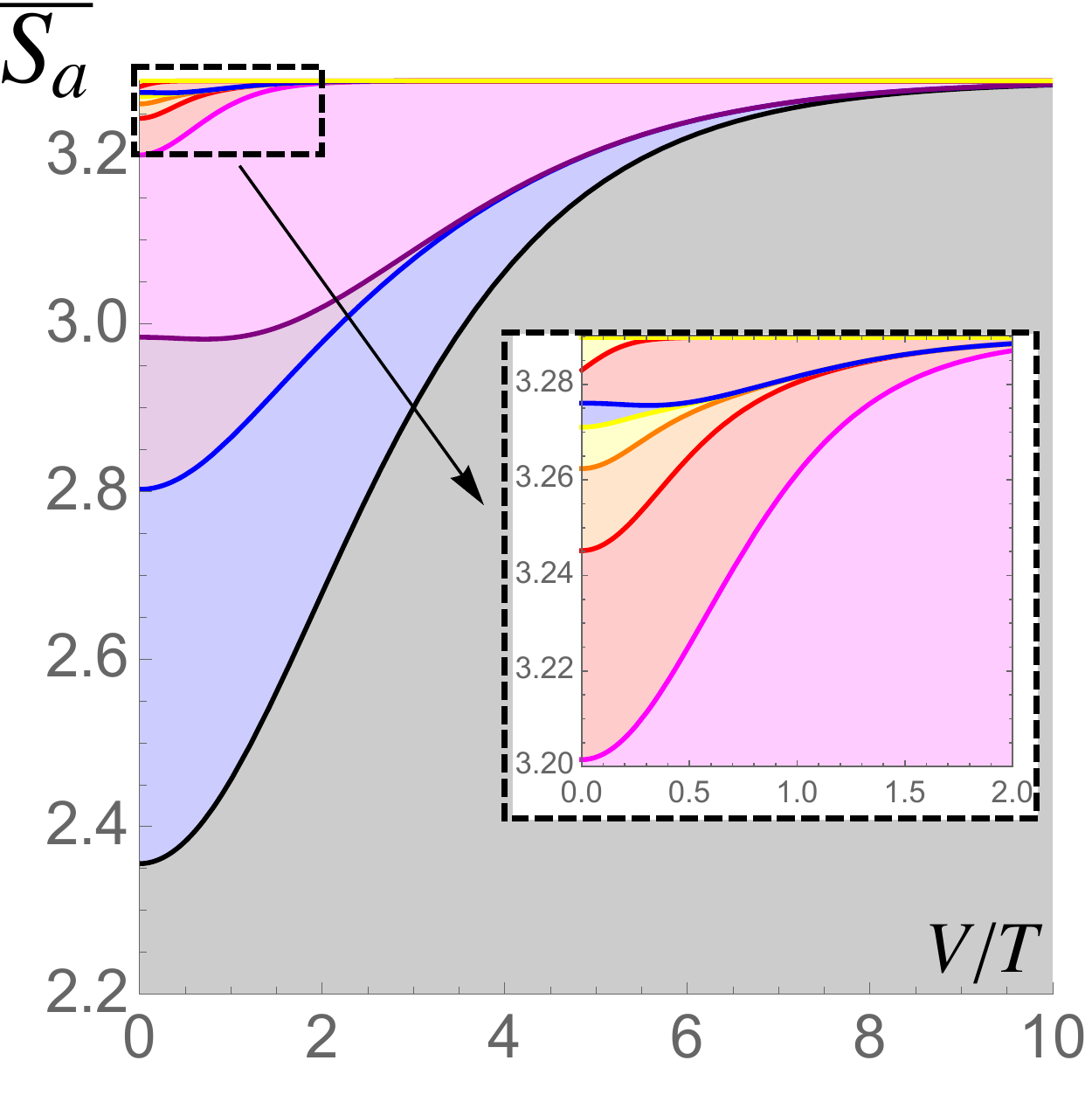}
\end{center}
\caption{Cumulative plot of the dimensionless entropy per unit length $\overline S_a=\frac{h v_\f}{\kb^2 T}\frac{S_a}{L}$ carried  by the 
$10=3+6+1$ QPs ($a=s,1,...,7,c^-,c^+$ from bottom to top) as a function of $V/T$, 
for $\lambda=\frac6{19}=\big(3+\frac16\big)^{-1}$
 ; the inset is a zoom. The total entropy $\sum_a\overline  S_a=\frac{\pi^2}{3}$ matches that of the original free boson gas. Note the vanishing of $\overline S_{a\neq s}$ in the low temperature limit where strings  freeze and the soliton carries all the entropy.
}
\label{figEntropy}
\end{figure}

The complex structure of the QP spectrum when $\lambda$ is varied is illustrated
in Fig.\ref{figCumulCond} where we display the fraction $x_a(\lambda,\frac VT)=\frac{N_a(\lambda,\frac VT}{\sum bN_b(\lambda,\frac VT)}$ of occupied QPs in the finite $T$ description (see Appendix \ref{AppendixCalculationNa} for the explicit calculation of the number of occupied particles $N_a(\lambda,\frac VT)$).
 It is sometimes advocated that the strings
are not QPs in the strict sense. We believe that this statement applies to the dressed description,
but not to the bare one.
The dressing operation, although drastically simplifying the structure of both the scattering and  the 
charge carried by strings, comes with a (double) price: first, the TBA equations have a non-symmetric kernel,
and second, the charge density of the  dressed modes $\AA_a(\theta)$ (even of the neutral ones)
is not localized  but has some spreading in rapidity space (see Appendix \ref{AppendixChargeDressing}),
making the modes $\AA_a(\theta)$ rather unconventional.
However, in the \emph{bare} soliton/strings\ basis, the modes $B_{k_j}(\theta)$ have diagonal, symmetric scattering, and  possess all the features of "usual" QPs: they are localized in momentum space, carry some entropy (but oddly no energy).
Ultimately, the physical meaning of these strings is clear: they are "book-keeping" particles (hence their entropic character)
that are necessary to account for the diffusion in momentum space experienced in the original (anti)soliton basis.

\section{Impurity scattering}
\label{SectionImpurity}

In this section we derive the structure 
of the impurity scattering matrix for the QPs.
We note that it was derived, in the particular (off-diagonal) case where $\lambda^{-1}$ is an integer, in Ref. \citeOR{FendleySaleur94},
but in the different case where the boson $\phi$ has fixed boundary conditions. We need 
here the general case  $\lambda\in\mathbb{Q}$ and free boundary conditions, which, as far
as we know, has not been considered.
After constraining the structure of the $R-$matrix in Section \ref{SectionStructR}
we will relate the scattering in the original soliton/antisoliton, and in the dressed QP basis
in Section \ref{SectionRmat} and derive the exact transmission probability of charged strings 
when crossing the impurity.

The SG model with impurity being integrable\citeOR{Ghoshal94}, it results that
the scattering over the impurity in the BSG model (\ref{HBSG}) factorizes, i.e.
it can be decomposed as a product of elementary one QP processes.
Formally, this is described by introducing the (many-body) impurity scattering matrix $\Cal R$. 
This object relates (incoming) states on the left of the impurity, built out of the modes $\AA_a^\ind{in}(\theta)$,
to (outgoing) states living on the right of the impurity and built out of the modes $\AA_a^\ind{out}(\theta)$.
Then, factorization means that a generic incoming state  $\ket{\psi}_\ind{in}=\prod_{i=1}^N \AA_{a_i}^\ind{in}(\theta_i)\ket{0}$ 
($\ket{0}$ is the many-body vacuum) will evolve  into a new state
 $\ket{\psi}_\ind{out}=\Cal R\ket{\psi}_\ind{in}=\sum_{b_1,b_2,...}\prod_{i=1}^N R_{a_i,b_i}(\theta_i-\theta_\B)\AA_{b_i}^\ind{out}(\theta_i)\ket{0}$ where $R_{ab}$ is the one-body impurity scattering
 matrix, and the dependence on the energy scale $\TB$ generated by the impurity interaction
 is simply encoded in
 $\theta_\B$, the impurity rapidity defined by $\theta_\B=\log(\TB/T)$. 
The one-body impurity scattering matrix $R_{ab}$, describing the fate of a single incoming QP with
 quantum number $a\in\{+,-\}$
 is well known in the soliton/antisoliton basis. 
Since the impurity interaction term $H_\B$ in Eq. (\ref{HBSG}) does not
conserve charge, it is a $2\time 2$ matrix over the soliton-antisoliton space (at fixed rapidity) ; 
charge conjugation symmetry
implies that it has only two distinct elements $R_{\pm\pm}=P$
and $R_{\pm\mp}=Q$, that are
 given by
\citeOR{Ghoshal94,FendleySaleurWarner}:
\bea
P^{(\lambda)}(\theta) &=& \frac{e^{\lambda\theta}}{1+ie^{\lambda\theta}}\;e^{i\varphi_\B^{(\lambda)}}(\theta)
\nl 
Q^{(\lambda)}(\theta)&=&\frac{i}{1+ie^{\lambda\theta}}\;e^{i\varphi_\B^{(\lambda)}}(\theta)
\label{RmatSol}
\\
\varphi_\B^{(\lambda)}(\theta) &=& C_\lambda+\int_{-\infty}^\infty\frac{dt}{2t}\sin\Big(\frac{2\lambda\theta t}{\pi}\Big)\,\frac{\sinh((1-\lambda)t)}{\cosh(\lambda t)\sinh(2t)}
\nonumber
\eea
with $C_\lambda$ a constant that will not be needed here.

\subsection{Structure of the impurity scattering}
\label{SectionStructR}

However, what is needed here is the impurity scattering matrix $\Cal R$ in the 
string/soliton basis. It is not obvious at all how to derive this object, since the strings
are built out of the magnon operator $B(\theta)$ (flipping an antisoliton into a soliton)
and not directly out of the soliton/antisoliton modes $\SS{\pm}(\theta)$. 
To this aim we compute a Loschmidt echo in both
QP basis. We will see that this allows to determine the structure of $\Cal R$,
that also factorizes in the new string/soliton basis.
\newcommand{\footnoteDoubly}{This 
does not break the basic requirement that the ZF pseudo particles cannot by
more than singly occupied, since we have here two different 
species}

Integrability in the presence of an impurity demands that
the impurity scattering $R$
is compatible with the bulk scattering $S$, i.e. the side of the impurity where the bulk scattering takes place does not matter. This requirement, the 
boundary Yang Baxter equation\citeOR{Ghoshal94}, reads:\vspace*{0.6mm}\\
\begin{minipage}{0.59\linewidth}
where in this graphical representation  the impurity is figured by the red dashed line, and an empty
\end{minipage}\hfill
\begin{minipage}{0.4\linewidth}
\hspace*{1mm}\raisebox{-3mm}{\includegraphics[width=.98\linewidth]{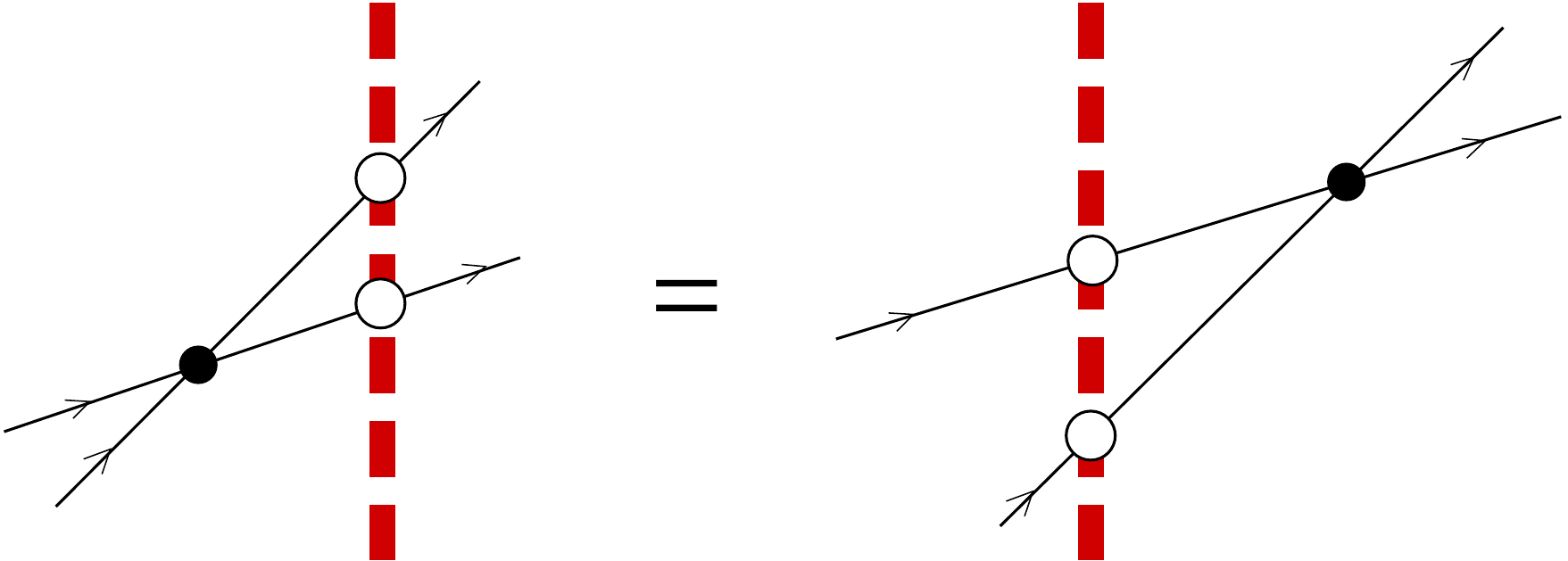}}
\vspace*{0mm}
\end{minipage}
 (respectively  full) 
circle stands for an 
insertion of the impurity scattering matrix $R$ (respectively bulk scattering matrix $S$). 
Repeated applications of the BYBE equation allow to derive 
a self consistency equation for the monodromy matrix $T_a^{\;b}$.
 It is more difficult to make mistakes when drawing this equation:
\begin{center}
\includegraphics[width=5cm]{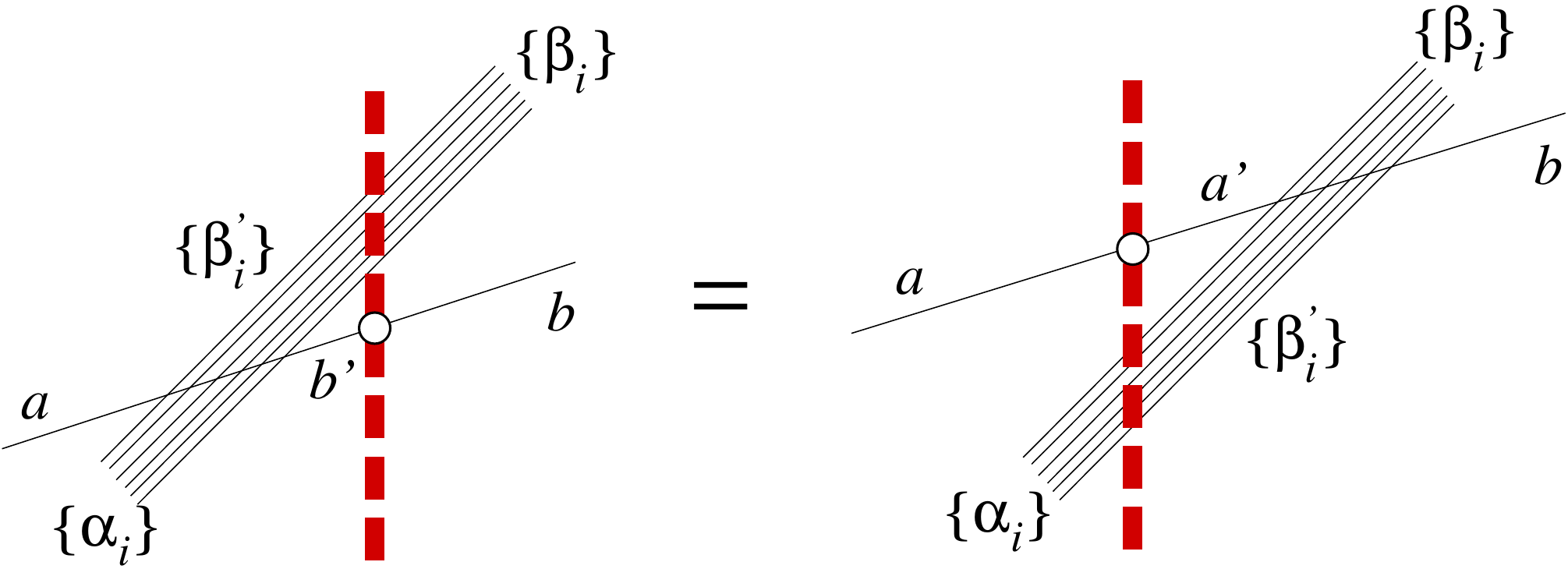}
\end{center}
than when writing it down:
\bea
[T_{a}^{\;\;b'}(\theta)]_{\{\alpha_i\}}^{\;\;\{\beta_i'\}}
\Cal R_{\{\beta_i'\}}^{\;\;\{\beta_i\}}(\theta_\B)
R_{b',b}(\theta-\theta_\B)
=
\nl
R_{a,a'}(\theta-\theta_\B)\Cal R_{\{\alpha_i\}}^{\;\;\{\alpha'_i\}}(\theta_\B)[T_{a'}^{\;\;b}(\theta)]_{\{\alpha'_i\}}^{\;\;\{\beta_i\}}.
\label{boundaryTYBE}
\eea
Now from (\ref{boundaryTYBE}) we can derive four relations (by fixing the free indices $a,b$) which, combined together, yield 
$[A+D,\Cal R]=\frac{P}{Q}[\Cal R,B+C]=\frac{Q}{P}[\Cal R,B+C]$ $\forall \,(\theta,\theta_\B)$, implying that all those commutators vanish. Remembering that 
$(A+D)(\theta)$ is nothing but the transfer matrix  defined in (\ref{deftaus}), we arrive at:
\be
[\tau^{(s)}(\theta),\Cal R(\theta_\B)]=0.
\ee
A direct consequence of this equation is that all non-degenerate eigenstates of $\tau^{(s)}$ are eigenstates of $\Cal R$. The non accidental degeneracies of the eigenvalues of $\tau^{(s)}$, for two different configurations $\conf\neq \conf'$, are actually linked to the U(1) charge symmetry of the problem: 
it turns out that
   $S_{\nst -1,a}(\theta)S_{\nst ,a}(\theta)=1$, $\forall\,\theta\;,\forall\, a=s,1,...,\nst $ (see Appendix \ref{AppendixScatPair}). 
Hence  a \emph{pair} of the last two (bare) strings, $B_{k_{\nst -1}}(\theta)B_{k_{\nst }}(\theta)$, does not produce any scattering at all,
i.e. it can be added or removed without affecting the quantization condition of the other QPs.
 More precisely, let us consider an allowed configuration $\conf=\{\theta^{(a)}_r\}$, $a=s,1,...,\nst $; 
 $r=1...N_a$ with $N_s$ solitons, and $N_j$ insertions of strings of type $j$ in the state (\ref{EigenVect}).
 Consider then a given rapidity $\theta_0$ that is doubly occupied\citeOR{footnoteDoubly}
 for the last two strings, 
i.e. both strings   $B_{k_{\nst }}(\theta_0)$ and    $B_{k_{\nst -1}}(\theta_0)$
appear in the state (\ref{EigenVect}). We can immediately deduce that the configuration $\conf'$
obtained from $\conf$ by \emph{removing} the pair $B_{k_{\nst -1}}(\theta_0)B_{k_{\nst }}(\theta_0)$
from the state (\ref{EigenVect}),
is (i) an allowed configuration and (ii) degenerate with $\conf$ in the sense that they share the same $\tau^{(s)}(\theta)$ eigenvalue.

Moving towards the dressed picture and performing the particle-hole transformation, we can conclude that
eigenvalues of the transfer matrix $\tau^{(s)}(\theta)$ are degenerate $\forall\,\theta$ for configurations
that are obtained from each other by the conversion of an arbitrary number of occupied positive strings $\AA_{c}^+$ into (initially empty) negative strings $\AA_{c}^-$, and \emph{vice versa}.
From this analysis we can conclude that the structure of the one-particle dressed $R-$matrix 
 is diagonal for all the neutral sector $a\in\{s,1,...,\nst -2\}$, and consists of a $2\times 2$
block for the charged sector:
\bea
R_{a,a}(\theta)&=&e^{i\xi\vph_a(\theta)}\quad{\rm if\;\;} a\neq c^\pm,
\nl
R_{c^\pm,c^\pm}(\theta)&=&P_\drs(\theta),
\nl
R_{c^\pm,c^\mp}(\theta)&=&Q_\drs(\theta).
\label{structRdressed}
\eea

\subsection{Impurity scattering matrix for strings and solitons}
\label{SectionRmat}


We now   determine the scattering data in (\ref{structRdressed}), by translating the details of the impurity scattering (\ref{RmatSol}) for solitons/antisolitons into the dressed ABA picture involving 
strings and solitons. To this end, we study the system with impurity  chosing a particular
many-body incoming state, namely the fully polarized state $\Ket \Omega $ defined in (\ref{defOmega}).
This situation physically corresponds to setting the temperature to zero with a finite negative voltage bias.
 
Let us prepare the system in the reference state $\ket\Psi_0=\ket\Omega\lin$, then evolve it in time until it has crossed the impurity so that it now consists of a many-body coherent superposition of outgoing modes (ending up in state $\ket\Psi_\infty=\Cal R\ket\Omega\lin$, where arbitrarily many $\SS{-}$ have been converted in $\SS{+}$ on the impurity), and finally project it onto the reference state 
$\ket\Omega\lout$. 
The overlap is a Loschmidt echo, and we define: 
\be
\Cal L=-\ln\big[\;
\vph\lout\!\langle\Omega\big|\Psi\rangle_{\infty}
\big].
\label{defLoschmidt}
\ee
Using the explicit form for the matrix $\Cal R$ we want to evaluate $\Cal L$ in the original soliton/antisoliton
basis and in the dressed basis, therefore relating the impurity scattering in the two descriptions.  In the original basis, since the incoming state is made of $N_s$ antisolitons, it is straightforward
to evaluate the overlap in (\ref{defLoschmidt}): it selects only the $R_{--}=P$ processes
yielding:
  \be
  e^{-\Cal L}=\prod_{j=1}^{N_s}P(\theta_j-\theta_\B).
  \label{LoschmidtBare}
  \ee
Let us now turn to the dressed description. We first need to express the state $\ket\Omega$ in the new basis:
at the bare level all strings are empty, so that after dressing by the particle-hole transformation, all
QP modes $\AA_a$ with $a\neq s, \nst $ are fully occupied and all modes $\AA_c^+$ are empty: $\ket\Omega=\prod_{i=1}^{N_s}\AA_s(\theta_i^{(s)})\prod_{j=1}^{\nst -1}\prod_{r=1}^{N_j}\AA_j(\theta_r^{(j)})\ZFS
$
where the dressed vacuum $\ZFS$ is defined in Eq.(\ref{defZFS}).

After scattering on the impurity, the state $\ket\psi_\infty$ still has all QP states of type $j<\nst -1$ fully occupied, but some QPs $\AA_{c}^-$ have been converted into $\AA_{c}^+$. Projecting onto $\ket\Omega\lout$ selects the events  $R_{c^-,c^-}$ for all $\theta_r^{(\nst -1)}$ so that we
 get from (\ref{structRdressed}) the following alternate form for  the Loschmidt echo:
 \bea
 e^{-\Cal L}=\prod_{r=1}^{N_{\nst -1}}P_\drs(\theta_r^{(\nst -1)}-\theta_\B)
 \nl
 \times \prod_{j=1}^{\nst -2}\prod_{r=1}^{N_{j}}e^{i\xi_j(\theta_r^{(j)}-\theta_\B)}  \prod_{i=1}^{N_s}
 e^{i\xi_s(\theta_i^{(s)}-\theta_\B)}.
 \label{LoschmidtDressed}
 \eea
  In the thermodynamical limit the system is described by the density of occupied antisolitons $\rho_{s^-}(\theta)$ in the bare description, or moving to the ABA basis, by densities of occupied solitons and strings $\rho_j(\theta)$.
 Note that since there are only antisolitons in $\ket{\Omega}$, one has $\rho_{s^-}=\rho_s$.
 It is actually more convenient to evaluate $\chi\vph_\B(\theta_\B)\equiv -\frac{i}{L} \p_{\theta_\B}\Cal F$:
 introducing
  $\Phi_\B^{(\lambda)}=-i\p_\theta\ln P^{(\lambda)}$,   $\Phi_\B^\drs=-i\p_\theta\ln P_\drs$, and  Fourier transforming with respect to $\theta_\B$  we obtain:
\bea
 \hat\chi_\B\vph(\omega)&=&\hat\rho_s(-\omega)\hat\Phi_\B^{(\lambda)}(\omega)
 \label{EqChiB}
 \\
&=&  \hat \rho_{c^-}(-\omega)\hat \Phi_\drs^\B(\omega)
+i\omega\!\!\!\!\!\!\sum_{a=s,1,...,\nst -2}\!\!\!\!\!\!
\hat\rho_a(-\omega)\hat\xi_a(\omega). 
\nonumber
  \eea
String bands being either  full and empty  (hence the entropy is totally fully carried by antisolitons, see
fig. \ref{figEntropy}) in the state $\ket \Omega$, the string densities in the continuum limit satisfy  
$\rho_j(\theta)=P_j(\theta)$ $\forall j\notin\{ s,\nst \}$ and $\rho_{\nst }(\theta)=0$.
Moreover, they are not independent of each other: from (\ref{contBAbare}) and using the 
relationship between bare and dressed densities, $\tilde\rho_j=P_j-\rho_j$ ($j\notin\{ s,\nst \}$) one gets
$-2\pi\,\eta_j\,P_j=\Phi_{s,j}\star \rho_s$  ($j\neq s$) so that finally:
\be
\hat \rho_j (\omega)= -\eta_j\,\hat\Phi_{s,j}(\omega) \, \hat\rho_s(\omega)\quad (1\leq j<\nst )
\label{DensitiesTEq0}
\ee
where $\Phi_{s,j}$ is the log derivative of the  \emph{bare} scattering (\ref{SmatSolString}).

Now focusing for a moment on the imaginary part of $\chi_\B$, that is responsible for the decay of the Loschmidt
echo, it receives contributions only from
those entries of the $R-$matrix that have modulus $|R_{ab}|\neq 1$, that is,
in the first  line of Eq. (\ref{EqChiB}), from the modulus of the soliton transmission amplitude, 
and in the second line from the term $j=\nst -1$ involving the charged QP  $\AA_c^-$.
The explicit form for the bare scattering $\hat\Phi_\B^{(\lambda)}$ that one deduces from (\ref{RmatSol}):
  \be
  \hat\Phi_\B^{(\lambda)}(\omega) =-\frac{i\lambda}2\,\delta(\omega)+ \frac{1-\tanh\frac{\pi\omega}2}{4\sinh\frac{\pi\omega}{2\lambda}},
\label{TFPSG}
\ee
combined with
the relationship between densities (\ref{DensitiesTEq0}) allow to derive a
remarkable relation,  relating
the imaginary part of $\chi_\B$ in the original soliton/antisoliton basis and in
the dressed string/soliton basis:
\be
\hat \rho_{s}(-\omega)\hat\Phi_\B^{(\lambda)}(\omega)
= \hat \rho_{c^-}(-\omega)\hat\Phi_\B^{(\pL)}(\omega)
\label{EqImpScat}
\ee
where one has used the soliton-charged string
bare scattering $\hat \Phi_{s,c^-} =\hat \varphi_{-k_{\nst -1}}^{(\varepsilon_{\nst -1})}
=(-1)^{\alpha}\frac{\sinh\frac{\pi\omega}{2\pL}}{\sinh\frac{\pi\omega}{2\lambda}}$ (as obtained from Eqs.(\ref{SmatSolString},\ref{phikomega}) 
and the explicit parities for the charged strings, see Appendix \ref{AppendixScatPair}).

Plugging (\ref{EqImpScat}) into Eq. (\ref{EqChiB}),  we conclude that the dressed $R-$matrix
can be written as:
 \bea
 R_{a,a}(\theta) &=& e^{i\xi_a(\theta)}\quad (a\neq c^\pm),
 \nl
 P_\drs(\theta) &=& P^{(\pL)}(\theta)\;e^{i\xi_c(\theta)},
 \label{DressedRMat}\\
 Q_\drs(\theta) &=& Q^{(\pL)}(\theta)\;e^{i\xi_c(\theta)},
 \nonumber
 \eea
where $P^{(\pL)}(\theta)$ and $Q^{(\pL)}(\theta)$ are defined in (\ref{RmatSol}) with the replacement $\lambda\to \pL$, i.e. the dressed $R-$matrix in the off-diagonal BSG model, up to phases, 
 is essentially that of a \emph{diagonal} BSG model: 
as far as impurity scattering is concerned, everything happens
as if the SG parameter were \emph{renormalized} to an integer
 value, $\lambda=\frac \pL \qL  \too \pL$.
 The phases $\xi_a(\theta)$, which we will not need in the following, 
 are furthermore constrained by the relation:
 \be
\sum_{a=s,1,...,\nst -2,c}\!\!\!\!\!\!
\eta_a\;\hat\Phi_{a,s}(\omega)\;\hat\xi_a(\omega) = 0.
 \ee

\subsection{Rate equation for the current}
\label{SectionRateEqu}

With the dressed scattering (\ref{DressedRMat}) at hand, we can derive the exact transmission probabilities
for the QPs, as a function of the rapidity and the impurity temperature $T_\B$. As enforced by the degeneracies of the eigenvalues of the transfer matrix $\tau_s$, all neutral particles 
have transmission one (including the eventual breathers in the attractive case, see Appendix \ref{AppendixAttractive}), while the charged strings $\AA_c^\pm(\theta)$ are transmitted with probability $\Cal T_\lambda(\theta)=|R_{c^\pm,c^\pm}(\theta-\theta_\B)|^2$ (and are scattered as the anti-QP $\AA_c^\mp$ with probability $1-\Cal T_\lambda(\theta)$):
\be
\Cal T_\lambda(\theta)=\frac{1}{1+\big( \overline T\,e^\theta\big)^{-2\hspace*{.01cm}\pL}}
\label{TransmissionStrings}
\ee
Note that the apparent temperature dependence (we still have $\overline T=\frac{T}{T_\B}$)  observed in (\ref{TransmissionStrings}) is just an 
artefact of our choice of the arbitrary scale $E_0=T$ parametrizing momentum $p =T e^\theta$.
A formula for the current can be established by evaluating the rate of change 
of the charge due to the scattering QPs, following closely
the diagonal solution\citeOR{FLS-PRB} but with charged strings replacing solitons/antisolitons. 

The charge $\delta Q=\delta Q_++\delta Q_-$ flowing through the impurity during a time interval $\delta t$, receives contributions from positive strings $\AA_c^+(\theta)$ as well as negative strings $\AA_c^-(\theta)$ incoming towards the impurity at rapidity $\in[\theta,\theta+d\theta]$,  
that are both transmitted with probability $\Cal T_\lambda(\theta)$.
The number of incoming strings $\AA_c^\pm(\theta)$ during $\delta t$ being $\delta n^\pm(\theta)= (v_\f\delta t)\frac{\kb T}{h v_\f}\rho_c^\pm(\theta)d\theta$, the transmitted charge reads $\delta Q_\pm =(\pm\qL e)\times\delta n^\pm\times \Cal T_\lambda $
leading to $\frac{\delta Q_\pm}{\delta t} =\pm\frac{\qL e\kb T}{h}\rho_c^\pm$, 
so that the total current  $I=(\delta t)^{-1}\int d\theta (\delta Q_+-\delta Q_-) $ finally evaluates to  Eq. (\ref{CurrentTBA}).


\section{Discussion}
\label{sectionDiscussion}

\begin{figure}[h]
\renewcommand{\figurename}{\textbf{Figure}}
\renewcommand{\thefigure}{\textbf{\arabic{figure}}}
\begin{center}
\includegraphics[width=6.5cm]{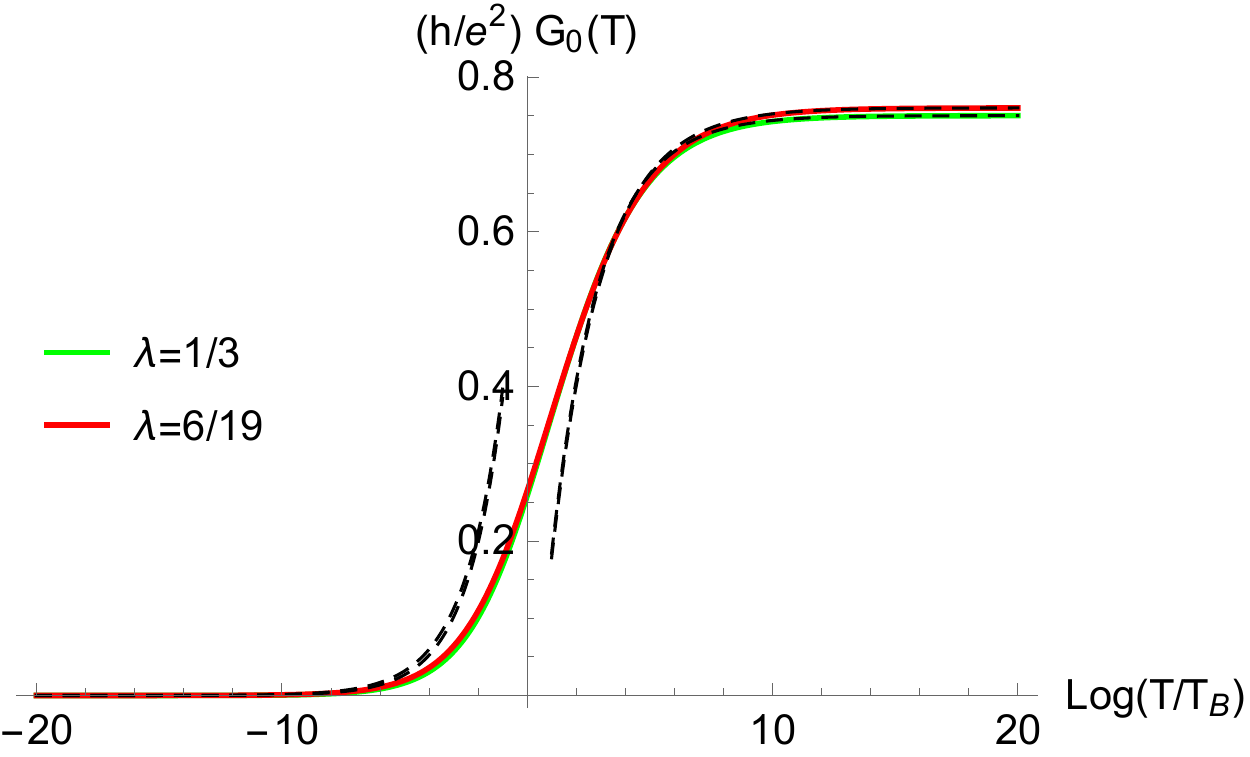}\\
\vspace*{0.4cm}
\includegraphics[width=4.cm]{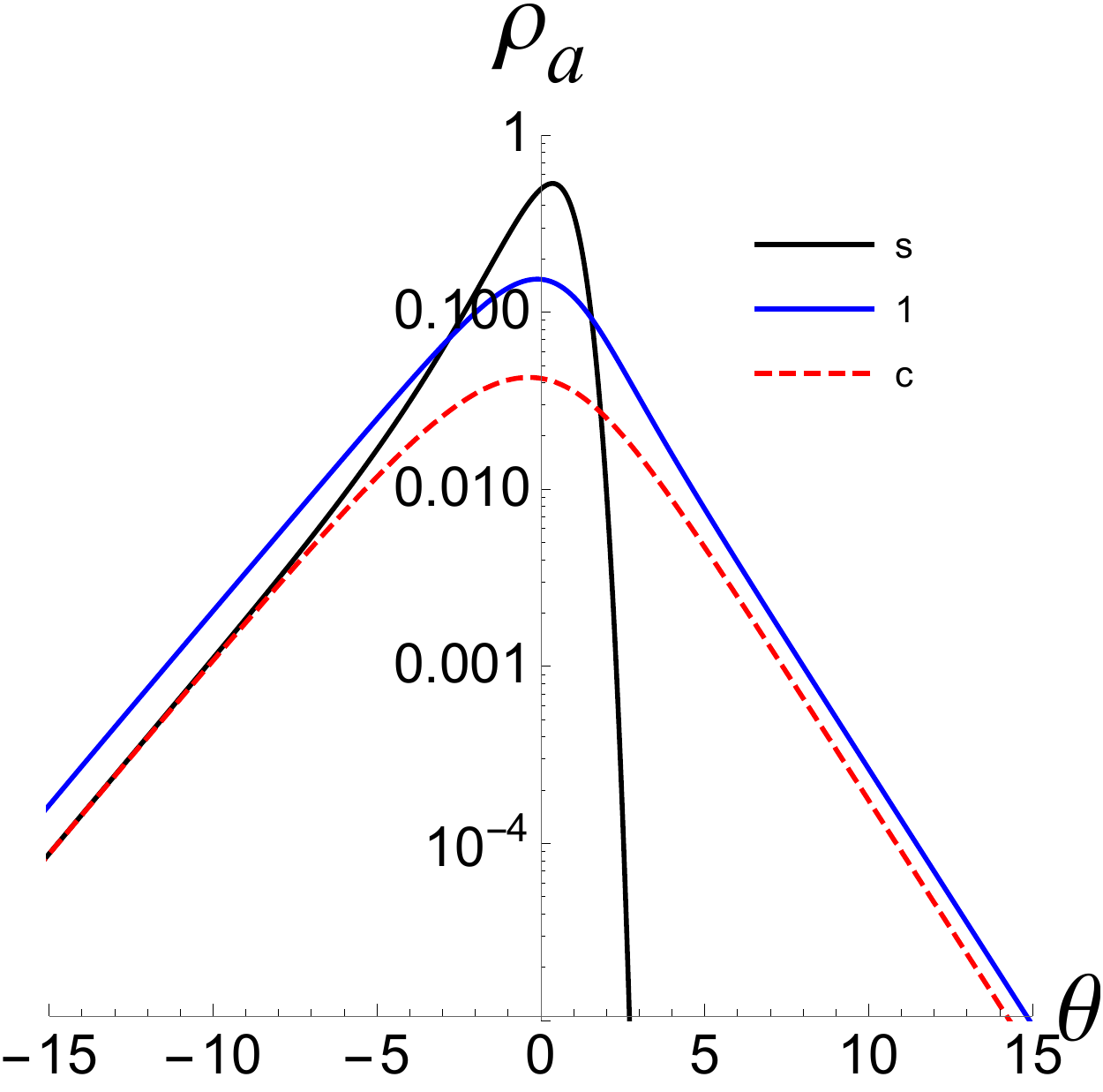}
\hspace*{-.4cm}
\raisebox{1.5cm}{\includegraphics[width=.6cm]{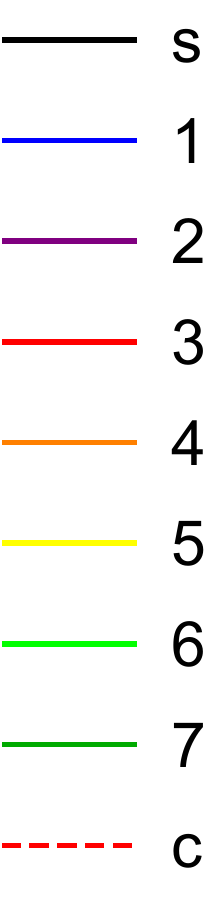}}
\hspace*{-.4cm}
\includegraphics[width=4.1cm]{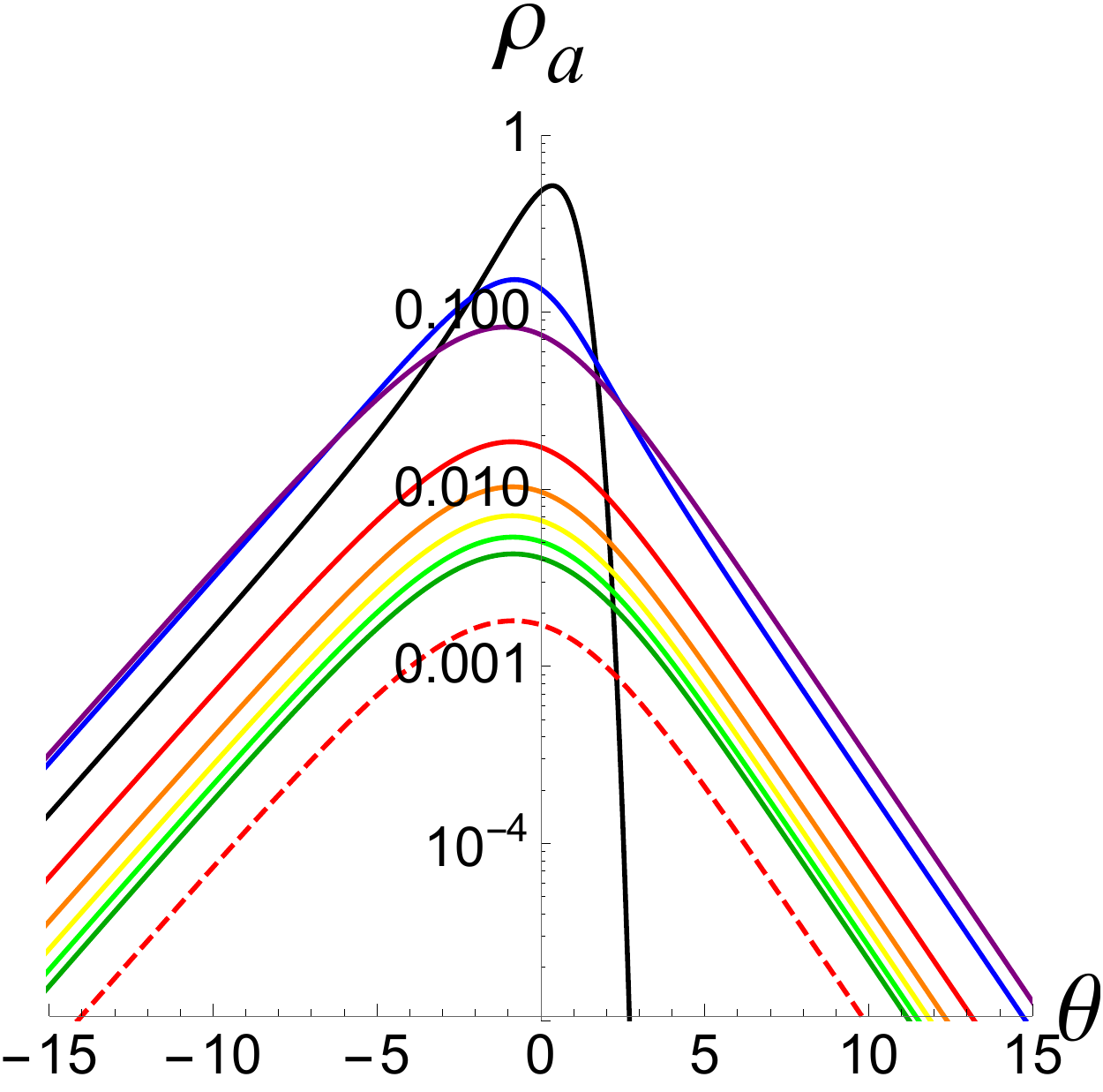}
\end{center}
\caption{{\bf[Top]} Universal scaling curves for the (reduced) linear conductance $\frac h{e^2}G_0(T/\TB)$ 
obtained from Eq. (\ref{G0TBA}), and 
shown here for  $\lambda_1=\frac13$ and $\lambda_2=\frac6{19}$,  illustrating  the  smoothness of the conductance  when 
$\lambda$ is varied.
The dashed lines are the 
exact asymptotics (\ref{G0AsymptIR}) and (\ref{G0AsymptUV}).
{\bf[Bottom]} QP densities as a function of $\theta$, shown here at $V=0$, for the case $\lambda_1$ [Left] and $\lambda_2 $ [Right].
 For those two close values at relative distance $\frac{\lambda_1-\lambda_2}{\lambda_1}\simeq 0.053$, the QP contents from which the conductance is evaluated are largely different (see also Figs.\ref{figPseudoElambdaEq1over3},\ref{figPseudoElambdaEq6over19} in Appendix \ref{AppendixExampleLambda}). Note in particular the different scales in the values of $\rho_c$, which decreases when the denominator $\qL$ increases. 
}
\label{figG0lambdaEq6over19}
\end{figure}

Our prediction (\ref{CurrentTBA}) for the finite voltage and temperature universal form of the current $I(V,T)$
allows for a  certain number of checks.
First, at vanishing temperature, as discussed in Section \ref{SectionThermoABA},
the strings freeze ; a possible description involving antisolitons only becomes
possible, and our prediction coincides with the $T=0$ predictions of Ref. \citeOR{FLS-PRB} exploiting
the diagonal character of the scattering. 

At finite temperature strings enter the stage and our prediction for the current is novel. A fist available benchmark is the high energy ($T\gg\TB$) limit of the
 the linear conductance $G_0(T)=\lim _{V\to 0}(\p I(V,T)/\p V)$, $G_\ind{max}=\frac{e^2}{h}\frac1{\lambda+1}$. From Eq. (\ref{G0TBA}) one gets $
\frac{h}{e^2}G_\ind{max} = (-1)^\alpha \qL^2 [ f_c]_{-\infty}^\infty$ and using the asymptotic values of the 
pseudoenergies, we
show  in Appendix \ref{AppendixDeltaFc}
that
$[ f_c]_{-\infty}^\infty=\frac{(-1)^\alpha}{\qL ^2(\lambda+1)}$. A second benchmark are
the  asymptotic regimes (large and small $\frac T{T_\B}$)  for the conductance,
that read 
(see Appendix \ref{AppendixKeldyshPT}):
\bea
\frac{G_0(T)}{G_\ind{max}}&\underset{T\ll \TB}{\sim}&\alpha\vph(\lambda)\,\Big(B({\textstyle \frac12,1+\frac1{2\lambda}})\frac T{\TB}\Big)^{2\lambda},
\label{G0AsymptIR}
\\
\frac{G_0(T)}{G_\ind{max}}&\underset{T\gg \TB}{\sim}&
1-\alpha\vph({\textstyle\frac{-\lambda}{\lambda+1}})\,\Big(B({\textstyle \frac12,1+\frac1{2\lambda}})\frac T{\TB}\Big)^{{\textstyle \frac{-2\lambda}{\lambda+1}}}
\label{G0AsymptUV}
\eea
where the coefficients read $\alpha\vph(\lambda)=
\frac{(\lambda+1)\Gamma(\lambda+1)^2 B\big(\lambda+1,\frac12\big)}{2}$
and  $B(a,b)=\frac{\Gamma(a)\Gamma(b)}{\Gamma(a+b)}$ is the Euler beta function.

The linear conductance $G_0(T)$ is shown in the top panel of Fig.\ref{figG0lambdaEq6over19} for two close values
of the SG parameter $\lambda_1=\frac 13$ and $\lambda_2=\frac 6{19}$
While the conductance varies smoothly between the two cases,
the QP spectrum changes drastically. One observes that the last band $\epsilon_c$ corresponding to the charged particle flattens when the denominator $\qL $ increases and consequently $P_c$ and $\rho_c$ decrease. For large $V/T$ a simple charge counting
argument leads to a reduction  of the number of  charged QP states  scaling as $\qL^{-1}$, 
and at small $V/T$ we show  in Appendix \ref{AppendixDeltaFc} that the number of charged QPs is even further reduced by an additional factor and scales as $\frac{1}{\qL^2}$, for large $\qL$ and typical fractions $\lambda$. This of course makes the approach to any given value $\lambda$ by a rational series $\lambda_\ell\underset{\ell\to\infty}{\too} \lambda$ highly singular, the charge $\qL \vph_{\ell}$ becoming infinitely large while the number of excited charged particles $N_c^\pm$ (at fixed $V,T$) goes to zero.

 The number of such strings, or ``book-keeping" particles (they keep track of the diffusion of charge in momentum space), turns out to be finite ($=\nst$) when $\lambda=\frac \pL \qL $ is a rational number, simply because then any scattering matrix element is periodic: $S(\theta+i\qL\pi)=S(\theta)$, implying 
 that strings 
 above a certain length (that scales as $\qL$) are not to be taken into account in the QP spectrum (or more precisely the description of their effects are incorporated into the first $\nst$ strings). 
 One can interpret this fact in the following way: when the denominator $\qL$ of $\lambda$ grows,
 it becomes more and more demanding to impose that all the diffusion interferes destructively, requiring the
 introduction of more and more species of book-keeping particles.

\begin{figure}[ht]
\begin{center}
\hspace*{-.8cm}
\includegraphics[width=8.4cm]{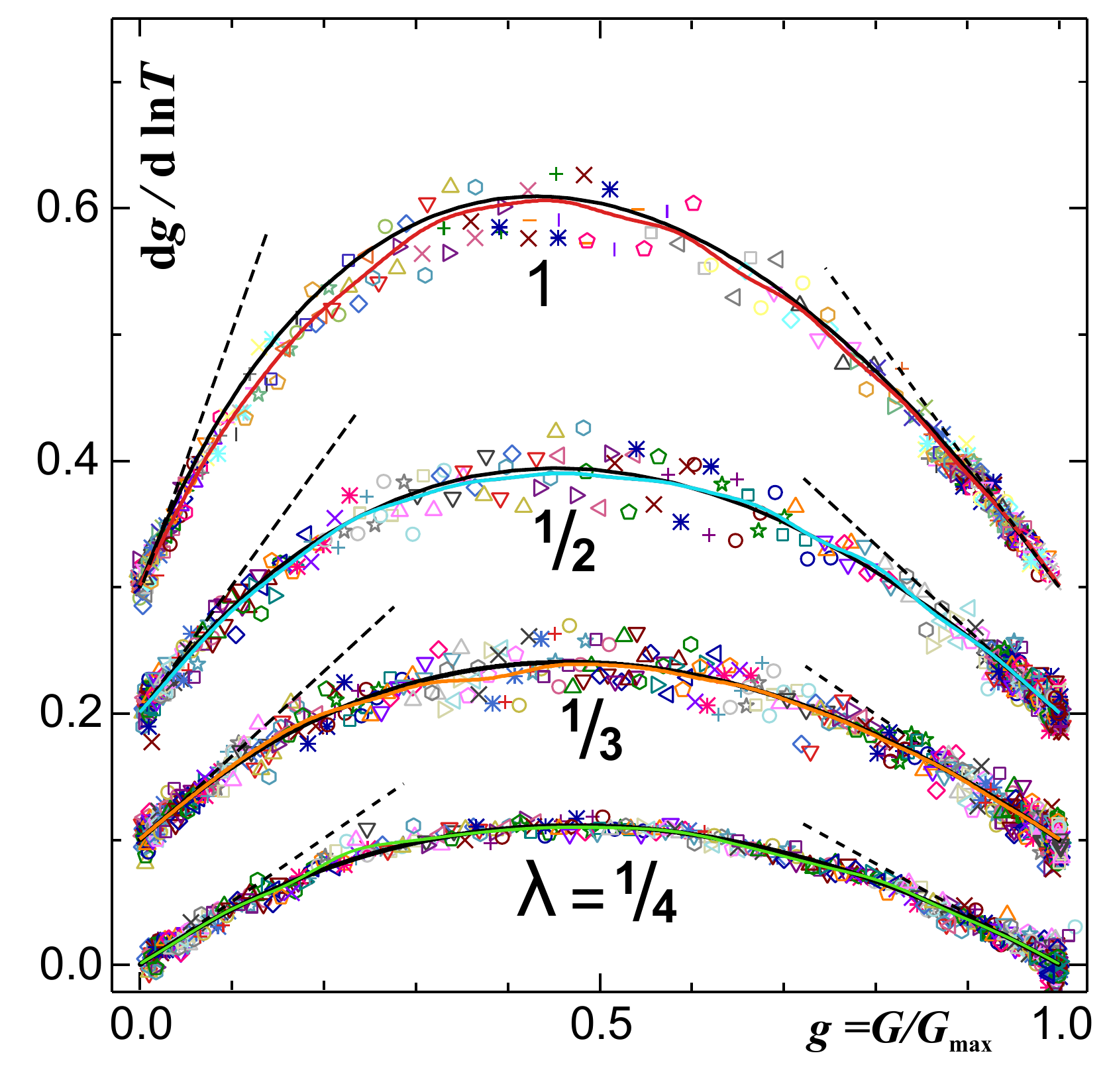}
\end{center}
\caption{
Transport measurements in the off-diagonal BSG model: universal comparison. 
Plotted here is
the dimensionless and universal quantity 
$\beta_{\rm RG}=\frac{\p g}{\p\ln T}$
as a function of $g=\frac{G_0(T)}{G_{\rm max}}=\frac{(\lambda+1)hG_0}{e^2}$.
The symbols display the raw experimental data  extracted from conductance measurements  in the experimental setting of Ref.\cite{Anthore18}, that is  described at low energy by the BSG model.
The SG parameter is fixed at $\lambda=\{1,\frac12,\frac13,\frac14\}$.
A vertical shift of 0.1 is applied for clarity.
The  averaged and noise-filtered raw data (coloured curves) are compared to the theoretical predictions (full black lines) for free fermions 
 ($\lambda=1$) and in the off-diagonal case by the TBA equations (\ref{epsilonTBA},\ref{G0TBA})  ($\lambda<1$). 
Dashed  straight lines at $g=0$ and $g=1$ indicate the asymptotic behaviors (\ref{G0AsymptIR},\ref{G0AsymptUV}). Note that there are no fitting parameters.
}
\label{figdGdLogT}
\end{figure}

Last, we compare our predictions to recent measurements of the $I(V,T)$ characteristics of a device
implementing the tunnelling in a resistive environment,  with a resistance $R_n=\frac{1}{n}\frac h{e^2}$ 
that can be tuned with high precision to the values $n=1,2,3,4$ \citeOR{Anthore18}. The universal  low energy regime  ($E\ll E_c$ with $E_c$ the charging energy of the environment)  is described\citeOR{SafiSaleur} by the sine-Gordon model out-of-equilibrium with $\lambda=\frac1n$.
It is possible to  get rid of the non-universal scale $T_\B$ by considering $\beta_{\rm RG}=\frac{\p g}{\p\ln T}$  with  the dimensionless conductance   $g=\frac{G_0}{G_{\rm max}}$ and $G_{\rm max} = \frac{e^2}h\frac1{\lambda+1}$.
The universal quantity $\beta_{\rm RG}(g)$ is the renormalization group $\beta$-function for the conductance ; it vanishes at the low energy fixed point where the impurity is at full opacity ($g=0$) and at the high
energy fixed point where the systems reaches its maximal conductance $g=1$. Note the remarkable agreement (Fig.\ref{figdGdLogT}), at the level of this universal quantity and with no fitting parameter, between  our predictions and the averaged experimental data.

\section{Conclusion}

In this paper we have shown that the boundary sine-Gordon model subject to a constant
voltage bias bears an exact solution
for arbitrary rational value of the sine-Gordon coupling $\lambda$. 
This gives  explicit access to the universal scaling function for the current $I(V,T)$
at arbitrary voltage and temperature, and virtually for arbitrary real value of $\lambda$
by successive approximations.
This yields amongst other the exact solution to the problem of  tunnelling
between Tomonaga L\"uttinger liquids with arbitrary TLL parameter $K=\frac1{\lambda+1}$, and  of
a tunnelling junction coupled to a resistive environment with resistance  $R=\lambda\frac{h}{e^2}$.
We believe that our results can also be of  interest to the analysis of 
transport in the fractional quantum Hall effect, whose analysis sometimes
resorts to an extrapolation of the results of Ref.\citeOR{FLS-PRL,FLS-PRB},
obtained for $\lambda$ integer, to non-integer values \citeOR{Chang03}.

As soon as $\lambda$ is not an integer, 
the exact solution requires the use of the string solutions of the Algebraic Bethe Ansatz,
that are additional quasiparticles, carrying  entropy but no kinetic energy, 
with a quasiparticle spectrum displaying an astonishingly 
complex structure as a function of $\lambda$.  

There is an adage, that can be phrased as "tunnelling selects the basis",
applying to all interacting quantum  impurity models that could be solved so far. It means that 
the ``good" basis of many-body states that solves the problem is built out of the quasiparticles that scatter one by one at
the impurity, without particle production (in technical terms, factorizing the scattering on the impurity).
In the off-diagonal sine-Gordon model,   although solitons/antisolitons
do factorize the impurity scattering, they don't constitute  the good basis:
charge transport, that is ballistic in the diagonal case, 
becomes diffusive in momentum space in the off-diagonal case. The ``good" basis is that of the entropic string QPs, that factorize the impurity
scattering \emph{and} diagonalize the bulk scattering.

Although strings are sometimes considered as ``fictitious" particles, we show that taking  seriously their existence
and using them to compute the electrical current $I(V,T)$, fits with excellent quantitative agreement
experiments carried for $\lambda\in\{\frac12,\frac13,\frac14\}$. Although this does  not a constitute
a proof of ``existence" of the string QPs, in particular because the current $I(V,T)$ being a continuous function of $\lambda$ is a blind observable to the internal string QP spectrum,
we can at least assess that their formal mathematical existence leads 
to prediction of physical quantities. 
It would be very interesting to have experimental investigations
at more complicated fractions than $\lambda^{-1}$ integer, where ($i$) the QP charge differs from the simple form $\qL=\lambda^{-1}$ and  ($i\!i$) the transmission probability $\Cal T(\theta)$ displays the non-trivial dependence on the numerator $\pL\neq 1$ of  $\lambda$.
It  seems fractions of the type $\frac56=\frac12+\frac13$, or $\frac7{12}=\frac13+\frac14$ would be the simplest candidates obtained by  lining up two resistors $\frac{h}{ne^2}$ and $\frac{h}{n'e^2}$ in series in the setting of Ref.\citeOR{Anthore18}.

The fluctuations of the charge transferred across the impurity is under investigation,
but is expected on general ground to be a continuous function of $\lambda$  therefore
 not revealing the charge $\qL$ of the dressed strings QPs -- we know at least that in the limit $V/T\to\infty$ this is the case. We believe this is consistent with the fact that the \emph{bare}
 strings basis, in which the QPs' charge is localized in momentum space, is then the correct basis.
 The noise will therefore involve contributions from  \emph{all} bare strings $j$ (carrying charge $2k_j$)
 compatible with a net effect washing out the charge $\qL$.

If ``existence" is taken in the strong sense of directly observable, one could think of modifying
the probing conditions of the system (e.g. by AC forcing, different temperatures, quenches...) so as to observe strings.
It might also be that any such attempt destroys their existence in the sense
that the breaking of  the integrable structure would lead to an immediate blurring of the complex structure of the string QPs spectrum.

{\bf Acknowledgments}. The author thanks A. Anthore, B. Doyon and F. Pierre for valuable discussions.

\section{Supplementary Material}

\subsection{Practical implementation of the TBA by the example: $\lambda=\frac13$ and $\lambda=\frac6{19}$}

\label{AppendixExampleLambda}

We present the QP spectrum $\AA_a$ and the TBA equations determining the pseudo energies $\epsilon_a(\theta)$. The TBA equations are coupled  equations of the form $\epsilon_a=\Cal F_a(\{\epsilon_{b}\})$
where $\Cal F_a$ is a non-linear integral operator taking as arguments all pseudoenergies.
A numerical solution can be obtained by successive approximations, starting from an initial guess $\epsilon_a^{(n=0)}$ (e.g. $\epsilon_a^{(n=0)}=\delta_{a,s}e^\theta$) and iterating $\epsilon_a^{(n+1)}=\Cal F_a(\{\epsilon_b^{(n)}\})$ until convergence is reached. 

Once the pseudo-energies have been computed, they give explicit access  to the  finite $V,T$
densities of occupied QPs per unit length $ \frac{\kb T}{h v_\f}\rho_a$
and total densities (occupied+empty) QPs per unit length $ \frac{\kb T}{h v_\f}P_a$
as $P_a=\eta_a\p_\theta\epsilon_a$ and $\rho_a=P_af_a=-\eta_a\p_\theta L_a$ with $L_a=\log\big(1+e^{\mu_a-\epsilon_a}\big)$ and $f_a^{-1}=1+e^{\epsilon_a-\mu_a}$.

In the following we use the notation $\overline V=\frac{eV}{\kb \TB}$ and $\overline T=\frac T{\TB}$.

\subsubsection{$\lambda=\frac13$}
\label{AppendixExampleLambda1over3}

In this case the continued fraction decomposition ({\ref{deffraccont}) terminates immediately
so $\alpha=1$, $\nu_1=m_\alpha=3$ and we have $m_\alpha+1=4$ QPs:
\begin{tabular}{rl}
\hspace*{.4cm}
$\bullet$ &  the soliton $\AA_s$, carrying energy and entropy,
\\$\bullet$ & the first string $\AA_1$, carrying  entropy only,
\\$\bullet$ &  and two charged strings $\AA_{c}^-=\AA_2$, $\AA_{c}^+=\AA_3$ carry-
\\& ing entropy and charge $\pm\qL=\pm3$.
\end{tabular}
The TBA diagram encoding the scattering kernel is shown in Fig. 
\ref{figPseudoElambdaEq1over3}{\bf.(a)}

The non-vanishing chemical potentials are $\mu_c^\pm=\pm 3\frac{eV}{2\kb T}$. 
There are $m_\alpha=3$ independent pseudo-energies $\epsilon_s$, $\epsilon_1$ and $\epsilon_c$, and a single family $\Cal F_1$ so the kernel $\kernel_{i,j}$ 
 involves a single function $\kernelem_{p_1}=\frac1{\cosh\theta}$ and a sign $\eta_1=1$.  The finite $V,T$ 
 dimensionless densities $P_a,\rho_a$ are determined by the TBA equations:
 \bea
 \epsilon_s&=&e^\theta -\frac1{2\pi}\frac1{\cosh\theta}\star L_1
 \nl
  \epsilon_1&=& -\frac1{2\pi}\frac1{\cosh\theta}\star \big(L_s+L_{c^+}+L_{c^-}\big) 
 \nl
  \epsilon_c&=& -\frac1{2\pi}\frac1{\cosh\theta}\star L_1
  \label {TBA1over3}
  \eea
 The denominator $\pL=1$ of $\lambda$ fixes the form of the transmission probability:
 \be
 \Cal T_{\frac13}(\theta) =   \frac{1}{1+\big(\overline  T e^\theta\big)^{-2}}
 \ee
so the reduced current $\overline I= \frac{h}{e\kb T} I$ reads:
\be
 \overline I_{\frac13}\vph(\barV,\barT)=3\int_{-\infty} ^\infty \!\!\! d\theta \;\Cal T_{\frac13}(\theta)\;\frac{\p}{\p\theta}
 \ln\frac
 {1+e^{-\epsilon_c(\theta)-\frac{3\barV}{2\barT}}}
 {1+e^{-\epsilon_c(\theta)+\frac{3\barV}{2\barT}}}
 \ee
\begin{figure}[h]
\renewcommand{\figurename}{\textbf{Figure}}
\renewcommand{\thefigure}{\textbf{\arabic{figure}}}
\begin{center}
\includegraphics[width=3.8cm]{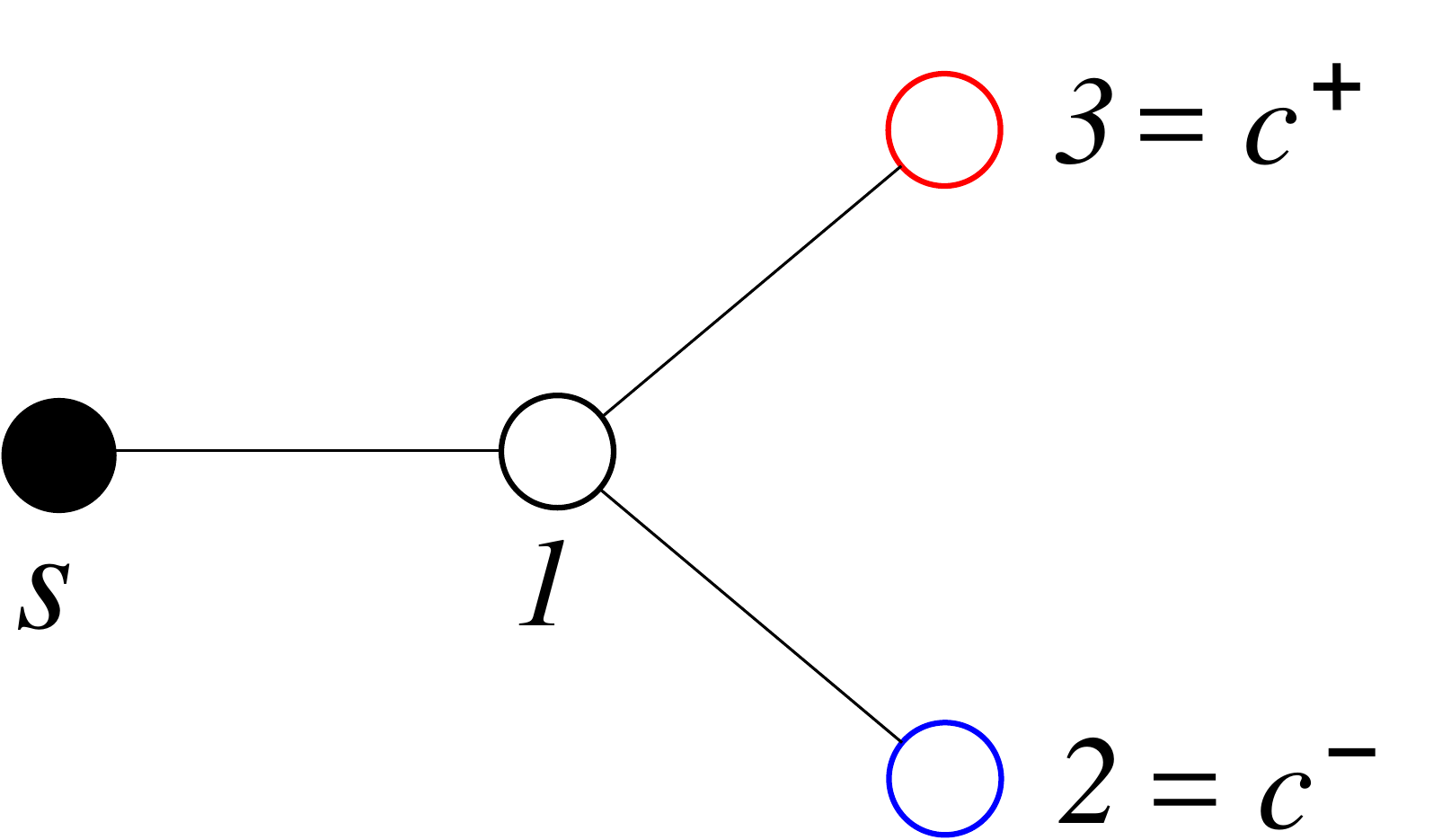}
\vspace{.8cm}\\
\includegraphics[width=4.5cm]{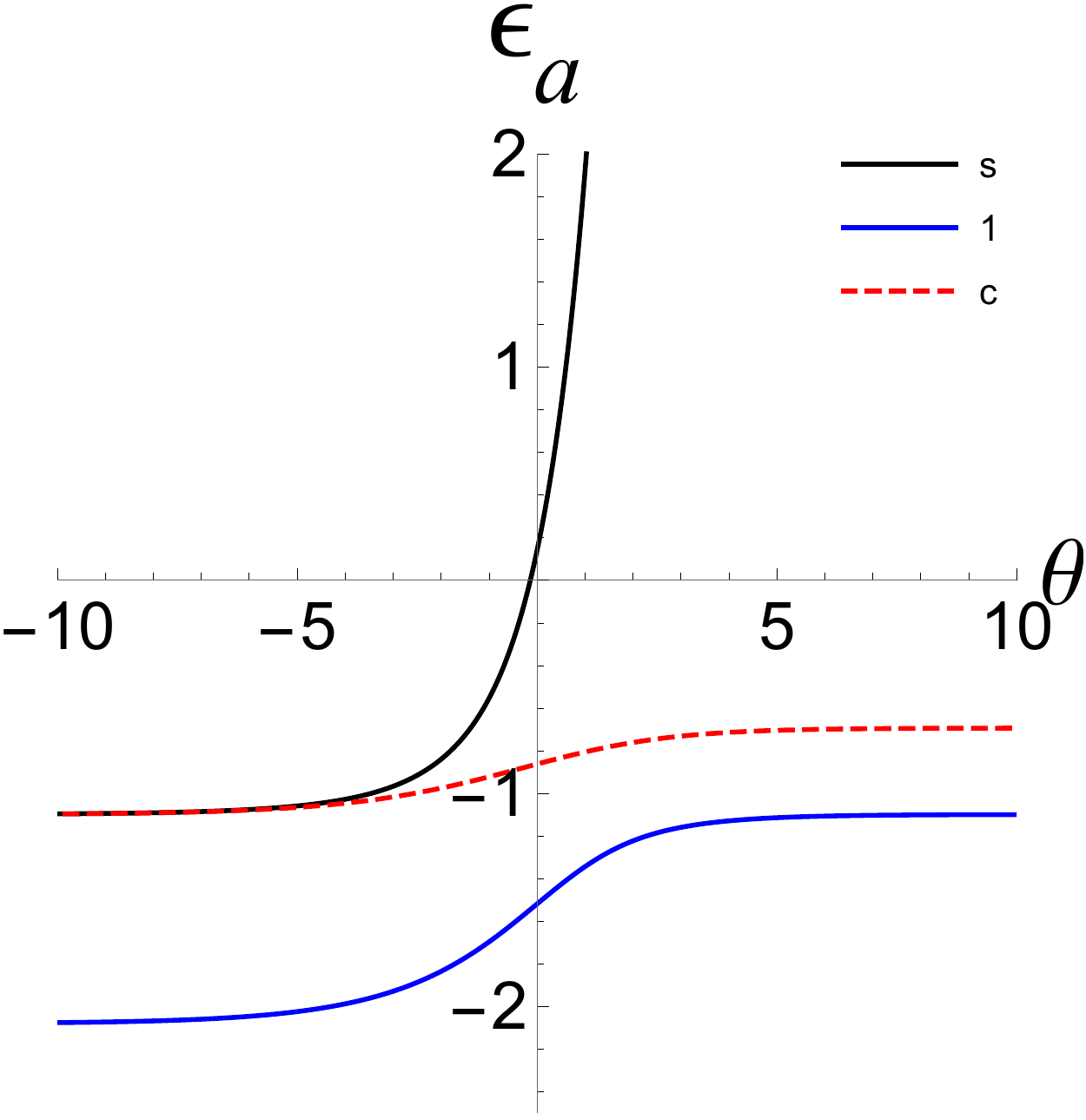}
\end{center}
\caption{
    {\bf[Top]} TBA diagram for  $\lambda=\frac13$. The black node is "massive" (i.e. the associated QP, namely the soliton, carries energy) whereas empty nodes are associated to strings.
  {\bf[Bottom]} QP spectrum for $\lambda=\frac13$: the pseudo-energies $\epsilon_a(\theta)$ as a function of $\theta$, for  $a\in\{s,1,c\}$, obtained by numerical integration of the TBA equations (\ref{TBA1over3}) at $V=0$.
}
\label{figPseudoElambdaEq1over3}
\end{figure}
Numerical integration of the  TBA equations (\ref{TBA1over3}) at fixed $\frac {\barV}{\barT}$ yields the pseudo-energies $\epsilon_a(\frac {\barV}{\barT};\theta)$ that are displayed for $\frac {\barV}{\barT}=0$ in Fig. \ref{figPseudoElambdaEq1over3}{\bf.(b)}.

\subsubsection{$\lambda=\frac6{19}$}
\label{AppendixExampleLambda6over19}

The continued fraction decomposition $\lambda=\frac1{3+\frac16}$ has $\alpha=2$ terms $\nu_1=3$, $\nu_2=6$, so from the numbers (Eq. (\ref{defmi})) $m_1=3$,  $m_\alpha=9$ we see that 
we have $10=m_\alpha+1$ QPs, grouped in two families $\Cal F_1=\{\AA_s,\AA_1,\AA_2\}$, $\Cal F_2=\{\AA_3,...,\AA_7,\AA_{c}^-=\AA_8,\AA_{c}^+=\AA_{9}\}$:
\begin{tabular}{rl}
\hspace*{.4cm}$\bullet$ & the soliton $\AA_s$, carries energy and entropy,
\\$\bullet$ & the two charged strings $\AA_{c}^\pm$ carry entropy and 
\\ & charge $\pm\qL=\pm19$,
\\$\bullet$ & all the other strings $\AA_{a=1,...,7}$ carry  entropy only.
\end{tabular}
The TBA diagram encoding the kernel is shown in Fig. 
\ref{figPseudoElambdaEq6over19}{\bf.(a)}.
The non-vanishing chemical potentials are $\mu_c^\pm=\pm \frac{19 \overline V}{2\overline T}$. 
There are $m_\alpha=9$ independent pseudo-energies $\epsilon_s$, $\epsilon_{1,...,7}$ and $\epsilon_c$.
The kernel has  nearest neighbours  entries involving two independent functions  $\overline\Phi_1=\frac1{2\pi}\kernelem_{p_i}$ with   
with $p_1=1$, $p_2=\frac16$:  
$$\overline\kernelem_1(\theta)=\frac1{2\pi}\frac{1}{\cosh\theta}\qquad ; \qquad \overline\kernelem_2(\theta)=\frac1{2\pi}\frac{3}{\cosh3\theta}$$
 as well as a self-interaction entry involving the function
$\frac1{2\pi}{\rm TF}^{-1}\big[ \frac{\hat\kernelem_{1} \hat\kernelem_{1/6} }  {\hat\kernelem_{5/6}}  \big]$:
$$\overline\kernelem^\ind{self}_{1,2}(\theta)=\frac1{4\pi}
\int_{-\infty}^\infty d\omega\, e^{i\omega\theta}\, 
\frac{
\cosh\frac{5\pi\omega}{12}
}
{
\cosh\frac{\pi\omega}{2}    \cosh\frac{\pi\omega}{12}.
}
$$

\begin{figure}[h]
\renewcommand{\figurename}{\textbf{Figure}}
\renewcommand{\thefigure}{\textbf{\arabic{figure}}}
\begin{center}
\raisebox{1.3cm}{(a)}
\includegraphics[width=8.cm]{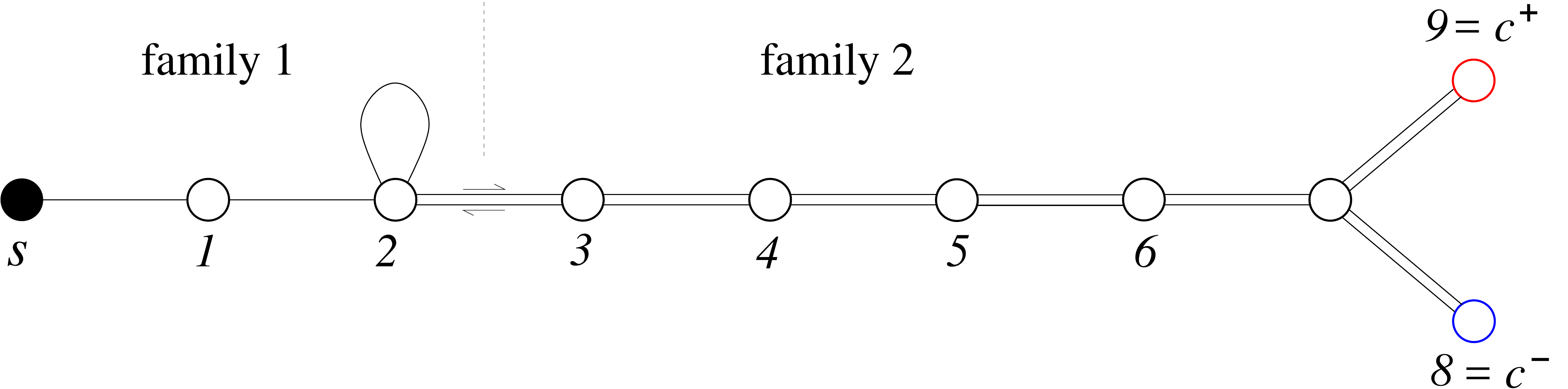}
\vspace*{.2cm}
\raisebox{1cm}
{\includegraphics[width=4.55cm]{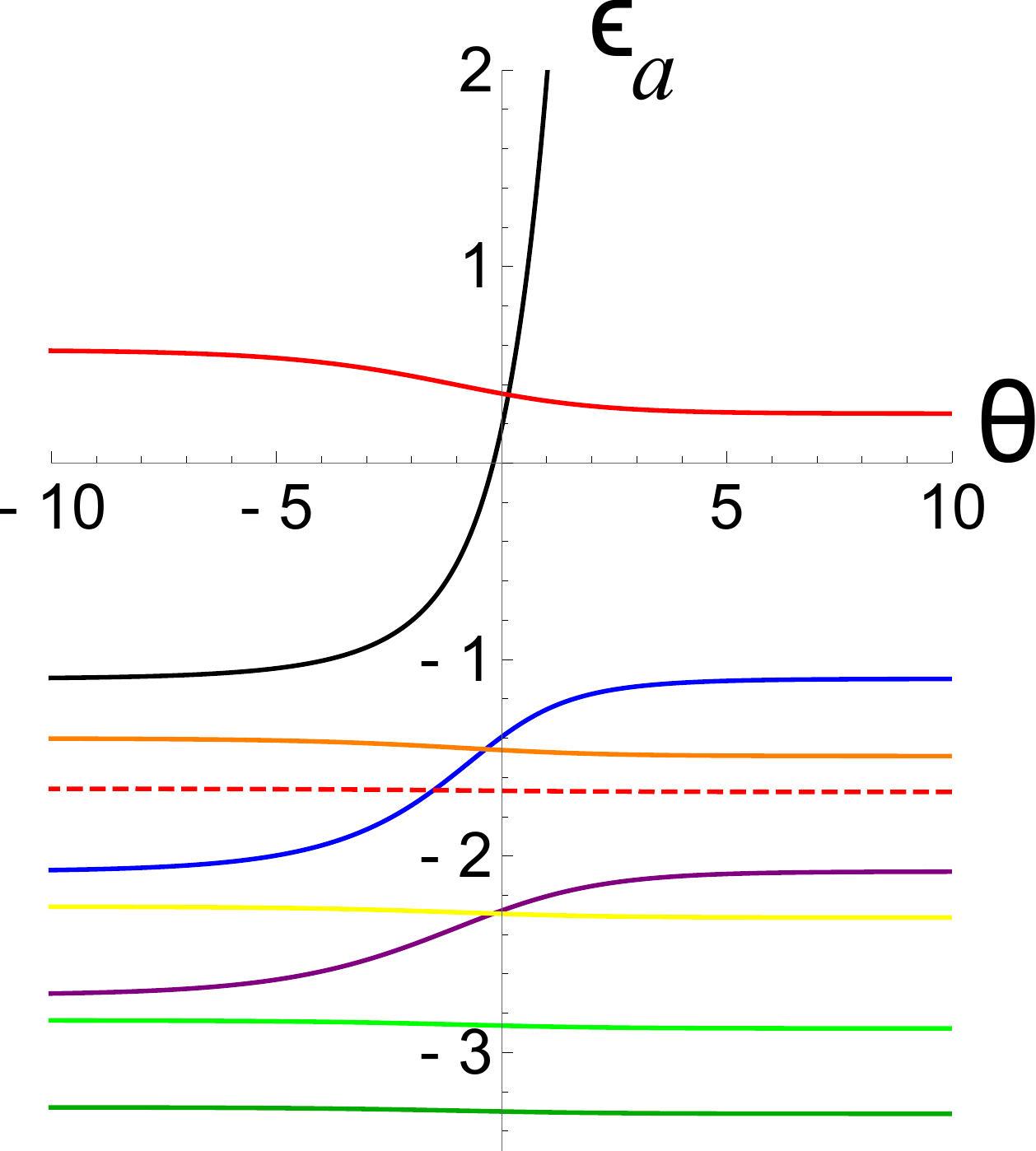}
\hspace*{.3cm}
\raisebox{1.1cm}{\includegraphics[width=.95cm]{figTBALegend9over29V0.pdf}}
}
\end{center}
\caption{
  {\bf[Top]} TBA diagram for  $\lambda=\frac6{19}$. The black node is "massive" (i.e. the associated QP, namely the soliton, carries energy) whereas empty nodes are associated to strings.
  {\bf[Bottom]} QP spectrum for $\lambda=\frac6{19}$: 
the pseudo-energies $\epsilon_a(\theta)$ as a function of $\theta$, for  
$a\in\{s,1,2...,7,c\}$, obtaned by numerical integration of the TBA equations (\ref{TBA6over19}) at $V=0$. 
 Compare the additional complexity in the spectrum with respect to the close value $\lambda=\frac 13$ in Fig. \ref{figPseudoElambdaEq1over3}.
Compare also the scales, in particular  for the flatness of  the last band (charged QPs $\AA_c^\pm$) when the denominator $\qL$ increases.
}
\label{figPseudoElambdaEq6over19}
\end{figure}

The  signs of the two families are  $\eta_1=$ and $\eta_2=-1$.  
The finite $V,T$ 
 dimensionless densities $P_a,\rho_a$ are determined by the TBA equations (\ref{epsilonTBA}) which become:
 \bea
  \epsilon_s&=&e^\theta -\overline\kernelem_1\star L_1
 \nl
  \epsilon_1&=& -\overline\kernelem_1\star \big(L_s+L_2\big) 
\nl
  \epsilon_2&=& -\overline\kernelem_1\star L_1
  -   \overline\kernelem^\ind{self}_{1,2}\star L_2 
  -\overline\kernelem_2\star L_3
 \nl
  \epsilon_3&=& -\overline\kernelem_2\star \big(-L_{2}+L_{4}\big)
\nl
  \epsilon_a&=& -\overline\kernelem_2\star \big(L_{a-1}+L_{a+1}\big)
  \qquad a=4,5,6
\nl
  \epsilon_7&=& -\overline\kernelem_2\star \big(L_6+L_{c^+}+L_{c^-}\big) 
 \nl
 \epsilon_c&=& -\overline\kernelem_2\star L_7
\label {TBA6over19}
  \eea
 The denominator $\pL=6$ of $\lambda$ fixes the form of the transmission probability:
 \be
 \Cal T_{\frac6{19}}(\theta) =   \frac{1}{1+\big(\barT  e^\theta\big)^{-12}}
 \ee
so the reduced current $\overline I= \frac{h}{e\kb T} I$ reads:
\be
 \overline I_{\!\frac6{19}}\vph(\barV,\barT)=19\!\int_{-\infty} ^\infty \!\!\! d\theta \;\Cal T_{\!\frac6{19}}(\theta)\;\frac{\p}{\p\theta}
 \ln\frac
 {1+e^{-\epsilon_c(\theta)-\frac{19\overline V}{2\overline T}}}
 {1+e^{-\epsilon_c(\theta)+\frac{19\overline V}{2\overline T}}}
 \ee
Numerical integration of the  TBA equations (\ref{TBA6over19}) at fixed $\frac {\overline V}{\overline T}$ yields the pseudo-energies $\epsilon_a(\frac {\overline V}{\overline T};\theta)$ that are displayed for $\frac {\overline V}{\overline T}=0$ in Fig. \ref{figPseudoElambdaEq6over19}{\bf.(b)}.


\subsection{Bethe equations for solitons and strings}

\label{AppendixABA}

In this Appendix one gathers some technical details about the Algebraic Bethe Anstatz (ABA).
One first starts (section \ref{AppendixBareABA}) by presenting the  discrete ABA equations
constraining the 
 allowed  configurations $\conf$ of QPs in terms of bare strings $B_{k_j}(\theta)$
 (defined in Eq.(\ref{defStrings})),  and their continuum limit
 when the number of QPs goes to infinity.
We then describe the dressing operation   in section \ref{AppendixDressedABA},
 implementing a transformation on the ensemble of solitons $\AA_s(\theta)$ and ``bare" strings $B_{k_j}(\theta)$ to reproduce (a variant of) the quantization equations
for solitons and strings first derived (for the string sector only) by Takahashi \citeOR{Takahashi72,TakahashiBook}. 
Finally the effect of dressing on the charge is presented in section \ref{AppendixChargeDressing}.

\subsubsection{Bare BA equations}
\label{AppendixBareABA}

We  consider a configuration $\conf$ comprising  $N=N_s$ solitons at rapidities $\{\theta^{(s)}_i\}$ $(i=1,...,N_s)$
and $N_j$ occupied strings of type $j$ at rapidities $\{\theta^{(j)}_i\}$ $(i=1,...,N_j)$:
$\conf=\{\{\theta^{(s)}_1,...,\theta^{(s)}_{N_s}\},\{\theta^{(1)}_1,...,\theta^{(1)}_{N^{(1)}}\},...,\{\theta^{(\nst )}_1,...,\theta^{(\nst )}_{N_{\nst }}\}\}$.
The requirement that the system be left invariant when an arbitrary particle is moved through 
all other particles and brought back to its initial position, for a  system of finite length $L$, leads to the
fundamental equations that define an \emph{allowed} configuration:
\bea
e^{2i\pi\Cal N_s(\theta)} &=& \tau^{(s)}_\conf (\theta)\,e^{i\pi \delta_s} \,e^{-iLE_0 \,e^\theta}\nl
 e^{2i\pi\Cal N_j(\theta)}&=& \tau^{(j)}_\conf (\theta) \,e^{i\pi \delta_j}
\quad (j=1,...,\nst )
\label{defEqBADiscr}
\eea
where the function $\Cal N_a(\theta)$ evaluates to integers when $\theta=\theta^{(a)}_i$ is an occupied rapidity
for species $a=s,1,...,\nst $ (i.e. $\theta^{(a)}_i\in\conf$), 
 $\delta_a\in\{0,1\}$ is a constant (depending on $\conf$) that will not be needed here, and where, in addition to the transfer matrix for solitons $\tau^{(s)}(\theta)$, one has introduced the
transfer matrices for strings $\tau^{(j)}(\theta)$:
\be
\tau_\conf^{(j)}(\theta) = \prod_{i=1}^{N_s} S_{j,s}(\theta^{(s)}_i-\theta) \prod_{j'=1}^{\nst } \prod_{i=1}^{N_{j'}} S_{j,j'}(\theta^{(j')}_i-\theta),
\label{defTau}
\ee
the scattering matrices being given in Eqs. (\ref{SmatSolString},\ref{SmatStringString}).
Note the presence of the factor $e^{ipL}$ in Eqs.(\ref{defEqBADiscr}) for solitons \emph{only},
showing that the strings are fictitious particles carrying no momentum and no kinetic energy.
The set of rapidities for which the right-hand side of Eqs.(\ref{defEqBADiscr}) evaluates to one
is actually larger than $\conf$: this defines additional, empty states  called holes, at rapidities
$\conf^h =\{  \theta^{(a),h}_i \}$. Thus, the total set of allowed rapidities (or ``Bethe roots") is $\conf\cup\conf^h$.

Introducing the (reduced) total density of Bethe roots per unit length, $P_a=-\eta_a\,  (-1)^{\delta_{a,s}+\delta_{a,\nst}}\,\frac{1}{E_0L}\frac{d \Cal N_a}{d\theta}$
(the signs $\eta_a$ are necessary to obtain a positive density\citeOR{Takahashi72})
as well as the (reduced) density of occupied Bethe roots per unit length $\tilde\rho_a^{}$, and taking 
the derivative of the logarithm of Eqs.(\ref{defEqBADiscr}) yields in the thermodynamical limit $L\to\infty$: 
\be
-\eta_a \, (-1)^{\delta_{a,s}+\delta_{a,\nst}}\,P_a = \delta_{a,s}e^\theta + \frac1{2\pi}\Phi_{a,b}\star \tilde\rho_b^{}
\label{contBAbare}
\ee
with $\Phi_{ab}(\theta) = -i\p_\theta \log S_{ab}(\theta)$ and the scattering matrix elements are given in Eqs. (\ref{SmatSolString},\ref{SmatStringString}). Since all particles scatter non-trivially on  all others, the bare kernel $\Phi_{ab}(\theta)$ has a complicated
structure. It can be made explicit in Fourier space using the elementary function
$\hat \varphi_k^{(\varepsilon)}(\omega) = \frac1{2i\pi} \int d\theta e^{-i\omega\theta} \p_\theta\log g^{(\varepsilon)}_k(\theta)$
(the function $g_k^{(\varepsilon)}$ is defined in (\ref{defgfunction})),
that reads
\bea
\hat\varphi_k^{(+)}(\omega)&=&
(1-\delta_{\tilde k,0})\;\;\frac{\sinh((1-\tilde k\lambda)\frac{\pi\omega}{2\lambda})}{\sinh(\frac{\pi\omega}{2\lambda})}
\label{phikomega}
\\
\hat\varphi_k^{(-)}(\omega)&=&\hat\varphi_{k+\lambda^{-1}}^{(+)}(\omega)
\nonumber
\eea
where $\tilde k =\frac{2}{\lambda}\big]\frac{k\lambda}{2}\big[$ and $]x[=x-[x]$ is the fractional
part of $x$. 

\subsubsection{Dressed BA equations}
\label{AppendixDressedABA}

It is very convenient, to simplify the continuous Bethe equations  (\ref{contBAbare}), to perform 
a particle-hole transformation by defining $\rho_j = \rho_j^{(-)}$ (and $\tilde\rho^{(+)}=P_j-\rho_j$)
 on all strings of type $j<\nst $. It means that we describe an allowed,
  but empty Bethe root $\theta^{(j)}_i\in\conf^h$ -- i.e. $\theta^{(j)}_i$ is solution of the 
  Bethe equation $\tau^{(j)}(\theta^{(j)}_r)=e^{i\pi\delta_j}$, but there is no insertion of the
   corresponding operator $B_{k_j}(\theta^{(j)}_r)$ in the product  (\ref{EigenVect}) -- 
    as an \emph{occupied} state for a new particle $\AA_j(\theta)$ at $\theta=\theta^{(j)}_i)$,
     whereas an initially occupied $B_{k_j}(\theta^{(j)}_i)$ string is described as an 
     allowed but empty $\AA_j(\theta^{(j)}_i)$ state. Thus the new particles $\AA_j$  are 
     holes in an  (interacting) sea of  strings. This sea could be termed the Zamalodchikov-Fateev sea ; it is the new vacuum, and   formally reads: 
 \be
 \ZFS=
 \prod_{j=1}^{\nst -1}
 \;\;\prod_{\theta^{(j)}_i\in\conf\cup\conf^h}
 B_{k_j}(\theta^{(j)}_i)\;\;\Ket{\Omega}.
 \label{defZFS}
 \ee 
 We also introduce  $\AA_{\nst }(\theta)=B_{k_{\nst }}(\theta)$, with density $\rho_{\nst }=\tilde\rho_{\nst }$, and define $\rho_{s}=\tilde\rho_{s}$ since the soliton mode is unaffected by dressing.
The new  reference state $(\ref{defZFS})$,
is a highly interacting object, causing the excitations $\AA_j(\theta)\ZFS$ above
it to be dressed by their interaction with the sea. This dressing affects both the scattering and the charge densities.

The effect of the particle-hole transformation is described by  introducing 
  the projector  $\mathbb{P}$ onto strings that are affected by the particle-hole transformation
(i.e. $\mathbb P_{ab}=\delta_{ab}(1-\delta_{a,\nst }-\delta_{as})$), the diagonal 
matrix ${\boldsymbol \eta}$ with entries $\eta_{j,j'}=\delta_{j,j'}\eta_j\vph$,  
 the vector $\hat{\boldsymbol  P}$ in Fourier 
space
 with components
  $\hat P_j$, the bare kernel $\hat {\boldsymbol \Phi}$ with components $\hat \Phi_{j,j'}$,
  and the vector $\hat {\boldsymbol \Phi}^{(s)}$ with components $\hat\Phi_{s,j}$. 
  Using $\tilde{\boldsymbol \rho}^{(+)} = \mathbb{P}{\boldsymbol P} + (1-2\mathbb{P}){\boldsymbol \rho}$,
  we have from (\ref{contBAbare}):
   $-{\boldsymbol\eta}\hat{\boldsymbol P}=\hat{\boldsymbol \Phi}
  \Big(\mathbb{P}\hat{\boldsymbol P}+(1-2\mathbb P)\hat{\boldsymbol \rho}\Big)
  + \hat {\boldsymbol \Phi}^{(s)}\hat\rho_s$
 which can be recast as 
$\hat{\boldsymbol P}=\hat{\boldsymbol  \kernel}\hat{\boldsymbol \rho} + \hat{\boldsymbol  \kernel}^{(s)}
\hat\rho_s$, defining the dressed string-string kernel
$\hat{\boldsymbol \kernel}=({\boldsymbol \eta}+\hat{\boldsymbol \Phi}\mathbb P)^{-1}\hat{\boldsymbol \Phi}(2\mathbb P-1)$ and the dressed string-soliton kernel $\hat{\boldsymbol  \kernel}^{(s)} = - 
({\boldsymbol \eta}+\hat{\boldsymbol \Phi}\mathbb P)^{-1} \hat {\boldsymbol \Phi}^{(s)}$.
We have checked explicitly that the resulting dressed kernel for the  QPs $\AA_a(\theta)$
has the minimal simple structure given in Fig. \ref{figDynkinGen}), leading to
the continuous dressed BA equations Eq. (\ref{BAdensities}).
The dressed kernel $\kernel$ is non symmetric but rather $\kernel_{ba}=\eta_a\eta_b\kernel_{ab}$.

\subsubsection{Charge dressing}
\label{AppendixChargeDressing}

Here we discuss how the charge of the QPs is  affected by dressing.
The presentation (\ref{BAdensities}) of the BA equations is very close to that 
initially obtained by Takahashi \citeOR{Takahashi72,TakahashiBook}
with the difference of the explicitation of the ZF-sea (\ref{defZFS}) 
(the dressed vacuum).
After the particle-hole transformation it is not so clear \emph{a priori} how we should, if we can, 
attach definite charges to the new QPs: indeed, they are 
 holes in a filled \emph{interacting} ZF-sea, and
 the  structure of this filled sea itself (depth, density of states, etc...) 
 depends on the whole configuration ; we furthermore  expect a non-trivial
  effect of the scattering of the new QPs $\AA_j$ on the (configuration-dependent) background seas. 
This question is answered by considering the   charge density 
whose Fourier transform can be expressed at the bare level as $\hat\rho_Q=\sum_{a=s,1,...,\nst }\tilde Q_a\widehat{\tilde\rho}^{(+)}_a$ where $\tilde\rho_a^{(+)}$ are the bare densities entering (\ref{contBAbare}),
with the bare components of the charge vector reading $\tilde Q_s=-1$ (the state $\ket{\Omega}$ is filled with antisolitons) 
and $\tilde Q_j=+2k_j$ with $k_j$ given by Eq. (\ref{defkj}) (since the string $B_{k_j}$  consists in  $k_j$ magnons 
each contributing twice the elementary soliton charge). We then get the dressed charge vector $Q_a$ 
defining the charge in the dressed basis as (we use $ \kernel_{ba}=\eta_a\eta_b\kernel_{ab}$):
\bea
\rho_Q(\theta)&=&\frac1{2\pi}(Q_a\star\rho_a)(\theta)
\label{rhoQphFrac}
\\
\hat {\boldsymbol Q}(\omega)&=&(\boldsymbol {\eta \kernel}(\omega){\boldsymbol \eta}\mathbb P+1-2\mathbb P)\tilde{\boldsymbol Q}
\eea
Observe in Eq. (\ref{rhoQphFrac}) how the convolution with the function $Q_a(\theta)$ 
describes a  spreading of the charge  due to scattering of the dressed QPs with the ZF-sea:
writing $\hat Q_a(\omega)=2\pi\delta(\theta)\hat Q_a(0)+(\hat Q_a(\theta)-2\pi\delta(\theta)\hat Q_a(0))$ 
the second term
 shows that indeed charge is not local in rapidity space. 
 But due to the sort-range nature of the matrix elements of
  $\kernel_{ab}(\theta)$ (they all have exponential decay), for quantities
 that are obtained by integration over a sizeable ($\gg 1$) 
 $\theta-$range one can replace $\kernel_{ab}(\theta)\to(\int\kernel_{ab})\delta(\theta)$. 
  In particular  when doing thermodynamics in the grand 
  canonical ensemble one only needs the \emph{total} charge of the configuration 
  $Q_\conf\vph=\vph\int_{-\infty}^\infty d\theta\; \rho_Q(\theta)$ 
  so  the aforementionned spreading is washed 
  out by $\theta-$averaging. We check explicitly that
  \be
\hat{\boldsymbol Q}(\omega)\underset{\omega\to 0}{\too} (0,...,0,-\qL,\qL)
\ee
with $\qL$ the denominator of $\lambda$ so that finally:
 \be
Q_\conf\vph =\qL\;\Big(N_{\nst }-N_{\nst -1}\Big)=\qL\Big(N_{c}^+-N_{c}^-\Big),
\label{ChargeConfigABAFrac}
 \ee
establishing that effectively  all QPs are neutral 
except for the last two strings carry charge $\pm \qL$, and motivating
 the introduction of the final  notation $\AA_c^+ \equiv \AA_{\nst }$
 and $\AA_c^- \equiv \AA_{\nst -1}$
for the last two QPs.
 This readily  fixes the chemical potentials 
 $\mu_a=\frac{\qL V}{2T}(\delta_{a,c^+}-\delta_{a,c^-})$ for the description
of the finite bias $V$ situation.


\subsubsection{Alternative dressing}
\label{AppendixDressedBis}

In this section one gives a last change of basis (an alternative kind of dressing)
leading to yet another 
presentation of the TBA equations
sometimes used in the literature.
It can be obtained, starting from the dressed basis used in the main text,  by performing a particle-hole transformation on those dressed QPs having $\eta_j=-1$. It leads to a modified dressed vacuum,
and affects the scattering in a way that we will shortly describe, as well as the charge that now reads: $Q=\eta_\alpha \qL(\tilde N_{\nst}-\tilde N_{{\nst}-1})$: the QPs $A_{\nst}$ now has charge $\eta_\alpha\qL$.
Defining $\tilde\epsilon_a=\eta_a\epsilon_a$, $\tilde L^{(\pm)}_a=\ln(1+e^{\pm\big(\mu_a-\epsilon_a)}\big)$,
as well as the densities $\tilde\rho_a^{(\pm)}$ of occupied and empty QP states as  $\tilde\rho_a^{(\pm)} = \rho_a$ if $\eta_a=\pm 1$, $\tilde\rho_a^{(\pm)} = P_a-\rho_a$ if $\eta_a=\mp 1$. This leads to a new presentation
of the BA equations (\ref{BAdensities}) and of the TBA equations (\ref{epsilonTBA}):
\bea
P_a &=& \delta_{as}\,e^\theta + \frac1{2\pi}\big(\tilde\kernel_{ab}\star\tilde\rho^{(\eta_b)} \big)(\theta)
\nl
\tilde\epsilon_a &=& \delta_{as} e^\theta - \frac1{2\pi}\big(\tilde\kernel_{ab}\star \tilde L^{(\eta_b)}\big)(\theta)
\eea
The new BA/TBA system has a \emph{symmetric} dressed Kernel  $\tilde\kernel_{ab}=-\eta_a\kernel_{ab}$,
this nevertherless comes with the price that the BA/TBA system has a less homogeneous form since it 
now involves explicitly
the signs $\eta_a$. This has the physical meaning, that when a QP $a$ goes through the gas of other QPs,
it accumulates a phase-shift that is determined by the  \emph{particles} $b$ it crosses when $\eta_b=1$,
or by the  \emph{holes}  it crosses when $\eta_b=-1$. 
 

\subsection{Asymptotics of the TBA equations}

\label{AppendixAsymptotics}

In this Appendix one gives the explicit form of the asymptotics of the pseudo energies $\epsilon_a(\theta)$
in the limits $\theta\to+\infty$ (UV limit) and  $\theta\to-\infty$ (IR limit), starting from
the TBA equations (\ref{epsilonTBA}).
It is convenient to work with the quantities $X_a = e^{-\epsilon_a(+\infty)}$ and
$Y_a = e^{-\epsilon_a(-\infty)}$ (one also defines $X_c\equiv X_{\nst -1}=X_{\nst }$,
$X_0\equiv X_s$,
and $Y_c\equiv Y_{\nst -1}=Y_{\nst }$, $Y_0\equiv Y_s$).
The analysis in the  diagonal case can be found e.g. in Ref.   
\citeOR{Klassen90} and we give here its off-diagonal generalization.

In the UV limit, the source term $\delta_{as}e^\theta$ in (\ref{epsilonTBA}) dominates and fixes the behavior 
$\epsilon_s(\theta)\sim e^\theta$, implying $X_s=0$. On the other hand, the string pseudo-energies have finite limits.
Using $\frac1{2\pi}\int_\RR d\theta \kernel_{ij}(\theta)=\frac12 I_{ij}$ where $I_{ij}$ is the (signed) incidence matrix of the TBA diagram (see Fig. \ref{figDynkinGen}) yields $\nst -1$ coupled equations for the quantities $X_a\big(\lambda,\frac VT\big)$: 
\be
\begin{array}{rll}
&&\hspace*{-1.5cm} a\notin \{m_{i}-1,m_{i},\nst -2,c\} :
\\
X_a^2&= & (1+ X_{a-1})(1+ X_{a+1})
\\
&&\hspace*{-1.5cm} i<\alpha:
\\
X_{m_i-1}^{2}&= & (1\!+\! X_{m_i-2})(1\!\!+\!\! X_{m_i-1})(1+ X_{m_i}),
\\
X_{m_i}^{2}&= & \frac{1+ X_{m_i+1}}{1+ X_{m_i-1}},
\\
 X^{2}_{\nst -2} &=& (1\!\!+\!\! X_{\nst -3})(1\!\!+\!\! X_{c} e^{\qL V/2T})(1\!\!+\!\! X_c e^{-\qL V/2T})
\\
X_{c}^{2}&=&1+ X_{\nst -2}
\end{array}
\ee
The solution is found to be 
\bea
 X_s&=&0  
\nl
(a\neq s,c)  \qquad\quad  X_a&=&  
\Big(\frac{
\sinh((y_{\ii(a)-1} +k_a )\frac V{2T} )
}{\sinh(y_{\ii(a)-1} \frac V{2T})
}\Big)^2-1
\nl
 X_c&=& \frac{
\sinh((\qL-y_{\alpha-1})\frac V{2T})
}{\sinh(y_{\alpha-1}\frac V{2T})}
\label{EqSolUV}
\eea
where the integer $\ii(a)$ is defined by the fact that  QP $a$
belongs to the family $\Cal F_{\ii(a)}$.

In the IR limit $\theta\to-\infty$, the source term $\delta_{as}e^\theta$ in the TBA equations (\ref{epsilonTBA}) vanishes and the $``s"$ node behaves as if it where a massless node added to the family $\Cal F_1$. All pseudo-energies have finite limit and one readily obtains 
\be
\{Y_s,Y_1...Y_{\nst -2},Y_c\} = \{\tilde X_1,\tilde X_2...\tilde X_{\nst -1},\tilde X_c\}
\label{EqSolIR}
\ee
 where $\tilde X_a=X_a\big(\tilde \lambda,\frac \qL {\tilde \qL }\frac VT\big)$ and $\tilde \lambda=\frac\lambda{\lambda+1}=\frac{\tilde \pL }{\tilde \qL }$ is the rational number with continued fraction $\{\nu_1+1,\nu_2,..,\nu_\alpha\}$.


\subsection{Bulk thermodynamical quantities}

\label{AppendixThermo}

In this Appendix we present the expressions of the thermodynamical quantities
for the gas of interacting dressed QPs $\AA_a(\theta)$.

\subsubsection{Free energy}
\label{AppendixFreeEnergy}

As a fundamental check of the consistency of the approach, one can compare the free energy
of the interacting QP gas at finite $(V,T)$, to that of a free boson under the same conditions 
-- the massless limit of the SG
model being indeed a very complicated representation of a free massless boson.

Writing the  free energy  of the bulk massless SG  model as $F=\frac{LT^2}{2\pi}\bar F$, one decomposes
the free energy as
 $\bar F=\bar E-T\bar S-\sum_a \mu_a\bar N_a$ with the reduced energy $\bar E=\int d\theta \,\rho_s(\theta)\,e^\theta$
 (remember that only the soliton carries energy),
 the particle numbers  $\bar N_a=\int d\theta \rho_a(\theta)$ and the entropy  $-T\bar S = \int d\theta P_a(\theta)\left[f_a(\theta)\ln f_a(\theta)+(1-f_a(\theta))\ln(1-f_a(\theta))\right]$. After a few manipulations
it can be recast as: 
\be
\bar F = \bar E- \widetilde{\bar E} - \sum_a\int d\theta\,P_a L_a
\label{freeEnergyInterm}
\ee
where the last term can be expressed as an ordinary definite integral on the variables $\epsilon_a$ using $P_a=\eta_a\p_\theta\epsilon_a$ and one has introduced the (reduced) ``total pseudo energy" of the system
$\widetilde{\bar E} = \sum_a \int d\theta\, \rho_a(\theta)\,\epsilon_a(\theta)$.

Using a remarkable trick \citeOR{FreeEnergyCalc}, the difference $\bar E- \widetilde{\bar E} $ can be expressed in a closed form involving only 
the limiting values $\epsilon_a(\pm\infty)$: $\bar E- \widetilde{\bar E} =\int  \rho_a(\kernel_{ba}\star L_b) = -\eta_a\int L_a'(\kernel_{ba}\star L_b) 
= -\eta_a[L_a(\kernel_{ba}\star L_b) ]_{-\infty}^\infty+\eta_a\int L_a(\kernel_{ba}\star L_b')$ where the last term of this last form can be shown to be nothing but $-(\bar E- \widetilde{\bar E} )$ using $\int f(g\star h) = \int h(g\star f)$ valid for arbitrary functions $f,h$ and any even function $g$. Finally we obtain 
$\bar E- \widetilde{\bar E} =-\frac12\eta_a [L_a(\kernel_{ba}\star L_b) ]_{-\infty}^{+\infty}$.

Collecting all terms 
one ends up with
the following explicit expression for the free energy per unit length:
\be
\frac{F(V,T)}{L}= \frac{ T^2}{2\pi}\;
 \!\!\sum_{a=s,1,...,\nst } \!\!\!\!
\eta_a 
\Big[\Cal F_{\mu_a}(x)\Big]_{x=X_a}^{x=Y_a},
\label{freenergyGeneral}
\ee
where the function $\Cal F_{\mu}$ is given by
\be
\Cal F_{\mu}(x)=\mbox{Li}_2\big(-xe^{\mu}\big)
+\frac12 \ln(x)\ln(1+xe^{\mu}).
\ee
Here $\mbox{Li}_2(x)=-\int_0^x \frac{dt}{t}\,\ln(1-t) $ is the second polylogarithm function and $(X_a,Y_a)$ are respectively the UV and IR limits of $e^{-\tilde\epsilon_a}$
given in (\ref{EqSolUV}) and (\ref{EqSolIR}), and one reminds that the chemical potential
are given by $\mu_a=\frac{\qL  V}{2T}(\delta_{a,c^+}-\delta_{a,c^-}) = \frac{\qL  V}{2T}(\delta_{a,\nst }-\delta_{a,\nst -1})$.

 The interacting QP gas should be equivalent at the level of the free energy
to a free massless boson with Hamiltonian $H_\ind{free}=\int_0^L \left[  (\p_x\phi(x))^2 - \frac{V}{2} Q(x) \right]$ and periodic boundary conditions, or
\be
H_\ind{free}=\int_0^L \big[  (\p_x\phi(x))^2 - \frac{V}{\sqrt{2\pi(\lambda+1)}}\, \p_x\phi(x) \big]
\label{Hfreeboson}
\ee
where one has used the normalization (\ref{chargedef}) of the charge.
The computation of the free energy is standard and one finds (see e.g. \citeOR{BookDifrancesco}):
\be
\frac{F_{\ind{free}}}{L}
=-\frac{T^2}{2\pi}\bigg(\frac{\pi^2}6+\frac1{\lambda+1} \Big(\frac{V}{2 T}\Big)^2\bigg).
\label{freenergyFB}
\ee

 It is not obvious that the expressions (\ref{freenergyGeneral}) and (\ref{freenergyFB}) do actually coincide for all $\lambda$.
 We checked that it is indeed the case with numerical accuracy $\sim 10^{-15}$ and for all rational $\lambda$'s with lengths $m_\alpha$ up to 15: this expresses the equivalence of the partition functions of the free bose gas and of the interacting QP gas (thanks to a infinity of polylogarithm identities).

\subsubsection{Entropy}
\label{AppendixEntropy}
The entropy carried by each particle can be computed directly using the same arguments. For species $a=s,1,...,\nst $, defining the dimensionless entropy $\overline S_a=\frac{h v_\f}{L \kb^2 T} S_1$, one finds: 
\be
\overline S_a = -\eta_a\Big[  2 \Cal F_{\mu_a}(x) +\mu_a \ln(1+xe^{\mu_a}) \Big]_{X_a}^{Y_a}
\label{defEntropy}
\ee

In the large voltage limit, using (\ref{EqSolUV}, \ref{EqSolIR}) we get that for all $a\neq s$, $X_a$ and $Y_a$ are exponentially large.
Using the limit $ \Cal F_0(x)\underset{x\to\infty}{\too}-\frac{\pi^2}6$, we conclude immediately that $S_a\too 0$. Taken care of the additional chemical potential term, we also get $S_c\to0$: this expresses the freezing of the string entropies in the limit $V/T\to 0$. In the means time the solitons carry all the entropy $\overline S_s=\frac{\pi^2}{3}$ that coincides with the (voltage independent) entropy of the original free thermal boson.

On the other hand, in the low voltage limit, strings do carry a finite  entropy.
The part of the entropy carried by the solitons can be estimated using (\ref{EqSolUV}, \ref{EqSolIR}) : we have $X_s=0$ and $Y_s=3$. Note that these limiting values are independent of $\lambda$. Plugging those values into (\ref{defEntropy}), using that $\sum_a \overline S_a=\frac{\pi^2}{3}$ and after a little algebra,
we get the universal formula (\ref{univEntropy}).

\subsubsection{Particle number}
\label{AppendixCalculationNa}

The reduced particle number for species $a$, $\overline N_a=\frac{2\pi}{LT}\VM{\hat N_a}_{V,T} = \int d\theta \rho_a(\theta)$, can be easily related to the asymptotic values $X_a,Y_a$ Eqs.  (\ref{EqSolUV},\ref{EqSolIR}) of the function $e^{\epsilon_a(\theta)}$
at $\theta=\pm\infty$. Since $\rho_a=-\eta_a\p_\theta L_a(\theta)$ we get 
\be
\overline N_a\Big(\lambda;\frac{\barV}{\barT}\,\Big) = \eta_a \ln\bigg(\frac{
1+Y_a e^{\mu_a}
}{1+X_a e^{\mu_a}}\bigg).
\label{defNa}
\ee

\subsubsection{Average charge}

The average charge accumulated in the system at finite $(V,T)$ can be of course  obtained by differentiating
the free energy (\ref{freenergyGeneral}) with respect to $V$. But it can also more directly be computed
by just counting ($\qL $ times) the difference in the occupation number of charged QPs  $\overline  Q\equiv\frac{2\pi}{LT}\Vm{\hat Q}_{V,T}=\qL \int d\theta (\rho_c^+-\rho_c^+)$. Using $\rho_c^\pm=-\eta_\alpha \p_\theta\vph L_c^\pm$  one arrives at:
\bea
\overline  Q&=&\qL \,\eta_\alpha\Bigg[\ln\bigg(\frac{1+Y_ce^{\frac{\qL V}{2T}}}{1+X_c e^{\frac{\qL V}{2T}}}\bigg)-\;\ln\bigg(\frac{1+Y_c e^{-\frac{\qL V}{2T}}}{1+X_c e^{-\frac{\qL V}{2T}}}\bigg)\Bigg]
\nl
&=&\frac{1}{\lambda+1}\;\frac VT
\label{chargeTBAGen}
\eea
where the last equality uses (\ref{EqSolUV},\ref{EqSolIR}). This is of course consistent with
a direct calculation within the free boson  model
(\ref{Hfreeboson}).


\subsection{Some properties of the continued fraction decomposition}

\label{AppendixNumberThy}

In this Appendix one presents some basic useful elements about the continued fraction defining the 
rational SG parameter $\lambda<1$ (in the repulsive case)  
\be
\lambda=\frac \pL  \qL =\frac{1}{\nu_1+\frac{1}{\nu_2+...\frac{1}{\nu_\alpha}}}\equiv \{\nu_1,\nu_2,...,\nu_\alpha\}.
\ee
The recursive nature of the continued fraction representation can be elucidated by associating to $\lambda$
the function $F_\lambda(x)$
defined by the continued fraction $F_\lambda(x)=\{\nu_1,...,\nu_\alpha,x\}=\frac1{\nu_1+\frac1{\nu_2+...+\frac1{\nu_\alpha+x}}}$. One has $F_\lambda(0)=\lambda$
and $F_\lambda(\infty)=\breve \lambda=\{\nu_1,...,\nu_{\alpha-1}\}$. 
Of course this function can be obtained repeated applications of  elementary functions 
$F_{\frac1\nu}(x)=\frac1{\nu+x}$, i.e. $F_\lambda(x)=F_{\frac1{\nu_1}}\circ...\circ F_{\frac1{\nu_\alpha}}(x)$. 

A connexion with $GL(2,\ZZ)$, the 
 group of $2\times 2$ unimodular matrices,  emerges if one associates to $F_\lambda$ 
a matrix $M_\lambda$ with positive integer coefficients such that 
$F_\lambda(x)=\frac{(M_\lambda)_{11}x+(M_\lambda)_{12}}{(M_\lambda)_{21}x+(M_\lambda)_{22}}$. 
For elementary functions $M_{\frac1\nu}=\Big({\small \begin{array}{cc}0&1\\1&\nu\end{array}}\Big)$ 
and it is easy to check that the matrix associated to $\lambda$ is  $M_\lambda=M_{\frac1{\nu_1}}
M_{\frac1{\nu_1}}...M_{\frac1{\nu_\alpha}}$. 
The function $-F_\lambda(-x)$ is an element of the modular group, and  the matrix 
$M_\lambda$ is  an element of $GL(2,\ZZ)$. Since $\mbox{Det}(M_{\frac1\nu})=-1$, one has immediately  $\mbox{Det}(M_\lambda)=(-1)^\alpha$.
Furthermore, by looking at the definition of the integers $y_i$ (defined in Eq. (\ref{defyi})), one can express 
explicitly the function $F_\lambda$ (and hence the matrix $M_\lambda$) as 
$F_\lambda(x)=\frac{\breve y_{\alpha-2}x+\breve y_{\alpha-1}}{y_{\alpha-1}x+y_\alpha}$ or
\be
F_\lambda(x)=\frac{\pL'x+\pL}{\qL 'x+\qL }
\ee
where the last equality defines the integers $\pL '[\lambda]=\breve y_{\alpha-2}$ and $\qL '[\lambda]=y_{\alpha-1}$. 
From this we deduce the relation 
\be
\pL '\qL -\pL \,\qL '=(-1)^\alpha. 
\label{EqSL2Z}
\ee

This gives a (not so practical) way to obtain the value of the integer $\qL' [\lambda]=y_{\alpha-1}$ starting 
from the integers $(\pL ,\qL )$: 
it is (up to a sign $ (-1)^{\alpha+1}$) the multiplicative inverse of $\pL $ in  the cyclic group 
$\ZZ_\qL =\mathbb Z/(\qL \ZZ)$ (this inverse exists since $\pL \wedge \qL =1$), so that
denoting by $\star_{\qL }$ the multiplication in $\ZZ_\qL $ one has 
$\pL \star_{\qL } \qL '=(-1)^{\alpha+1}$.

\subsubsection{Scattering matrix for the pair of strings $B_{k_{\nst }}(\theta)B_{k_{\nst -1}}(\theta)$}
\label{AppendixScatPair}

Here we gives details about the representation of the U(1) charge symmetry at the level
of bare ABA: the particle-hole operation exchanges the two last strings $B_{\nst }$ and $B_{\nst -1}$.
We first check that the pair of strings $B_{k_{\nst }}(\theta)B_{k_{\nst -1}}(\theta)$
is transparent in the sense that 
\be
S_{a,\nst }(\theta)S_{a,\nst -1}(\theta)=1\quad \forall\,\theta,\quad \forall\, a=s,1,...,\nst.
\ee
To see this, in view of Eq. (\ref{SmatSolString}), it is enough to show that $S_{1,\nst }(\theta)S_{1,\nst -1}(\theta)=1$ $\forall\theta$.  Let us compute the parities: using (\ref{EqSL2Z}) we get $\varepsilon_{\nst }=(-1)^{\pL '}e^{i\pi[((-1)^{\alpha+1}-\pL )/\qL ]}=(-1)^{\pL '+1}$ (the exceptional case $\alpha=1$ has to be treated separately and yields the same result) and similarly $\varepsilon_{\nst -1} 
=(-1)^{\pL +\pL '+1}$ so that $\varepsilon_{\nst }\varepsilon_{\nst -1}=(-1)^{\pL }$. 
When $\pL $ is even, the strings have same parities and the imaginary shifts $\lambda k\frac\pi2$ in the functions $g_k$ are such that $\lambda (k_{\nst }+k_{\nst -1})$ is
an even integer so that $S_{1,\nst }S_{1,\nst -1}=1$. 
On the other hand when $\pL$ is odd, the strings have opposite parities and the scattering matrix of the negative parity string is built on $g^{(-)}_k =g^{(+)}_{k+\lambda^{-1}}$ :  the imaginary shifts are then such that $\lambda (k_{\nst }+k_{\nst -1})+1$ is
again an even integer so that again $S_{1,\nst }S_{1,\nst -1}=1$. 

An  immediate consequence is that (see the definition of the transfer matrices for strings, Eqs. (\ref{defTau}) in Appendix \ref{AppendixABA}) 
 the spectrum for the last two strings is degenerate, 
i.e. a given rapidity $\theta$ is an allowed rapidity for the last string $B_{k_{\nst }}$ 
if and only if it is an allowed
rapidity for the one-but-last string $B_{k_{\nst -1}}$.
 This degeneracy is of course related to 
charge conjugation symmetry, and results in the last two pseudo-energies coinciding, $\epsilon_{c^+}=\epsilon_{c^-}$, in the continuous  limit.

\subsubsection{Computation of $G_0\propto [f_c(\theta)]_{-\infty}^{+\infty}$, and $N_c^\pm\propto [L^\pm_c(\theta)]_{-\infty}^{+\infty}$}
\label{AppendixDeltaFc}

Here we compute the the quantity $\Delta f_c = f_c(+\infty)-f_c(-\infty)$
that determines the high temperature linear conductance. Taking the limit $V\to 0$ in Eqs.  (\ref{EqSolUV}) and (\ref{EqSolIR}), one has $X_c =\frac{\qL -\qL '}{\qL '} $ and $Y_c=\frac{\tilde \qL -\tilde \qL '}{\tilde \qL '}$. 
From this we deduce $f_c(\infty)=\frac{1}{1+X_c^{-1}} = 1-\frac{\qL '}\qL $ and a similar expression for 
$f_c(-\infty)$ so that $\Delta f_c=\frac{\tilde \qL '}{\tilde \qL }-\frac{\qL '}{\qL }$. Since $\tilde \lambda=\frac{\lambda}{\lambda +1}$, we immediately have $\tilde \qL = \pL +\qL $, while $\tilde \qL '$ is determined by the relation
 $\pL \star_{\pL +\qL } \tilde \qL '=(-1)^{\alpha+1}$: but this does not give an explicit value for $\tilde \qL '$. 
To proceed, we consider the function $F_{\tilde \lambda}(x)$: one has 
$F_{\tilde\lambda}=\frac{1}{1+\frac1{F_\lambda}}=\frac{\pL 'x+\pL }{(\pL '+\qL ')x+\pL +\qL }$ 
showing that $\tilde \qL '=\pL '+\qL '$. After a little algebra we conclude that 
\be
\Delta f_c=\frac{(-1)^\alpha}{\qL ^2(\lambda+1)}.
\label{DeltaFc}
\ee
This yields the expected result for the linear conductance $G_0(T)=\p_V I(V,T)\big|_{V=0}$
\be
\frac{h}{e^2}G_0(\infty)=-\,\eta_\alpha\, \qL ^2\Delta f_c=\frac1{\lambda+1},
\label{G0infinity}
\ee 
which is indeed a continuous function of $\lambda$.

Using the same tricks one calculates the number of  occupied charged strings per unit length,
$\frac{2\pi}{LT}N_c^\pm=\bar N_c^\pm = \eta_\alpha \rho_c^\pm = -\eta_\alpha[\ln\big(1+e^{-\epsilon_c\pm\qL L/(2T)}\big)]_{-\infty}^\infty = \eta_\alpha\ln\bigg(
\frac{1+e^{\pm\frac{\qL V}{2T}}Y_c}
{1+e^{\pm\frac{\qL V}{2T}}X_c}
\bigg)$. We give the expressions in  the two limits of  small and large voltage w.r.t. the temperature:
\be
\begin{array}{lll}
V\ll T&  &
{\displaystyle N_c^+=N_c^-=\eta_\alpha \;\ln\bigg(1+\frac{\eta_\alpha}{\qL(\pL'+\qL')}\bigg),
}
\vspace*{.1cm}
\\
V\gg T& &
{\displaystyle N_c^+ = \frac1{\qL}\;\frac{1}{\lambda+1}\,\frac VT
\;\;;\quad N_c^-=0.
}
\end{array}
\label{AsymptNcpm}
\ee
At large voltage the number of charged particles in the thermodynamical limit shows an expected  suppression $\propto \qL^{-1}$ (on the basis of charge counting) of the number of  charged particles. 
At small voltage and for large $\qL$
we get  $N_c^\pm\simeq\frac1{\qL\qL'}\;\frac1{\lambda'+1}$
where again $\lambda'=\{\nu_1,...,\nu_{\alpha-1}\}$,
so that  for  complex 
fractions $\lambda$ with large $\alpha$, 
the suppression $\propto\qL^{-2}$ of charged QPs  is even more pronounced.


\subsection{Attractive case}

\label{AppendixAttractive}

In this Appendix one considers the attractive case $\lambda>1$, 
where the analysis can be carried out in a similar way.
The main difference is that in addition to solitons/antisolitons, the spectrum contains
neutral boundstates, the breathers. There are $ \nu_0 =[\lambda]$ distinct breathers, 
whose mass parameter scales as the soliton/antisoliton mass $m_s$ according to 
$m_b=2m_s \sin\frac{\pi b}{2\lambda}$, $b=1,..., \nu_0 $.

The scattering data (\ref{defa}-\ref{defPhiZ}) has to be complemented with 
the breather-soliton and breather-breather scattering, which turns out to be diagonal
with \citeOR{Zamolodchikov79}:
\bea
S_{bs}&=&\frac{\sinh\theta+i\cos\frac{b\pi}{2\lambda}}{\sinh\theta-i\cos\frac{b\pi}{2\lambda}}
\;\prod_{b'=1}^{b-1}\Big(g^{\frac12}_{\frac{b-2b'}{\lambda}-1}(\theta)\Big)^2
\nl
S_{b,b'} &=& \frac{\sinh\theta+i\sin\frac{(b+b')\pi}{2\lambda}}{\sinh\theta-i\sin\frac{(b+b')\pi}{2\lambda}}
 \;\;\frac{\sinh\theta+i\sin\frac{(b-b')\pi}{2\lambda}}{\sinh\theta-i\sin\frac{(b-b')\pi}{2\lambda}}\nl
 &&\;\;\;\;\times\prod_{\ell=1}^{b'-1} \Big(
 g^{\frac12}_{\frac{b'-b-2\ell}{\lambda}}(\theta)     \,   g^{\frac12}_{\frac{b'+b-2\ell}{\lambda}-2}(\theta)
 \Big)^2
\label{SmatBreathers}
\eea
where $b\geq b'$ in the last line and the functions $g^{\frac12}_k$ are defined from the function $g^{(+)}_k$ in (\ref{defgfunction}) with the replacement $\lambda\to\frac12$.
Taking the Fourier transform of the log derivative of Eqs.(\ref{SmatBreathers}) yields
the bare scattering elements $\hat \Phi_{ij}=\hat \Phi_{ji}=\frac{1}{2i\pi}\int d\theta\,e^{-i\omega\theta}\p_\theta \ln S_{ij}(\theta)$, $i,j=1,2,..., \nu_0 ,s$ (note the tilda over the first entry, introduced for later convenience):
\bea
\widehat{\tilde\Phi}_{s,s} &=& 
\Cal A\;\frac{\sinh\big(\frac{\pi\omega}{2\lambda}(\lambda-1)\big)}
{2\cosh\big(\frac{\pi\omega}{2\lambda}\big)}
\quad ; \quad 
\Cal A= \frac{\coth\big(\frac{\pi\omega}{2\lambda}\big)}{\cosh\big(\frac{\pi\omega}{2}\big)}
\nl
\hat\Phi_{b,s} &=& - \Cal A\; \sinh\big(\frac{\pi\omega}{2\lambda}b\big)
\label{TFSmatBreathers}
\\
\hat\Phi_{b,b'}&=& \delta_{b,b'}- 2\Cal A\;\cosh\big(\frac{\pi\omega}{2\lambda}(\lambda-b)\big)\,\sinh\big(\frac{\pi\omega}{2\lambda}b'\big)
\nonumber
\eea
where in the last line $b\geq b'$ is assumed.

To proceed we now need to treat the off-diagonal scattering of solitons/antisolitons 
by introducing strings within the ABA.
Note, however, that since the scattering (\ref{SmatBreathers}) is diagonal,
it results that,  in the transfer matrix for solitons the part depending on the breathers
can be factorized and is a mere (scalar) prefactor, so that the string-breather
scattering is \emph{trivial}, $\Phi_{b,str}\equiv 0$. 

It is easy to show that the whole string structure is independent  of the presence of breathers.  Indeed, writing $\lambda= \nu_0 +\lambda_f$, where $\lambda_f<1$ is the fractional part of $\lambda$, 
the  scattering matrix $S_{1s}$ between solitons and the $1^\ind{th}$ string reads:  $S_{1s}^{(\lambda)}(\theta)=(-1)^{ \nu_0 }\,S_{1s}^{(\lambda_f)}(\frac{\lambda}{\lambda_f}\theta)$ (we mention explicitly, by a subscript, the SG parameter). 
Now $S_{1s}$  is the building block (see Eqs. (\ref{SmatSolString},\ref{SmatStringString})) for 
any $S_{a,a'}$ ($a,a'\in\{s,1...\nst \}$, $(a,a')\neq(s,s)$),
Hence, up to a rescaling of $\theta$, all the scattering  involving   strings  is that of the repulsive model with SG parameter $\lambda_f$, even after dressing, $\hat\kernel^{(\lambda)}_{a,a'}(\omega)=\hat\kernel_{a,a'}^{(\lambda_f)}\big(\frac{\lambda_f}{\lambda}\omega\big)$ from which
$\kernel^{(\lambda)}_{s,1}(\theta)=\kernelem_{\lambda_f/\lambda}(\theta)$. 

We now bring the TBA system to its final form, by dressing the breather sector. First, the string dressing operation is the same as in the $\lambda_f$ case, we arrive at the following equations for the
pseudo energies ($\tilde\epsilon_b$,$\epsilon_s$) of the breathers and  soliton  (the tilda is introduced for later convenience):
\bea
\tilde\epsilon_b^\ind{br} &=& m_be^\theta -\frac{1}{2\pi}\Phi_{bb'}\star \tilde L^\ind{br}_{b'} -\frac{1}{2\pi}\Phi_{bs}\star  L_{s}^{(+)}
\label{bareTBAbr}
\\
 \epsilon_s &=& m_s e^\theta-\frac{1}{2\pi}\Phi_{sb}\star \tilde L^\ind{br}_{b} -\frac{1}{2\pi}\Phi_{ss}\star   L_{s}  
-\frac{1}{2\pi}\kernelem_{\lambda_f/\lambda}\star L^{\ind{str}}_{1} 
\nonumber
\eea 
where the soliton-soliton scattering is renormalized by strings and reads 
 $\Phi_{ss}=\tilde\Phi_{ss}-\Delta_\ind{str}$ where the  shift $\Delta_\ind{str}$ due the dressing on the string sector can be easily identified without calculations: in the repulsive case, the final TBA system has a vanishing soliton-soliton scattering, so that $\Delta_\ind{str}^{(\lambda_f)} (\theta)= \tilde\Phi_{ss}^{(\lambda_f)}(\theta)$, yielding:
 \be
\hat \Phi^{(\lambda)}_{ss} = \widehat{\tilde\Phi}^{(\lambda)}_{ss}-\widehat{\tilde\Phi}^{(\lambda_f)}_{ss} = \Cal A 
 \;\frac{\sinh\big(\frac{\pi\omega}{2\lambda} \nu_0 \big)}
{2\cosh\big(\frac{\pi\omega}{2\lambda}\lambda_f\big)}
 \ee
Now defining the projectors $\mathbb{P}_\ind{br}$ on breathers and $\mathbb{P}_\ind{m}$ on massive particles (i.e. excluding strings), and introducing the matrix $\widetilde {\hat  \kernel}_{ij}(\omega) = \mathbb{P}_{\ind{br},i}\mathbb{P}_{\ind{br},j}(\delta_{i,j+1}+\delta_{i,j-1}) \;\hat \kernelem_{1/\lambda}(\omega)
+ \delta_{i, \nu_0 }\delta_{j, \nu_0 } \;\frac{\hat \kernelem_{\lambda_f/\lambda}(\omega)\hat \kernelem_{1/\lambda}(\omega)}{\hat \kernelem_{(1-\lambda_f)/\lambda}(\omega)}+(\delta_{i,  \nu_0  }\delta_{j,s}+\delta_{j, \nu_0 }\delta_{i,s})\;\hat \kernelem_{\lambda_f/\lambda}(\omega)
$, we have the following identity: $\mathbb{P}_\ind{m}(1-\widetilde {\boldsymbol{\hat \kernel}} \mathbb{P}_\ind{br})\boldsymbol{\hat \Phi}\mathbb{P}_\ind{m} = \mathbb{P}_\ind{m}\widetilde {\boldsymbol{\hat \kernel}} \mathbb{P}_\ind{m}$.  This identity simplifying the breather sector  was first used in the diagonal SG model \citeOR{Zamolodchikov91}, and is  in fact very close in nature  to that used in the string sector of the XXZ model\citeOR{Takahashi72}.
Multiplying both sides of Eqs. (\ref{bareTBAbr}) by $(1-\widetilde {\boldsymbol{\hat \kernel} }\mathbb{P}_\ind{br})$, and using $f(\theta)\star e^\theta = \hat f(\omega=-i)\,e^\theta$ valid  when the convolution is well defined, one obtains:
\bea
\tilde \epsilon^\ind{br}_b &=&  \frac{1}{2\pi}\widetilde \kernel_{bb'} \star (\tilde \epsilon^\ind{br}_{b'}+\tilde L^\ind{br}_{b'}) + \frac{1}{2\pi}\widetilde\kernel_{bs}\star L_{s} 
\nl
\epsilon_s &=& \frac{1}{2\pi}\widetilde \kernel_{bs} \star  (\tilde \epsilon^\ind{br}_{b}+ \tilde L^\ind{br}_{b})  
-\frac{1}{2\pi}\kernelem_{\lambda_f/\lambda}\star L^{\ind{str}}_{1} 
\nonumber
\eea
Note that the mass terms have disappeared from the equations ; they are however still present through the boundary conditions on pseudoenergies at $\theta\to+\infty$: $\tilde \epsilon^\ind{br}_b(\theta)\sim m_b e^\theta$ ($b=1,..., \nu_0$) and $\epsilon_s(\theta)\sim  e^\theta$.

We then define the final pseudo energies $\epsilon^\ind{br}_b = \eta_0 \tilde \epsilon^\ind{br}_b$ with the sign $\eta_0=-1$:  noticing  that   $\tilde \epsilon^\ind{br}_b +
\tilde L^\ind{br}_b = L^\ind{br}_b$ and considering the entire spectrum including the strings, we check that the full TBA equations in the attractive case  $\lambda=\lambda_f+  \nu_0  $ can be recast as $\epsilon_i= - \frac{1}{2\pi} \kernel^{(\lambda_R)}_{ji}\star L_j$, where  $\kernel^{(\lambda_R)}$
is the kernel (see Fig. \ref{figDynkinGen}) of the repulsive SG model with SG parameter $\lambda_R=\lambda^{-1}$.  Hence, the TBA system in the attractive case can be obtained from the 
repulsive case $\lambda_R=\lambda^{-1}$ in the following way: the first family of strings of the $\lambda_R$ case corresponds to the breathers of the $\lambda$ case, the $(i+1)^\ind{th}$ family of strings in the $\lambda_R$ case corresponds to the $i^\ind{th}$ family of strings in the $\lambda$ case, the mass term for the soliton is suppressed, and the kernel and signs read:
\be
\kernel^{(\lambda)}(\theta) = \kernel^{(\lambda_R)}(\theta)\quad;\quad \eta^{(\lambda)}_a=-\eta^{(\lambda_R)}_a.
\label{mappingBr}
\ee
Note also that one can obtain a presentation of the TBA system where the breathers' pseudonergies
are positive by moving to the alternate dressed basis introduced in 
Appendix \ref{AppendixDressedBis}.

Adding the impurity in the attractive case is done exactly in the same way as in the repulsive one:
due to the trivial breathers-strings scattering, the structure of the dressed one-particle impurity 
scattering matrix is the same:
all neutral particles scatter diagonally, i.e. $R_{aa}=e^{i\xi_a(\theta)}$ and $\xi_a$
a real odd function of rapidity, and the two charged strings scatter via a $2\times 2$
matrix.
 The Loschmidt echo can be written as $\Cal L=\Cal L_\ind{br}+\Cal L_f$
where the breather contribution reads the same in both the original soliton/antisoliton basis
and in the soliton/strings basis. 
 The other piece $\Cal L_f$  is the contribution 
of the antisolitons in the original soliton/antisoliton, and of the soliton and strings in the dressed basis,
so that
introducing again $\chi^{(\lambda)}_\B(\theta_\B)=-iL^{-1}\p_{\theta_\B}\Cal L_f$, one has
explicitly :
\be
\hat\chi^{(\lambda)}(\omega)=\hat\rho_s(-\omega)\hat \Phi_\B^{(\lambda)}=\hat \rho_{c^-}(-\omega)\hat\Phi_\B^{(\pL )}(\omega),
\label{EqImpAtt}
\ee
allowing to conclude that the impurity scattering of the charged strings coincides again (up to phases)
with that of  the soliton/antisoliton  of a \emph{diagonal} BSG model with integer SG parameter $\pL $, the numerator of $\lambda$. 
 
\subsection{Perturbative calculation}
\label{AppendixKeldyshPT}

In this Appendix one computes the leading behavior of the current at small and large $V,T$
using a Keldysh perturbation theory.

\subsubsection{Perturbative evaluation of the current}

Close to the high energy fixed point, the conductance vanishes, so the tunnelling term $H_\B=\gamma\cos\beta\phi(0)$ can be treated in perturbation. It describes a homogeneous system with conductance $G_{\rm max}=\frac{1}{2\pi(\lambda+1)}$, weak perturbed by back-scattering terms, so that the current can be written as $I=G_{\rm max}V-I_{\rm BS}$ defining the back-scattered current.
An expansion in powers of $\gamma$ will yield the asymptotical
form of the conductance at large $V,T$.
On the other hand, at small $V,T$, a similar calculation
can be carried on in the dual picture, by perturbing 
with  the operator $\tilde H_\B=\tilde \gamma\cos\tilde \beta\phi(0)$ with $\tilde\beta=\frac{8\pi}\beta$
being now irrelevant, and physically describing the tunnelling of electrons in the context of tunnelling 
between Tomonaga L\"uttinger liquids. 

Let us start with the high energy expansion. The high fixed point itself, incorporating the voltage bias, is described by a free boson $\tilde \phi(x,t)=\phi(x,t)-V\sqrt{\frac{1}{8\pi(\lambda+1)}}\,(x-t)$ with Hamiltonian $H_0[\tilde \phi]=\int_\mathbb{R} dx\,(\p_x\tilde\phi)^2$, and the total Hamiltonian reads $H=H_0+ H_\ind{T}$ where the tunneling term reads:
\be
H_T=\gamma \cos\big(\beta\tilde\phi(0,t)-V t\big) = \gamma \OT(t).
\ee 
We want to evaluate the current $
I=-\frac12\Delta_0 \langle Q(x)\rangle$
where $\Delta_0f(x)=f(0^+)-f(0^-)$
is the discontinuity 
at $x=0$ and  the charge density $Q(x)$ is defined in Eq. (\ref{chargedef}). 

In the framework of Keldysh perturbation theory, the average of the back scattered current bears an expansion in powers of $\gamma$ starting at order 2. Since $H_T$ is a relevant perturbation, the $\gamma$-expansion 
will have a meaning only close to the high-energy fixed point, i.e. for $\dr{max}(V,T)\gg \TB$. At the fixed point itself, $I=G_\ind{max} V = \frac{e^2}{h(\lambda+1)}V$ ; finite but small $\gamma$ gives rise to a weak back-scattered current $I_{\rm BS} = G_{\rm max} V-I$:
\bea
I_{\rm BS}=\frac{\gamma^2}{4}\Delta_0\vm{Q(x)}^{(2)}+\Cal O(\gamma^4)\nl
\vm{Q(x)}^{(2)}=\int_{\Ccurl_K} dt\,dt'\,\Cal G_\beta\vph(t,t',-x)
\label{Q2gen}
\eea
where the finite $V,T$ correlator
$
\Cal G_\beta\vph(t,t',-x) = \VM{\OT(t)\;\OT(t')\;  Q(x,t=0)}_{\!0}
$
is evaluated in the unperturbed theory ($\gamma=0$).
A building block of the correlator is 
$
h(t-t')\equiv e^{i\frac\pi {\lambda+1}\sgn(t-t')}\langle e^{\pm i\beta\tilde\phi(t)}e^{\mp i\beta\tilde\phi(t')}\rangle
=|t-t'|^{-\frac2{\lambda+1}}
$ at $T=0$, with the finite temperature expression obtained
 via the usual mapping\citeOR{BookDifrancesco} of the plane geometry onto the cylindrical with circumference  $T^{-1}$, $z\to\omega = \frac{1}{2i\pi T}\ln z$:
\be
h(t-t')= \frac{(\pi T)^\frac2{\lambda+1}}{\sinh(\pi T|t-t'|)^{\frac2{\lambda+1}}}.
\ee
Note that there will be short-distance divergencies when integrating over $t-t'$ for $\lambda\leq 1$.
As already argued\citeOR{FLS-PRB}, one can instead do the calculations at $\lambda>1$, where the integral converges, and then analytically continue to other values of $\lambda$ keeping in mind that the final results for physical quantities should be smooth, analytical functions of the parameter $\lambda$. 
The (half) Fourier transform of this function, $h^+(k)\equiv \int_0^\infty dt\,e^{-ikt}h(t)$, will also be useful:
\be
h^+(k) = 
(2\pi T)^{ 1-\frac{2\lambda}{\lambda+1}}\;\Gamma\Big(\frac{\lambda-1}{\lambda+1}\Big)\frac{\Gamma(\frac{1}{\lambda+1}+i\frac k{2\pi T})}{\Gamma(\frac{\lambda}{\lambda+1}+i\frac k{2\pi T})} 
\label{gbThetak}
\ee
Next, the zero temperature correlator
$\langle
e^{\pm i\beta\tilde\phi(t)}e^{\mp i\beta\tilde\phi(t')}\,\p_x\tilde\phi(-x)
\rangle
=\frac\beta{4\pi} \;(i(t-t'))^{-\frac{2\lambda}{\lambda+1}}\;\big(\frac1{i(t+x)}+\frac1{i(t'+x)}\big)
$
is brought to finite temperature by the usual mapping, and finally the Keldysh correlator in (\ref{Q2gen}) reads:
\bea
&\Cal G_\beta(t,t',-x) = -\frac{1}{2\pi}\,e^{-i\frac\pi {\lambda+1}{ {\rm sign}}(t-t')}\sin(V(t-t'))\;h(t-t')
\nl
&\hspace*{.1cm}\times
\Big[\frac{\pi T}{\tanh\big(\pi T(x+t+i\eta 0^+)\big)}-\frac{\pi T}{\tanh\big(\pi T(x+t'+i\eta'0^+)\big)}\Big]
\label{CorrCurrent2}
\eea
where  $(\eta,\eta')=\sgn(\Im(t,t'))\in\{\pm\}$ label the different Keldysh paths contributing to $\Ccurl_K$  in (\ref{Q2gen}). 
Since $Q$ is a conserved quantity, one has simple  $x$ dependence in (\ref{CorrCurrent2}), that
allows for an important simplification upon Fourier transforming w.r.t. $x$:
$\hat {\cal G}_\beta(t,t',k)=
\frac{i}{2\pi}e^{-i\frac\pi {\lambda+1}{\rm sign}(t-t')} \;\sin(V(t-t'))\;g(t-t')
 \Big[C_\eta(k)e^{ikt}  -C_{\eta'}(k)e^{ikt'}\Big]$ where
$C_\eta(k)
= \frac{\eta}{1+e^{-\eta k/T}}$.
 
The next step is to perform the integration over $t,t'$. The Keldysh integration can be split in four pieces, $\Ccurl=\bigcup\limits_{(\eta,\eta')\in\{\pm\}} \Ccurl_{(\eta,\eta')}$, where $\Ccurl_{(-,-)}$ 
(respectively $\Ccurl_{(+,+)}$) corresponds to the two times $t<t'$ lying  
on the forward (respectively backward) branch, $\Ccurl_{(-,+)}$ corresponds to half the piece with $t$ lying on the forward branch and $t'$ on the backward branch with $t<t'$, and last $\Ccurl_{(+,-)}$ corresponds to the other half of the mixed forward/backward contour, with $t>t'$. 
All those expressions can be combined  to yield: 
\bea
\int_{-\infty}^\infty dx\vm{Q(x)}^{(2)}e^{-ikx}&=& -i \sum_{\eta,\eta'} \eta\eta'\;e^{-i\eta\frac\pi {\lambda+1}}
\\
\times\; \int_{-\infty}^0 dt' e^{ikt'}
\int_{-\infty}^0 &d\tau&\;\sin (V\tau) h(\tau)[C_\eta e^{ik\tau}-C_{\eta'}] 
\nl
=-i
\sin\Big(\frac\pi {\lambda+1}\Big)\;(\hat h^+(V)
&-&
\hat h^+(-V))\;\times\int_0^\infty dx e^{-ikx} 
\nonumber
\eea
where the discontinuity appearing in $\vm{Q(x)}^{(2)}$ (the last term is the integral representation of the Heaviside function) is reminiscent from the breaking of translation symmetry by the impurity interaction term. After a little algebra this yields the following expression for the current:
\bea
I_{\rm BS}^{(2)}&=&
-\frac{\gamma^2}{2}\;
(2\pi T)^{1-\frac{2\lambda}{\lambda+1}}\sin\Big(\frac{\pi}{\lambda+1}\Big)
\Gamma\Big(\frac{\lambda-1}{\lambda+1}\Big)
\nl
&&
\times\; \Im\bigg[  \frac{\Gamma(\frac{1}{\lambda+1}+i\frac V{2\pi T})}{\Gamma(\frac{\lambda}{\lambda+1}+i\frac V{2\pi T})} \bigg].
\label{IBSUV}
\eea

\subsubsection{Conductance in the high energy limit}

The expression (\ref{IBSUV}) is valid as soon as $V\gg\TB$ or $T\gg\TB$ but the ratio $V/T$ is arbitrary. Two particular limits
are interesting:
\\$\bullet$ Low $V$ \emph{i.e.} $\frac VT\ll1$.
Defining $K=\frac{1}{\lambda+1}=\frac{\beta^2}{8\pi}$ (it has the physical meaning of the TLL parameter in the case where the BSG model describes tunnelling between TLL's) and using $\Im\big[  \frac{\Gamma(K+iy)}{\Gamma(1-K+iy)} \big]\underset{y\ll 1}{\simeq} -y\cos(\pi K)\Gamma(K)^2$ one gets the linear conductance 
\be
G_\ind{BS}(V=0,T) = \frac{\pi \gamma^2 {\rm B}(K,K) }4 (2\pi T)^{2K-2}
\ee
\\$\bullet$ Low $T$ \emph{i.e.} $\frac VT\gg1$. Using $\Im\big[  \frac{\Gamma(K+iy)}{\Gamma(1-K+iy)} \big]\underset{y\gg 1}{\simeq} -y^{2K-1}\cos(\pi K)$, the zero temperature non-linear conductance reads 
 \be
 G_\ind{BS}(V,T=0) = \frac{\pi\gamma^2(2K-1)}{4\Gamma(2K)} V^{2K-2}.
 \ee 
Therefore a universal  ratio (in the sense that it no longer depends on $\TB$)  can be defined, 
 \be
 {\mathscr R}_K\shS{=}{.4936}\frac{G_\ind{BS}(V\shS{=}{.6}0,T\shS{\gg}{.6}\TB)}{G_\ind{BS}(V\shS{\gg}{.6}\TB,T\shS{=}{.6}0)}=\frac{\Gamma(K)^2}{2K\shS{-}{1}1}\bigg(\shS{{ eV \over 2\pi\kb T }}{.6}\bigg)^{\!2-2K}
 \label{UnivRatioK}
 \ee
The non-universal relation between $\gamma$ and $\TB$ (it depends on the high regularization scheme) 
can be fixed in the scaling limit where $\TB$ is infinitely smaller than the bandwidth $W$ using the zero-temperature solution \citeOR{FLS-PRB}  : 
\be
\frac{G_\ind{BS}(V,T\shS{=}{.6}0)}{ G_\ind{max}}
\shS{
 \underset{V\gg \TB}{\simeq}
}{1}
\textstyle{\frac{K(2K-1)\,{\rm B}(K,\frac12)}{2}}
\big({\rm B}({\textstyle \frac 12,\frac12\shS{+}{1}\frac1{2\lambda}})\frac{eV}{\kb \TB}\big)^{2K-2}
\ee
 with $G_\ind{max}=\frac{e^2}{h(\lambda+1)}=G(V,\infty)=G(\infty,T)$ the high-energy maximal conductance. Using the universal ratio (\ref{UnivRatioK}) finally leads to the finite temperature $(T\gg \TB)$ conductance:
 \be
 \frac{G_{\ind{BS}}(V\shS{=}{.6}0,T)}{G_\ind{max}}
 \underset{T\gg \TB}{\simeq}
 {\textstyle \frac {K\Gamma(K)^2{\rm B}(K,\frac12)}{2}}
\big({\rm B}({\textstyle \frac 12,\frac12\shS{+}{1}\frac1{2\lambda}})
\textstyle\frac{2\pi T}{ \TB}\big)^{2K-2}
\ee
 which is nothing but Eq. (\ref{G0AsymptUV}).

\subsubsection{Conductance in the low energy limit}

 On the other hand, in the low energy limit $V,T\ll \TB$ the effective coupling constant $\gamma$ blows up and perturbation theory in $\gamma$ breaks down. In this limit the boson $\phi(x=0)$ tends to be pined to the minima of the boundary potential and it results that the current is small. 
 A valid perturbative approach is possible  in a dual picture \citeOR{LesageSaleur99}, where the system is again described by a free boson $\tilde\phi$ perturbed by a dual boundary term  $\tilde H_\B=\tilde\gamma\cos(\tilde\beta\tilde \phi(0))$
 with $\tilde\beta=\frac{8\pi}{\beta}$. This term is now irrelevant in the renormalization group sense,
 so that perturbation theory in $\tilde\gamma$ is now valid. All the details of the calculations are the same, except
 $K=\frac1{\lambda+1}=\frac{\beta^2}{8\pi}$ is replaced by $\tilde K=\frac 1K$ or $\tilde\lambda =-\frac\lambda{\lambda+1}$ so the current reads:
 \bea
I^{(2)}&=&
\frac{\tilde\gamma^2}{2}\;
(2\pi T)^{2\lambda}\sin(\pi\lambda)
\Gamma(-1-2\lambda)
\nl
&&
\times\; \Im\bigg[  \frac{\Gamma(\lambda+1+i\frac V{2\pi T})}{\Gamma(-\lambda+i\frac V{2\pi T})} \bigg].
\label{IIR}
\eea
 
 We again find a universal ratio relating the zero voltage and zero temperature limits:
 \be
 \widetilde{\mathscr R}_K=\hspace*{-0.105mm}
 \frac{G_\ind{}(V\shS{=}{.3} 0,T\shS{\ll}{.3}\TB)}{G_\ind{}(V\shS{\ll}{.3}\TB,T\shS{=}{.3}0)}
 =\frac{\Gamma\big(\frac1K\big)^2}{\frac2K-1}\bigg(\!{e V \over 2\pi \kb T }\!\bigg)^{2-\frac2K}
 \label{UnivRatioKtilde}
 \ee
Combining this last result with the zero temperature limit\citeOR{FLS-PRB}:
\be
\frac{G(V,T\shS{=}{.6}0)}{ G_\ind{max}}
\underset{V\ll\TB}{\simeq} \frac{{(\frac2K-1)\rm B}(\frac1K,\frac12)}{2K}\big({\rm B}({\textstyle \frac 12,\frac12\shS{+}{1}\frac1{2\lambda}})
\textstyle\frac{eV}{\kb \TB}\big)^{\frac2K-2}
\ee
we thus arrive at:
 \be
 \frac{G(V\shS{=}{.6}0,T)}{G_\ind{max}}
 \underset{T\ll \TB}{\simeq}
 {\textstyle \frac {\Gamma(\frac 1K)^2{\rm B}(\frac 1K,\frac12)}{2K}}
\big({\rm B}({\textstyle \frac 12,\frac12\shS{+}{1}\frac1{2\lambda}})
\textstyle\frac{2\pi T}{ \TB}\big)^{\frac2K-2}
\ee
which is nothing but the zero voltage limiting form (\ref{G0AsymptIR}).

\bibliographystyle{plain}

\begin{thebibliography}{9}




\bibitem{MoreIsDifferent}
P.W. Anderson,
Science  {\bf 177},  4047 (1972).

\bibitem{Woelfle08}
For a recent review see P. W\"olfle, 
Rep. Prog. Phys. {\bf81}, 032501  (2008).

\bibitem{FermiLiquids}
L.D. Landau and E.M. Lifshitz, 
{\it Statistical Physics}, Part 2, Pergamon, Oxford, 1980 ; 
P. Nozi\`eres, 
{\it Theory of interacting Fermi systems},
 Benjamin, New York, 1964.

\bibitem{BCS} 
J. Bardeen, L. Cooper, and J. Schrieffer, 
Phys. Rev. {\bf 106}, 162 (1957).

\bibitem{BookFetterWalecka}
A. Fetter and J.D. Walecka,
{\it Quantum theory of many-particle systems},
Dover Pub., New York, 2003.


\bibitem{Luttinger63}
J. Luttinger, 
J. Math. Phys. {\bf 4}, 1154 (1963).

\bibitem{BookGiamarchi}
T. Giamarchi,
{\it Quantum physics in one dimension}, Oxford University  Press, Oxford, 2003.

\bibitem{Auslaender02}
O. M. Auslaender, 
Science {\bf295}, 825  (2002).

\bibitem{Jompol09}
Y. Jompol, C.J.B. Ford, J.P. Griffiths, I. Farrer, G.A.C. Jones, D. Anderson, D.A. Ritchie, T.W. Silk, and A.J. Schofield, 
Science {\bf 325},  597 (2009).


\bibitem{Saminadayar1997} 
L.Saminadayar , Y. Jin, B. Etienne, and D.C Glattli, 
Phys. Rev Lett. {\bf 79}, 2526 (1997).

\bibitem{dePicciotto1997}
R. de Picciotto  \emph{et al.}, 
Nature {\bf 389}, 162  (1997).


\bibitem{BookEsslerHubbard}
F. H. Essler, H. Frahm, F. G\"ohmann, A. Kl\"imper, and  V.E. Korepin, 
{\it The one-dimensional Hubbard model},
 Cambridge University Press, Cambridge, 2005.


\bibitem{FRG-Review}
W. Metzner, M. Salmhofer, C. Honerkamp, V. Meden, and K. Sch\"onhammer,
Rev. Mod. Phys. {\bf84}, 299 (2012).

\bibitem{Meden08}
V. Meden, S. Andergassen, T. Enss, H. Schoeller, and K. Schoenhammer,
 New J. Phys. {\bf 10},  045012 (2008).

\bibitem{andergassen10}
S. Andergassen, V. Meden, H. Schoeller, J. Splettstoesser, and M.R. Wegewijs
Nanotech. {\bf 21},  272001 (2010). 

\bibitem{Schmitteckert14}
P. Schmitteckert, S.T. Carr, and H. Saleur,
Phys. Rev. B{\bf89}, 081401  (2014).

\bibitem{Aristov14}
D.N. Aristov and P. W\"olfle,
Phys. Rev. B {\bf 90},  245414 (2014).





\bibitem{KaneFisher92}
C.L. Kane and M.P. A. Fisher,
Phys. Rev. B {\bf46}, 15233 (1992).



\bibitem{FLS-PRL}
P. Fendley, A.W.W. Ludwig, and H. Saleur,
Phys. Rev. Lett. {\bf74}, 3005  (1995).


\bibitem{FLS-PRB}
P. Fendley, A.W.W. Ludwig, and H. Saleur,
Phys. Rev. B{\bf52}, 8934  (1995).



\bibitem{SafiSaleur}
I. Safi and H. Saleur,
Phys. Rev. Lett. {\bf93}, 126602  (2004).


\bibitem{Schmid83}
A. Schmid, 
Phys. Rev. Lett. {\bf51},  1506 (1983).

\bibitem{Guinea85}
F. Guinea, V. Hakim, and A. Muramatsu, 
Phys. Rev. Lett. {\bf54},  263  (1985).


\bibitem{BookWeiss}
U. Weiss, {\it Quantum dissipative systems}, 4th Edition, World Scientific, 2012.

\bibitem{Fisher85}
M.P.A. Fisher and W. Zwerger, 
Phys. Rev. B{\bf32},  4190 (1985).


\bibitem{Kane95}
C.L. Kane and M.P.A. Fisher,
Phys. Rev. B {\bf52},  17393  (1995).


\bibitem{Chang03}
A.M. Chang,
Rev. Mod. Phys. {\bf75}, 1449  (2003).





\bibitem{Lukyanov2007}
S. Lukyanov and P. Werner,
J. Stat. Mech. {\bf 2007}, P06002  (2007).

\bibitem{Callan90}
C. G. Callan and L. Thorlacius,
Nucl. Phys. B{\bf329}, 117 (1990).

\bibitem{Sen02}
A. Sen, 
JHEP {\bf 0204}, 048 (2002).



\bibitem{SchillerHershfieldFreeSystems}
A. Schiller and S. Hershfield,
Phys. Rev. Lett. {\bf77},  1821 (1996) ;
Phys. Rev. B{\bf 51}, 12896(R)  (1995);
{\bf58},  14978  (1998);
{\bf62}, R16271(R)  (2000).

\bibitem{KomnikGogolin03}
A. Komnik and A.O. Gogolin, 
Phys. Rev. Lett. {\bf90},  246403  (2003).

\bibitem{SelaAffleck09}
E. Sela and I. Affleck,
Phys. Rev. Lett. {\bf102}, 047201  (2009).



\bibitem{Bazhanov99}
V.V. Bazhanov, S.L. Lukyanov, and A.B. Zamolodchikov,
Nuclear Physics B {\bf 549 }, 529 (1999).


\bibitem{NoteSDIRLM} To be complete, one should add to the list the solution to the self dual interacting resonant level model \cite{Boulat08}, which is actually a (non-trivial) unitary deformation of the BSG model at the particular
 reflectionless value $\frac{8\pi}{\beta^2}=4$ 
where the BSG model enjoys an enlarged SU(2) symmetry.

\bibitem{Boulat08}
 E. Boulat, H. Saleur, and P. Schmitteckert
Phys. Rev. Lett. {\bf101}, 140601   (2008).



\bibitem{Anthore18}
A. Anthore, Z. Iftikhar, E. Boulat, F.D. Parmentier, A. Cavanna, A. Ouerghi, U. Gennser, and F. Pierre,
Phys. Rev. X {\bf 8}, 031075  (2018).


\bibitem{Takahashi72}
M. Takahashi and M. Susuki,
Prog. Theor. Phys. {\bf48}, 6,  2187 (1972).

\bibitem{TakahashiBook}
M. Takahashi, {\it Thermodynamics of one-dimensional solvable models},
Cambridge University Press, Cambridge, 2005.



\bibitem{NoteChargeNormalization}
\NoteChargeNormalization





\bibitem{Ghoshal94}
S. Ghoshal and A. Zamolodchikov, 
Int. J. Mod. Phys. A{\bf}09, 21,  3841 (1994).



\bibitem{Zamolodchikov79}
A.B. Zamolodchikov and A.B. Zamolodchikov,
 Annals of Physics {\bf120}, 253   (1979).
 
 
\bibitem{Zamolodchikov90}
A.B. Zamolodchikov,
Nuclear Physics B{\bf342}, 695 (1990).

\bibitem{Fendley92}
P. Fendley and K. Intriligator,
Nucl.Phys. B{\bf372}, 533  (1992).


\bibitem{KorepinBook}
V. E. Korepin, G. Izergin, and N.M. Bogoliubov, 
{\it Quantum inverse scattering method and correlation functions}, 
Cambridge University Press, Cambridge, 1993.

\bibitem{FaddeevHouches}
L. D. Faddeev, {\it How Algebraic Bethe Ansatz Works for Integrable Model}, 
p. 149 in {\it Sym\'etries Quantiques/Quantum Symmetries: Les Houches, Session LXIV},
A. Connes, K. Gawedzki, and J. Zinn-Justin eds, North-Holland Publishing, 1998.


\bibitem{ContinuedFractions}
L. Lorentzen and H. Waadeland, 
\emph{Continued fractions with applications Studies in computational Mathematics 3}, North-Holland, 1992.


\bibitem{BookDifrancesco}
P. Di Francesco, P. Mathieu, and  D. S\'en\'echal,
{\it Conformal field theory}, Springer, New York, 1997.



\bibitem{BookTsvelikEtAl}
A. Gogolin, A. Nersesyan, and A. Tsvelik,
{\it Bosonization and strongly correlated systems},
Cambridge University Press, Cambridge, 1998.



\bibitem{FendleySaleurWarner}
P. Fendley, H. Saleur, and N. Warner, 
Nucl. Phys. B{\bf430}, 577 (1994).



\bibitem{Fowler82}
M. Fowler and X. Zotos,
Phys. Rev B{\bf25}, 5806 (1982).

\bibitem{Chung83}
G. Chung and Y.C. Chang,
Phys. Rev. Lett {\bf50}, 791 (1983).


 
\bibitem{FendleySaleur94}
P. Fendley and H. Saleur,
Nucl. Phys. B{\bf428}, 681 (1994).

\bibitem{Tateo95}
R. Tateo,
Int. J. Mod. Phys. {\bf A10}, 1357 (1995).



\bibitem{Zamolodchikov91}
A. Zamolodchikov,
Phys. Lett. {\bf B253}, 391 (1991).


\bibitem{NoteCompleteness}
\NoteCompleteness



\bibitem{Hao13}
Wenrui Hao, Rafael I. Nepomechie, and Andrew J. Sommese,
Phys. Rev. E{\bf88}, 052113 (2013).


\bibitem{footnoteDoubly}
\footnoteDoubly



\bibitem{Klassen90}
T.R. Klassen and E. Melzer,
Nucl. Phys. B{\bf338}, 485 (1990).




\bibitem{FreeEnergyCalc}
G.E. Andrews, R.J. Baxter, and P.J. Forrester, J. Stat. Phys. {\bf 35}, 93  (1984) ; 
V.V. Bazhanov, A.N. Kirillov, and N.Y. Reshetikhin, Pisma ZhETF {\bf 46}, 500 (1987).


\bibitem{LesageSaleur99}
F. Lesage and H. Saleur, 
Nucl. Phys. B{\bf546}, 585 (1999).


\end{thebibliography}

\end{document}